\documentclass[cernpreprint, texlive=2020, USenglish, texmf, orcidlogo]{atlasdoc}
 
\usepackage{atlaspackage}
\usepackage{atlasbiblatex}

\usepackage{atlasphysics}

\graphicspath{{logos/}{figures/}}
 
\usepackage{atlasheavyion}
 
\usepackage{ANA-HION-2020-10-PAPER-defs}

\usepackage{placeins}
\usepackage{amsmath}
\usepackage{bm}
\usepackage{geometry}
\usepackage{graphicx}
\usepackage{multirow}
\usepackage{dcolumn}
\usepackage{mathtools}
\usepackage{bigfoot}
\usepackage[export]{adjustbox}
 
\AtlasTitle{
Measurement of muon pairs produced via $\gamma\gamma$ scattering in non-ultraperipheral
Pb+Pb collisions at $\sqrt{s_{_\mathrm{NN}}} = 5.02$ TeV with the ATLAS detector
}
 
\AtlasJournalRef{Phys.~Rev.~C~107~(2023)~054907}
\AtlasDOI{10.1103/PhysRevC.107.054907}
\AtlasAbstract{
Results of a measurement of dimuon photoproduction in non-ultraperipheral
Pb+Pb collisions at $\sqrt{s_{_\mathrm{NN}}} = 5.02$~TeV  are presented.
The measurement uses ATLAS data from the 2015 and 2018 Pb+Pb
data-taking periods at the LHC with an integrated luminosity
of $1.94~\mathrm{nb}^{-1}$.
The $\gamma\gamma \rightarrow \mu^+\mu^-$ pairs are identified via
selections on pair momentum asymmetry and acoplanarity.
Differential cross-sections for dimuon production are measured in
different centrality, average muon momentum and pair rapidity
intervals as functions of acoplanarity and
$k_\perp$, the transverse momentum kick of one muon relative to the other.
Measurements are also made as a function of the rapidity separation of
the muons and the angle of the muon pair relative to the second-order
event plane to test whether magnetic fields
generated in the quark--gluon plasma affect the measured muons.
A prior observation of a centrality-dependent broadening of the
acoplanarity distribution is confirmed.
Furthermore, the improved precision of the measurement reveals a depletion
in the number of pairs having small acoplanarity or $k_{\perp}$
values in more central collisions.
The acoplanarity distributions in a given centrality interval are
observed to vary with the mean-$p_{\mathrm{T}}$ of the muons in the pair,
but the $k_{\perp}$ distributions do not.
Comparisons with recent theoretical predictions are made.
The predicted  trends associated with effects of magnetic fields on
the dimuons are not observed.
}

\author{The ATLAS Collaboration}
 
\PreprintIdNumber{CERN-EP-2022-047}

\hypersetup{pdftitle={ATLAS document},pdfauthor={The ATLAS Collaboration}}
 
\begin{document}
 
\maketitle

\section{Introduction}

% The next lines are included from the .//tex/intro.tex input file
In relativistic heavy ion collisions, the Lorentz contracted electromagnetic
fields of the ions act as a source of high energy quasi-real
photons~\cite{Fermi1924,Weizsacker1934,Williams1935}.
In collisions involving Pb ions at the LHC, the photons emitted coherently
by the incoming nuclei can have energies of up to 80~\GeV\
in the center of mass frame~\cite{HION-2016-02}.
The processes involving the interactions of such coherently produced photons
are typically studied in ultra-peripheral collisions (UPC);
events where the contributions from strong interactions are absent
as they typically occur at impact parameters larger than twice the nuclear radius.
Such UPC \gamgam\ events can lead to exclusive final states with
dileptons~\cite{HION-2016-02,HION-2021-16,CMS-HIN-19-014}.
The \gamgam\ induced processes are also present in heavy-ion collisions
with hadronic interactions, but are more difficult to observe
due to the large amounts of background particles produced in such collisions.
 
Recent measurements of dilepton production via \gamgam\ scattering
in non-ultraperipheral (non-UPC) heavy-ion collisions~\cite{HION-2018-11,Adam:2018tdm}
have stimulated significant interest due to the possibility that such
pairs can be used as electromagnetic probes of the quark--gluon plasma
created in such collisions~\cite{HION-2018-11,Klein:2018fmp,Ye:2018jwq}.
In particular, the alignment of the two leptons in
the transverse plane and/or the lepton-pair transverse
momenta may be modified by the scattering of the leptons off constituents of the plasma or
through interaction with long-range magnetic fields generated by the
charge flow in the plasma.\footnote{See Refs.~\cite{Klein:2018fmp,Klein:2020jom} for arguments that magnetic fields
produced by the spectator charges do not play a significant role. Also see \cite{Wang:2021oqq} which argues that the magnetic field generated in heavy-ion collisions may be weaker than previously thought. }
 
Previous measurements~\cite{HION-2018-11} of the centrality dependence of dimuon acoplanarity
in non-UPC \PbPb\ collisions at the Large Hadron Collider (LHC) have shown
a systematic broadening of the acoplanarity distribution that is consistent
with expectations of electromagnetic scattering of one or both of the
muons in the quark--gluon plasma~\cite{Klein:2018fmp}.
Similarly, a measurement of electron-pair transverse momentum distributions in non-UPC \AuAu\
collisions at the Relativistic Heavy Ion Collider showed a broadening compared to theoretical
expectations for vacuum production~\cite{Adam:2018tdm}. That broadening was found to be consistent with the
expected deflection of the electrons in magnetic fields produced
during the collision~\cite{Ye:2018jwq}.
However, it has also been argued that the effects observed in non-UPC heavy-ion
collisions may result from physics associated with the
initial state, particularly a variation of the transverse momenta of
the initial photons~\cite{Zha:2018tlq} with the impact parameter
of the nuclear collision. That idea led to a prediction~\cite{Brandenburg:2020ozx} that the
acoplanarity of dilepton pairs produced in ultraperipheral \gamgam\
collisions may depend on the breakup of the incoming nuclei, due to
the dependence of the breakup probability on the impact
parameter.\footnote{The predicted effect relates to the ``core'' of the
acoplanarity distribution not the enhancement of the tail due to
dissociative processes seen in Ref.~\cite{HION-2016-02}.}
Such an effect was observed recently~\cite{CMS-HIN-19-014}.
 
A recent alternative  analysis \cite{Klein:2020jom}
proposes a quantum-mechanical extension of the equivalent
photon approximation  using the Wigner functions to describe, simultaneously, the initial-state photon
transverse coordinate and transverse momentum distributions.
Calculations using these photon Wigner functions~\cite{Klein:2020jom,Klusek-Gawenda:2020eja,} show effects
similar to those seen in the data due to  the modulation of the
photon Wigner distributions as a function of transverse momentum
and/or impact parameter with respect to the parent nucleus.
 
Specific tests, not possible with previous data, have been suggested
for evaluating the possible role of magnetic
effects~\cite{Klein:2018fmp, Klein:2020jom}. In particular, it was
proposed to study the dependence of the dilepton acoplanarity
and/or transverse momentum imbalance on the rapidity
separation of the leptons and on the dilepton momentum direction relative
to the impact parameter vector. Magnetic fields are predicted to affect
the leptons in a rapidly increasing way with increasing
rapidity separation between the leptons and to be largest for
leptons emitted perpendicular to the magnetic field, since their direction  in
the transverse plane is correlated with the direction of the impact
parameter vector~\cite{Skokov:2009qp,Deng:2012pc}. In noncentral heavy-ion collisions, the ``elliptic''
second-order event plane angle is well-correlated with the direction
of the impact parameter \cite{Qiu:2011iv}. Thus, a measurement of dileptons as a function of
the angle between the dilepton axis and the second-order event
plane may provide independent sensitivity to magnetic fields \cite{Asakawa:2010bu} that
is crucial to the understanding of possible chiral magnetic
effects \cite{Kharzeev:2013ffa,Kharzeev:2015znc,} in
nucleus--nucleus collisions.
 
In order to experimentally distinguish between the various
explanations of the observed effects in dilepton production
from \gamgam\ scattering in non-UPC heavy-ion collisions,
a larger data set, as well as a broader set
of measurements, is clearly needed. To this end,
this \papertype\ repeats and extends the measurements presented in
Ref.~\cite{HION-2018-11}, taking advantage of the increased integrated luminosity
of the 2018 \PbPb\ run at the LHC and improved trigger selections. It presents results from an ATLAS
measurement of non-UPC production
of \gamgammumu\ pairs in $\sqn = 5.02$~\TeV\ \PbPb\ collisions at the
LHC using a combination of 2015 and 2018 \PbPb\ data sets with a total
integrated luminosity of $1.93~\mathrm{nb}^{-1}$. This integrated luminosity is
about four times larger than that available for the previous
measurement. The results presented here
complement a recent ATLAS measurement~\cite{HION-2016-02}
of exclusive UPC dimuon production performed over a more restrictive
mass range but a factor of eight times larger range in
acoplanarity. Small contributions from QED final-state radiation and
dissociative processes studied in that measurement are both strongly
suppressed in this analysis by kinematic selections
(requirements on the $\alpha$, $\kp$ and $A$ variables discussed below) and obscured by
backgrounds generated in non-UPC \PbPb\ collisions.
The centrality of those collisions is characterized by the total
transverse energy within the acceptance of the ATLAS forward
calorimeters. The \gamgammumu\ measurement is
performed over a kinematic fiducial region: $|y| < 2.4$, $\pT >
3.7$~\GeV, and $\Mmumu < 45$~\GeV,
where $y$ and \pT\ are single-muon
rapidities and transverse momenta, respectively, and \Mmumu\ is the
dimuon invariant mass. Candidate muon pairs
are identified using selections on pair acoplanarity, \Aco, and
asymmetry, $A$, defined as \AtlasCoordinateDef\
\begin{align*}
\alpha & \equiv 1- |\phi_1-\phi_2|/\pi,\\
A & \equiv \left|{p_{\text T}}_1-{p_{\text T}}_2\right|/\left({p_{\text T}}_1+ {p_{\text T}}_2\right),
\end{align*}
where $\phi_{1,2}$ and ${p_{\text T}}_{1,2}$ represent
the azimuthal angles and  the transverse momenta of each of the two
muons, respectively.

The dominant background in this measurement, namely pairs of muons resulting from heavy-flavor
(HF) decays,  is suppressed through requirements on the pointing of the muons to the
primary vertex.
The remaining background is estimated using a template-fitting procedure
based on the combined distance of closest approach of the two muons
to the collision vertex. Potential non-negligible backgrounds from
Drell--Yan (DY) processes are estimated using Monte Carlo (MC) simulations.
 
Distributions of \Aco, and the associated transverse momentum scale \kp,
\begin{equation}
\kp \equiv  \frac{1}{2} \left({p_{\text T}}_1+{p_{\text
T}}_2\right)  \left(\,\pi - |\phi_1 - \phi_2|\, \right) = \pi \alpha \ptbar, \label{eq:kperp}
\end{equation}
are measured as a function of \PbPb\ collision centrality and the
average of the transverse momenta of the two muons,
$\ptbar \equiv\left({p_{\text T}}_1+ {p_{\text
T}}_2\right)/2$. It is shown later that the \Aco\ distributions vary
significantly with \ptbar, while the \kp\ distributions do
not. Thus, the \kp\ distributions are better suited for assessing
the centrality-dependent modifications of the dimuon
alignment. However, some of the theoretical calculations
are only available for acoplanarity, so results are presented
using both variables. Moments of the \kp\ distributions
are used to quantify the centrality-dependent modifications of the
dimuon alignment.
 
It was shown in Ref.~\cite{HION-2018-11} that the dimuon asymmetry
distributions are not sensitive to the
small transverse momentum scales associated with the observed modifications
of the dimuon alignment. However, the  \gamgammumu asymmetry distributions are
much narrower than those from background QCD processes. To further
strengthen an already robust demonstration that the observed signal
represents \gamgammumu pairs, the asymmetry distributions are
measured in different centrality bins and compared with the results
obtained from the \starlight~\cite{Klein:2018fmp} event generator.
 
To test predictions \cite{Klein:2018fmp,Klein:2020jom} that
magnetic broadening of the dimuon \Aco\ or \kp distributions
should depend on the rapidity separation of the two muons,
the \kp\ distributions are measured as a function of
$\Dy \equiv |y_1 - y_2|$, where $y_1$ and $y_2$ represent the
rapidities of the two muons. To test whether magnetic broadening effects have a directional dependence in
the transverse plane, the \kp distributions are also measured
as a function of $2\Delta \phi
\equiv 2\left|\phi_{\mu\mu} - \Psi_2\right|$, where $\phi_{\mu\mu}$,
defined in Section~\ref{sec:ktrapgapdphi},
represents the orientation of the dimuons in the transverse plane, and $\Psi_2$ is the
second-order event plane angle.

To allow tests of theoretical calculations, cross-sections for the production
of muon pairs are measured in different  centrality and \ptbar intervals.
The total cross-section  for \gamgammumu\ production,
including UPC contributions, within the  fiducial constraints of the
measurement is also obtained. Separate ``normalized pair yields''
representing the fraction of the total \PbPb\ \gamgammumu\ yield
measured in a given centrality and kinematic interval are also presented.
In these relative yields, some systematic uncertainties in the
measurement cancel out.
 
The remainder of this paper is structured as follows:
Section~\ref{sec:detector} describes the ATLAS detector;
Section~\ref{sec:data} describes the data and Monte Carlo samples
used in the measurement and the applied event and dimuon
selections; Section~\ref{sec:analysis} describes the corrections
for trigger and reconstruction inefficiency and the estimation and
subtraction of backgrounds; Section~\ref{sec:systematics} describes
the systematic uncertainties in the measurement;
Section~\ref{sec:results} presents the results; and
Section~\ref{sec:conclusion} presents a summary of the results and
conclusions.

% End of text imported from the .//tex/intro.tex input file

\section{ATLAS detector}
\label{sec:detector}

% The next lines are included from the .//tex/detector.tex input file
The ATLAS detector~\cite{PERF-2007-01} is composed of an inner
tracking detector (ID) inside a superconducting solenoid magnet, electromagnetic
and hadronic calorimeters, and a muon spectrometer (MS) with superconducting
toroid magnets, and has a high-speed trigger and data-acquisition system.
 
The ID, consisting of a silicon pixel detector, a silicon microstrip tracker,
and a transition radiation tracker, is immersed in a \SI{2}{\tesla}
axial magnetic field~\cite{IDET-2010-01}.
The ID provides charged-particle tracking in the pseudorapidity
range $|\eta| < 2.5$.
The high-granularity silicon pixel detector covers the interaction region and
typically provides four measurements per track.
The pixel detector is followed by the silicon microstrip tracker, which
typically provides measurements of four two-dimensional space points per track.
These silicon detectors are complemented by the transition radiation
tracker, which enables radially extended track reconstruction up to
$|\eta| = 2.0$, providing around 30 hits per track.
 
The calorimeter system consists of a liquid-argon (LAr) electromagnetic calorimeter
covering $|\eta|<3.2$, a steel/scintillator sampling hadronic calorimeter
covering $|\eta|<1.7$, a LAr hadronic calorimeter covering $1.5 < |\eta| < 3.2$,
and a forward calorimeter (FCal) covering $3.1 < |\eta| <4.9$.

The MS comprises separate trigger and high-precision tracking chambers
that measure the deflection of muons in the magnetic field of the superconducting
air-core toroids.
The precision chamber system covers the region $|\eta|<2.7$ with three layers of monitored
drift tubes complemented by cathode strip chambers in the forward region.
The muon trigger system covers the range $|\eta|<2.4$ with resistive plate chambers in the barrel
and thin gap chambers in the endcap regions.
 
Two zero-degree calorimeters (ZDCs), which measure neutrons emitted from the incident nuclei
at large absolute rapidities, are used for triggering and for offline
event selection.
The ZDCs are located symmetrically at a distance of $\pm 140$~m from the nominal interaction point
and cover $|\eta|>8.3$.
Each of the ZDCs consists of four modules, each containing slightly more than one interaction
length of tungsten absorber.
 
The ATLAS trigger system~\cite{TRIG-2016-01} consists of a first-level (L1)
trigger implemented using a combination of dedicated electronics and
programmable logic, and a software-based high-level trigger (HLT).
Muon triggers are formed using a combination of L1 triggers that find candidate
muons from the MS trigger chambers and HLT triggers that combine ID
and MS tracks.
The L1 muon trigger has lower acceptance ($|\eta|<2.4$) than the full MS ($|\eta|<2.7$).
 
An extensive software suite~\cite{ATL-SOFT-PUB-2021-001} is used in the reconstruction and analysis of real and simulated data, in detector operations, and in the trigger and data acquisition systems of the experiment.

% End of text imported from the .//tex/detector.tex input file

\section{Data sets, reconstruction, and event selection}
\label{sec:data}

% The next lines are included from the .//tex/data.tex input file
\subsection{Data Sets}
The \PbPb\ data used in this measurement were recorded in 2015 and 2018
with integrated luminosities of 0.49~\inb\ and 1.44~\inb, respectively.
Two dimuon triggers were used for this analysis.
The first trigger (L1Single) required a single muon with $\pt>4~\GeV$ at L1,
and two muons with $\pT > 4$~\GeV\ at the HLT.
The second trigger (L1Pair) required two muons with $\pt>4~\GeV$ at L1
and also at the HLT.
By requiring only one muon at L1, the first trigger has greater L1
trigger efficiency, but it was prescaled for a small portion of the
data-taking and thus did not sample the full luminosity.
A total of 4.7~million and 12.2~million events were obtained
from these triggers in 2015 and 2018, respectively.

Separate samples of \MinBias\ \PbPb\ events and \PbPb\ events
selected by a combination of single-muon triggers
are used to evaluate the efficiency of the muon triggers used in this analysis.
The \MinBias\ sample is built using a combination of three mutually exclusive triggers.
The first (second) trigger required the total transverse energy in the
calorimeters at L1, $E_{\mathrm{T}}^{\mathrm{L1}}$, to be greater than
600~\GeV\ (between 50~\GeV\ and 600~\GeV) without any additional requirements at the HLT.
The third trigger required $E_{\mathrm{T}}^{\mathrm{L1}}$ to be less than 50~\GeV\ with the
additional requirement of an energy deposition above the single-neutron threshold in either one of the ZDCs.
At the HLT, this trigger additionally required a reconstructed track having $\pt>0.2~\GeV$.
The single-muon triggered sample is built using a combination of three
triggers that require a muon with
$\pT > 4$, 6 and 8~\GeV\ at the HLT.

The performance of the ATLAS detector in reconstructing muon pairs is
evaluated using an MC sample obtained by overlaying \gamgammumu\ events
produced with the \starlight event generator
onto \MinBias\ \PbPb\ events simulated using the
\textsc{Hijing}~v1.383~\cite{PhysRevD.44.3501} event generator.
The detector response in the MC samples was simulated using \GEANT~\cite{Agostinelli:2002hh},
and the resulting events are reconstructed using the same algorithms that are
applied to the data~\cite{SOFT-2010-01}.
A total of 4 million such events are analyzed using the same methods as
applied in the data analysis.
The \starlight\ MC sample is also used for comparison with the measured observables.

Potential backgrounds from DY processes are estimated
using the \POWHEGBOX[v2]~\cite{Nason:2004rx,Frixione:2007vw,Alioli:2010xd}
generator interfaced to \PYTHIA[8] configured using parameter values set to the
AZNLO tune~\cite{STDM-2012-23,} and CTEQ6L1~\cite{Pumplin:2002vw} parton
distribution functions (PDFs). Separate samples of \pp, \pn, and \nn
events were generated and combined with appropriate isospin weights.
The \POWHEGBOX\ generator was configured to provide per-event weights for
five different nuclear PDF sets: nCTEQ15~\cite{Kovarik:2015cma},
EPPS16 NLO~\cite{Eskola:2016oht}, nNNPDF1.0
NNLO~\cite{AbdulKhalek:2019mzd}, nNNPDF2.0
NLO~\cite{AbdulKhalek:2020yuc}, and TUJU19
NNLO~\cite{Walt:2019slu}. Thus, separate evaluations of the DY
background in this measurement are obtained for all five PDF
sets. Variations of the renormalization and factorization scales are
performed using the nCTEQ15 set and are compared with a similar set of
variations performed in \powheg\ using the nucleon \CT[14nnlo]
set~\cite{Gao:2013xoa,}.

\subsection{Event and muon-pair selections}
\label{sec:eventsel}
Events used in the analysis are required to have been recorded during stable
running conditions of the LHC, to have no detector hardware or readout
error, and to have a reconstructed collision vertex.
Charged-particle tracks and collision vertices are reconstructed using
standard methods~\cite{PERF-2015-08} tuned for the conditions of
\PbPb\ collisions and assuming a single collision vertex per event.
In addition to track kinematic parameters, the ID reconstruction
also provides information about the minimum distances \dz\ and $z_0$ between the projected
track and the reconstructed vertex in the transverse and
longitudinal planes, respectively.
Muons are reconstructed by combining ID tracks with tracks reconstructed
in the muon spectrometer. The muons are required to pass the ``medium''
muon selection requirements described in Ref.~\cite{MUON-2018-03}.
 
Opposite-sign muon pairs passing the following preselections are used
for the analysis: each muon has $\pT > 3.7$~\GeV\footnote{This value
is less than the thresholds applied in the muon trigger to account
for differences in the muon momentum measurement between the trigger and
the offline reconstruction and to allow the maximum possible
acceptance for pairs having \ptbar near 4~\GeV.} and $|\eta|<
2.4$; the pair has a dimuon invariant mass less than 45~\GeV; and both
muons must be matched in angular space to HLT-reconstructed
muons. These kinematic selections are largely determined by the
acceptance of the MS; the mass restriction is applied to avoid
contamination from $Z$ boson decays.
 
\begin{figure}
\centering
\includegraphics[width=0.5\textwidth]{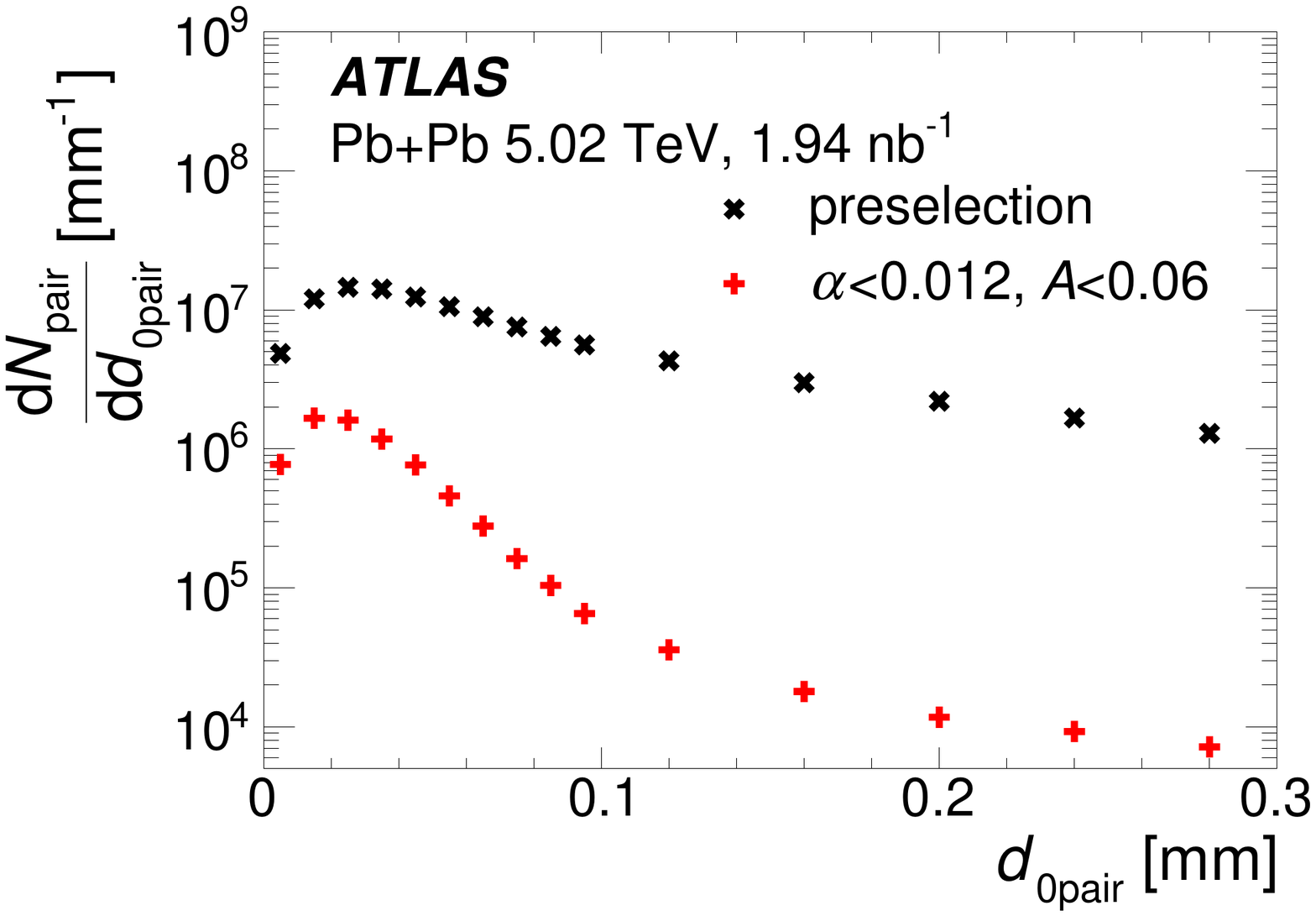}
\includegraphics[width=0.5\textwidth]{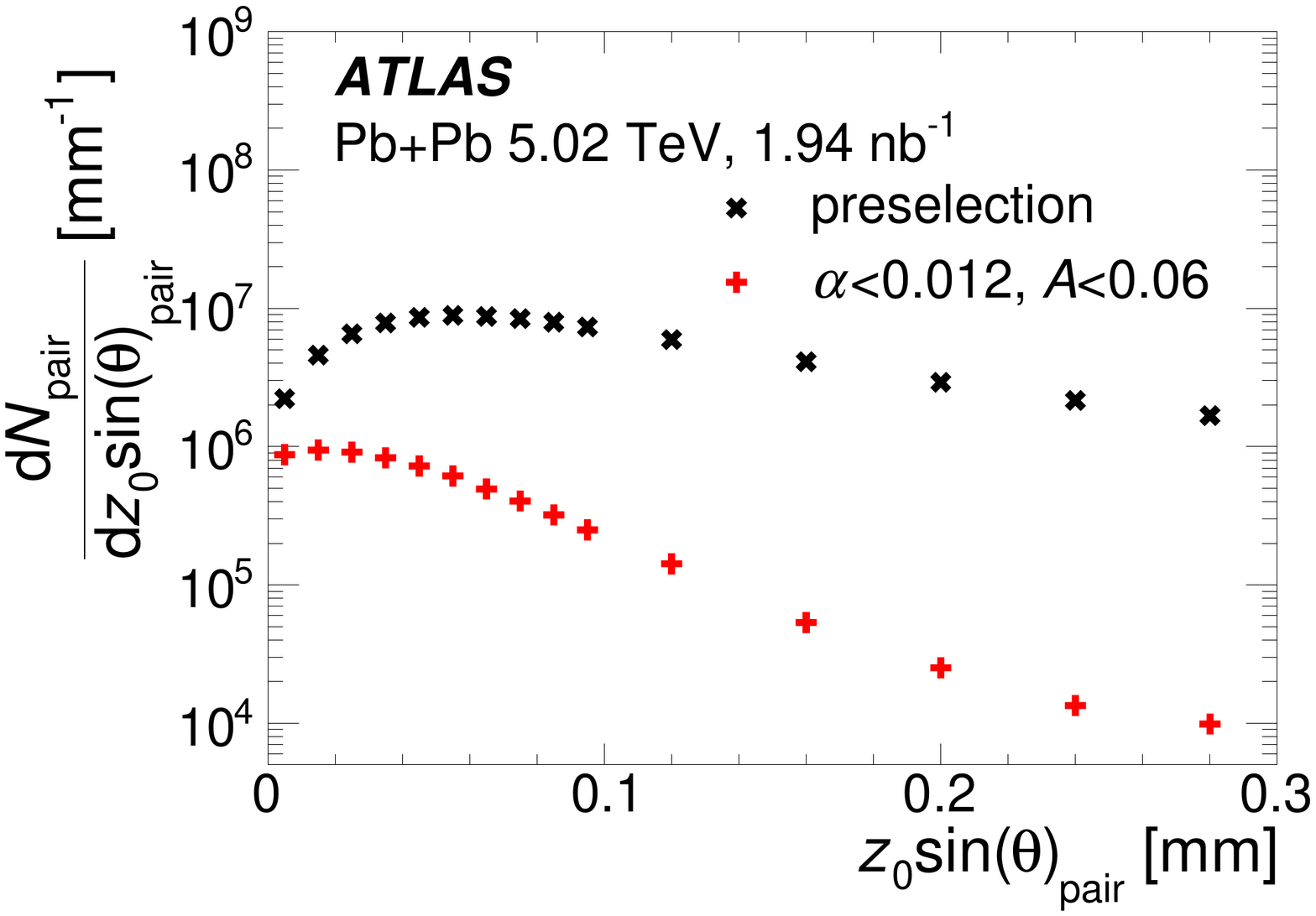}
\caption{
Distributions of \ensuremath{{d_0}_{\mathrm{pair}}}\xspace (top) and
\zzp\  (bottom) for muon pairs passing the
preselections (black), and for pairs additionally
passing the Fid-\Aco ($A<\AsymCutForAcop$, $\alpha<\AcopCutForAsym$) selection (red).
The error bars in both panels correspond to the statistical uncertainties,
and are typically too small to be seen.
The pairs passing the Fid-\Aco selection have
much smaller \dzp and \zzp values than pairs passing the
preselections, due to the large HF background in preselected pairs.
\label{fig:dzp_zzp_distributions}
}
\end{figure}

To reduce the background from semileptonic decays of heavy-flavor hadrons,
requirements are imposed on the pointing of the muons to the vertex
using a combination of the single-muon \dz\ and \zzst\ values (where $\sin\theta$ is the polar angle of the muon track):
\begin{align*}
\dzp &  \equiv \sqrt{{d_0}^2_1 + {d_0}^2_2},\\
\zzp &  \equiv \sqrt{\left({{z_0} \sin{\theta}}\right)_1^2 + \left({{z_0}\sin{\theta}}\right)_2^2}.
\end{align*}
Distributions of \dzp\ and \zzp\ for pairs passing the above
preselections  are shown in Figure~\ref{fig:dzp_zzp_distributions} in
the top and bottom panels, respectively. Also shown for comparison are
\dzp and \zzp\ distributions for pairs passing a kinematic fiducial
selection, described below, that suppresses the HF decay contribution.
The fiducial selection strongly suppresses the yield of pairs with
large \dzp and \zzp values that predominantly result from HF-decay
background pairs.
The following
selections are imposed on muon pairs used in the measurement:
\begin{equation*}
\dzp < 0.1~\text{mm}, \zzp < 0.2~\text{mm}.
\end{equation*}
These requirements reduce the yield of HF-decay pairs by a factor of $\approx$2
while introducing an inefficiency for \gamgammumu\ pairs of
${\lesssim}2\%$.

Following the methods of Ref.~\cite{HION-2018-11}, candidate
\gamgammumu\ pairs are obtained from those passing the preselection
and the \dzp\ and \zzp\ requirements by imposing stringent requirements on the
pair asymmetry and either the acoplanarity or the \kp\ value. For this paper,
two different fiducial selections are defined: $A < \AsymCutForAcop
\bm{\wedge} \Aco < \AcopCutForAsym$ or $A < \AsymCutForAcop \bm{\wedge} \kp <
\KpCutForAsym$, labeled Fid-\Aco and Fid-\kp, respectively.  Both
fiducial selections include the muon pseudorapidity and
\pT\ requirements and the pair mass
constraints included in the preselections. The separate fiducial
selections are motivated by the HF and DY subtraction that is
discussed later in Section~\ref{sec:backsub}. In particular, the
backgrounds are observed to be uniform as a function of \Aco and \kperp
as long as no requirement is imposed on the other variable.
However, because of the direct relationship between \Aco and \kperp
made explicit in Eq.~(\ref{eq:kperp}), a selection on \Aco introduces a
\ptbar-dependent constraint on \kp and {\textit vice versa}.
A single fiducial selection would thus distort the shapes of the HF backgrounds
and make those shapes sensitive to an applied \ptbar selection. For
that reason, separate fiducial regions are used. Specifically, the
Fid-\Aco selection is used for measurements of acoplanarity
distributions, and the Fid-\kperp selection is applied in measurements of
\kperp distributions.
 
Otherwise, for consistency with the
previous measurement~\cite{HION-2018-11}, the Fid-\Aco selection is
applied for many of the plots in Section~\ref{sec:analysis} that
document technical details of the analysis, while measurements of the production
cross-sections and related quantities are presented using only the
Fid-\kp selection.

\subsection{Centrality}
\begin{table}[!b]
\centering
\caption{
\label{tbl:fcalet}
The \sumETFCal\ intervals used for this measurement along with the
corresponding centrality ranges and \AvgTAA\ values.
No centrality calibration is available (see text) for the intervals
labeled ET0--ET3, which cover the 10\% most peripheral collisions.
}
\begin{tabular}{|c|c|c|}
\hline
FCal-\et range [TeV] & Centrality [\%] & $\langle T_{AA}\rangle $ [mb$^{-1}$]   \\
\hline
$>3.62$         &    0--5  & 26.0~~~~ \\
2.99--3.62      &  ~~5--10 & 20.4~~~~ \\
2.05--2.99      &   10--20 & 14.4~~~~ \\
1.37--2.05      &   20--30 & 8.77\\
0.875--1.37~~   &   30--40 & 5.09\\
0.525--0.875    &   40--50 & 2.75\\
0.290--0.525    &   50--60 & 1.35\\
0.144--0.290    &   60--70 & ~~0.601\\
0.0637--0.144~~ &   70--80 & ~~0.239\\
0.0240--0.0637  &   80--90 & ~~~~0.0815\\
0.0150--0.0240  & ET3      & - \\
0.0100--0.0150  & ET2      & - \\
0.0050--0.0100  & ET1      & - \\
$<$ 0.005       & ET0      & - \\
\hline
\end{tabular}
\end{table}
 
In ATLAS heavy-ion measurements, the \PbPb\ collision centrality is
characterized by the total transverse energy, \sumETFCal,
measured in the ATLAS forward calorimeters. The relationship between
\sumETFCal\ and the geometry of the \PbPb\ collisions is evaluated
using a Glauber model analysis \cite{Miller:2007ri,Loizides:2014vua}
following standard methods (see details presented in Ref.~\cite{ATLAS-CONF-2016-110}). That analysis also provides
values for the nuclear overlap parameter, \TAA, which describes the
effective nucleon--nucleon luminosity of a \PbPb\ collision. While the
\gamgammumu\ process may have a very different
dependence on \PbPb\ collision geometry than soft or hard
QCD scattering processes, the use of standard \PbPb\ centrality intervals
nonetheless remains useful, for example, in estimating backgrounds
from QCD processes. This analysis uses the following set of centrality
intervals, defined in terms of percentiles of the minimum-bias
\PbPb\ \sumETFCal distribution: ten-percent intervals
spanning the range 10--90\%, plus the two intervals 0--5\% and 5--10\%.
A set of larger, combined centrality intervals are used in specific
instances to reduce statistical uncertainties in some of the measured
quantities or distributions.  To cover the 10\% most peripheral
collisions, which have $\sumETFCal < 24$~\GeV\ and, for which,
centrality calibrations are not available,\footnote{Events with the lowest \sumETFCal\ values have
significant contamination from various kinds of ultraperipheral
collisions, so the \PbPb\ ``centrality'', which characterizes hadronic
\PbPb\ collisions, cannot be defined for these events.}
a set of \sumETFCal\ intervals, labeled ET0--ET3, are used. These have
lower \sumETFCal boundaries at integer multiples of 5~\GeV\ except for
ET0, which includes all events having $\sumETFCal < 5$~\GeV.
To suppress the contribution of ultraperipheral collisions for the
measurement of non-UPC \gamgammumu\ production, events included in the
above centrality intervals are required to have at least one neutron
in each ZDC. A separate set of exclusive UPC events which have zero
neutrons in one or both of the ZDCs and no additional charged
particles beyond the two muons, is used to evaluate the effects of the
\dzp\ and \zzp\ selections and for comparison with the non-UPC
results.
 
The inclusive cross-section for production of dimuons
in $\sqn = 5.02$~\TeV\ \PbPb\ collisions is measured for tests of theoretical calculations and for use in this paper.
For this purpose, no event selections are imposed beyond those described in
Section~\ref{sec:eventsel}. The inclusive measurement includes
pairs that pass neither the exclusive UPC requirement nor the ZDC
coincidence requirement applied to the non-UPC intervals. Such pairs
are referred to as ``unassigned''.
 
\begin{figure}[!tb]
\centering
\includegraphics[width=1.0\textwidth]{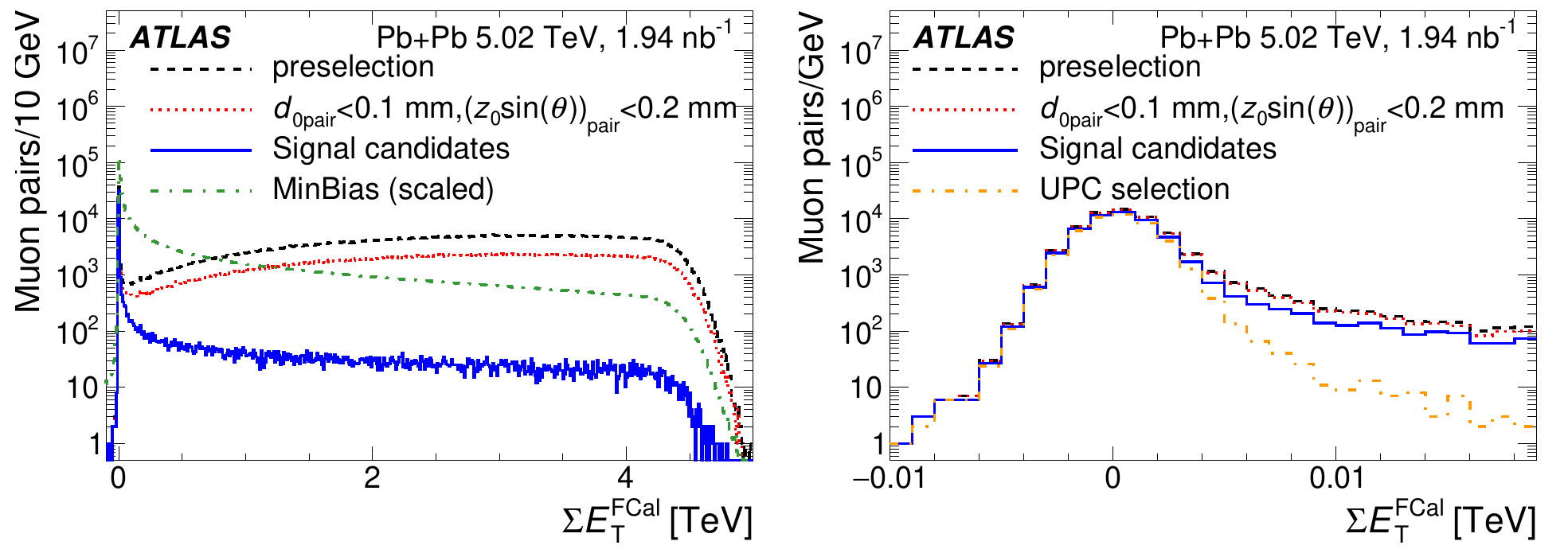}
\caption{
Left: distributions of \ensuremath{\Sigma E_{\text{T}}^{\text{FCal}}}\xspace for different muon pair selections:
pairs passing preselections only, additionally passing the $\dzp<0.1$~mm and $\zzp<0.2$~mm
requirements, and additionally passing the asymmetry and acoplanarity
($A<\AsymCutForAcop$, $\alpha<\AcopCutForAsym$) requirements (labeled
as ``Signal candidates''). The
\ensuremath{\Sigma E_{\text{T}}^{\text{FCal}}}\xspace distribution
for minimum-bias events (arbitrarily scaled)
is also shown for comparison.
Right: the same distributions, excluding the minimum-bias distribution,  plus
that for the UPC event selection, shown over a restricted range of
\ensuremath{\Sigma E_{\text{T}}^{\text{FCal}}}\xspace values. The
distributions extend to negative values due to electronic noise in
the calorimeter.
\label{fig:fcalet}
}
\end{figure}

Table~\ref{tbl:fcalet} summarizes the
\sumETFCal\ intervals used in this measurement, indicates the corresponding centrality range, where
available, and for those with a centrality calibration, provides the
corresponding \AvgTAA\ values.
Figure~\ref{fig:fcalet} shows, in the left panel, the
\sumETFCal\ distribution for events with muon pairs passing
different stages of pair selection: preselections only; also passing
the $\dzp<0.1$~mm and $\zzp<0.2$~mm requirements; and passing these
plus the Fid-\Aco\ selection. Also shown is the
\sumETFCal\ distribution for minimum-bias \PbPb\ collisions. The right
panel shows distributions for the different dimuon selections over a restricted range of
\sumETFCal\ values near zero;  it also shows the
\sumETFCal\ distribution for events passing the UPC
selection.
 
All of the \sumETFCal\ distributions show a strong
enhancement near $\sumETFCal = 0$ that results from the
geometric enhancement of peripheral collisions. For dimuon events, the
low-\sumETFCal\ peak results, primarily, from a large contribution of UPC
\gamgammumu\ events which (excluding dissociative photon-induced
processes) have no particles in the acceptance of the forward
calorimeters. Thus, the \sumETFCal\ distribution for these events
primarily reflects electronic
noise in the detector.  At larger \sumETFCal\ values, all
distributions show a long plateau that reflects the
nucleus--nucleus collision geometry. However, for pairs passing
the preselections or the preselections plus vertex requirements, the yield
increases with increasing \sumETFCal\ relative to that for
minimum-bias \PbPb\ events and the Fid-\Aco\ selection. This
behavior results from the geometric enhancement of QCD hard-scattering
processes -- particularly the production of heavy flavor -- the rates
of which are proportional to \TAA.  The application
of the Fid-\Aco\ selection yields a \sumETFCal\ distribution that is almost
flat at large transverse energies. However, even with the fiducial
selection, there remains a non-negligible and centrality-dependent HF-decay background that
must be subtracted. Thus, the apparent flatness of the \sumETFCal\ distribution is
at least partially accidental. However, as is seen below, even in the most central
collisions, the HF backgrounds are,  at most, comparable to the
signal pair yields, so \gamgammumu\ production is observed over the
full range of \PbPb\ collision centralities, including the most central collisions.

% End of text imported from the .//tex/data.tex input file

\section{Analysis}
\label{sec:analysis}

% The next lines are included from the .//tex/analysis.tex input file
\begin{table}[t]
\caption{
Numbers of muon pairs in each centrality interval for different
\ptbar\ ranges and the two fiducial selections.
Also listed are the total \gamgammumu\ yields.
The interval labeled as \textit{Unassigned} includes pairs that
contribute to the inclusive yield but
do not appear in any of the listed centrality intervals, as they do
not satisfy the ZDC coincidence requirement imposed on the non-UPC
centrality intervals, and  also do not satisfy the exclusive
UPC requirement.
}
\label{tbl:paircountscent}
\centerline{
\begin{tabular}{|l|rr|rr|rr|}\hline
Centrality & \multicolumn{2}{c|}{$4 < \ptbar < 5$~\GeV} &
\multicolumn{2}{c|}{$5 < \ptbar < 6$~\GeV} &
\multicolumn{2}{c|}{$\ptbar>$ 6 \GeV} \\
range & Fid-\Aco & Fid-\kp\ & Fid-\Aco & Fid-\kp\ & Fid-\Aco &
Fid-\kp\ \\ \hline
~~0--5\%   &    673 &    617 &    384 &    325 &    534 &    439  \\
~~5--10\%  &    562 &    530 &    318 &    276 &    498 &    418  \\
10--20\%   &    924 &    871 &    588 &    531 &    943 &    807  \\
20--30\%   &    771 &    732 &    515 &    472 &    807 &    707  \\
30--40\%   &    611 &    580 &    422 &    396 &    790 &    721  \\
40--50\%   &    463 &    447 &    396 &    381 &    763 &    720  \\
50--60\%   &    413 &    410 &    354 &    342 &    631 &    601  \\
60--70\%   &    370 &    366 &    308 &    302 &    595 &    564  \\
70--80\%   &    362 &    358 &    289 &    285 &    536 &    516  \\
80--90\%   &    374 &    369 &    269 &    266 &    510 &    492  \\
ET3        &    154 &    152 &    114 &    113 &    201 &    191  \\
ET2        &    129 &    129 &     95 &     95 &    184 &    175  \\
ET1        &    242 &    241 &    209 &    205 &    320 &    309  \\
ET0        &   1539 &   1531 &   1241 &   1224 &   2116 &   2027  \\
UPC        &  17090 &  17042 &  12070 &  12004 &  16753 &  16461  \\
Unassigned &    328 &    327 &    297 &    296 &    435 &    426  \\ \hline
All Events &  25005 &  24702 &  17869 &  17513 &  26616 &  25574  \\ \hline
\end{tabular}
}
\end{table}
 
A total of 69490 muon pairs pass the combination of
preselections, \dzp\ and \zzp\ requirements, and Fid-\Aco\
selection.
For comparison, 67789 pairs pass the full
selections and the Fid-\kp selection.
The smaller Fid-\kp\ yield  results from the \kp\ selection being
more restrictive than that for the acoplanarity.
Both fiducial selections reject about 10\% of the pairs in the UPC
sample that pass the preselections and pair vertex requirements.
The excluded events primarily result from hard QED radiation or
dissociative photon-induced processes
(see Ref.~\cite{HarlandLang:2012qz} and references therein) but may
also include background, non-\gamgammumu\ events.  For non-UPC
events the fiducial selections also suppress
QCD backgrounds which generate muon pairs that typically have much larger
acoplanarity and \kp values than  \gamgammumu\ pairs.
Table~\ref{tbl:paircountscent} lists the number of pairs passing the
Fid-\Aco\ and Fid-\kp\ selections in each centrality interval for
three intervals of \ptbar.

Dimuon mass and \ptbar\ distributions for all dimuons passing the
Fid-\Aco\ selection are shown in Figure~\ref{fig:mass_ptbar}.
The distributions for measured pairs are suppressed at low \Mmumu\ and
\ptbar\ values owing to inefficiencies in the muon trigger and offline
reconstruction.
Corrections for the resulting losses are discussed in the
following section.
An additional suppression at low \Mmumu\ results from the acoplanarity
and asymmetry requirements imposed as part of the fiducial selection.
A decrease observed in the \ptbar\ distribution at the highest values
results from the $\Mmumu < 45$~\GeV\ requirement.
 
\subsection{Trigger and reconstruction efficiency}
In the analysis described below, muon pairs are corrected for trigger and
reconstruction efficiency by application of a per-pair weight to data events:
\begin{equation*}
W = \frac{1}{\varepsilon^{\text{trig}} \, \varepsilon^{\text{rec}}_1
\,  \varepsilon^{\text{rec}}_2 \effvtx}\, ,
\end{equation*}
where $\varepsilon^{\text{trig}}$ is the pair trigger efficiency,
$\varepsilon^{\text{rec}}_1$ and $\varepsilon^{\text{rec}}_2$ are the
single-muon reconstruction efficiencies, and \effvtx is the vertex-pointing efficiency.
 
Since all dimuons used in this analysis are
back-to-back in azimuth and are thus well-separated in the detector,
the efficiency of the dimuon triggers is evaluated using independent,
single-muon trigger efficiencies.
The single-muon trigger efficiencies are evaluated
by testing whether offline-reconstructed muons
passing the preselections described in Section~\ref{sec:data} are
matched to a muon found by the trigger.
These efficiencies are evaluated as a function of muon \pT\ and
of the product of the muon charge and pseudorapidity, $q\eta$.
The single-muon trigger efficiencies are combined -- separately for the
L1Single and L1Pair triggers -- to produce per-pair
trigger efficiencies.
The trigger efficiencies are found to change only by a few percent
over the full centrality range of the measurement,  whereas the
detector occupancy varies by orders of magnitude.
 
\begin{figure}[!tb]
\centering
\includegraphics[width=1.0\textwidth]{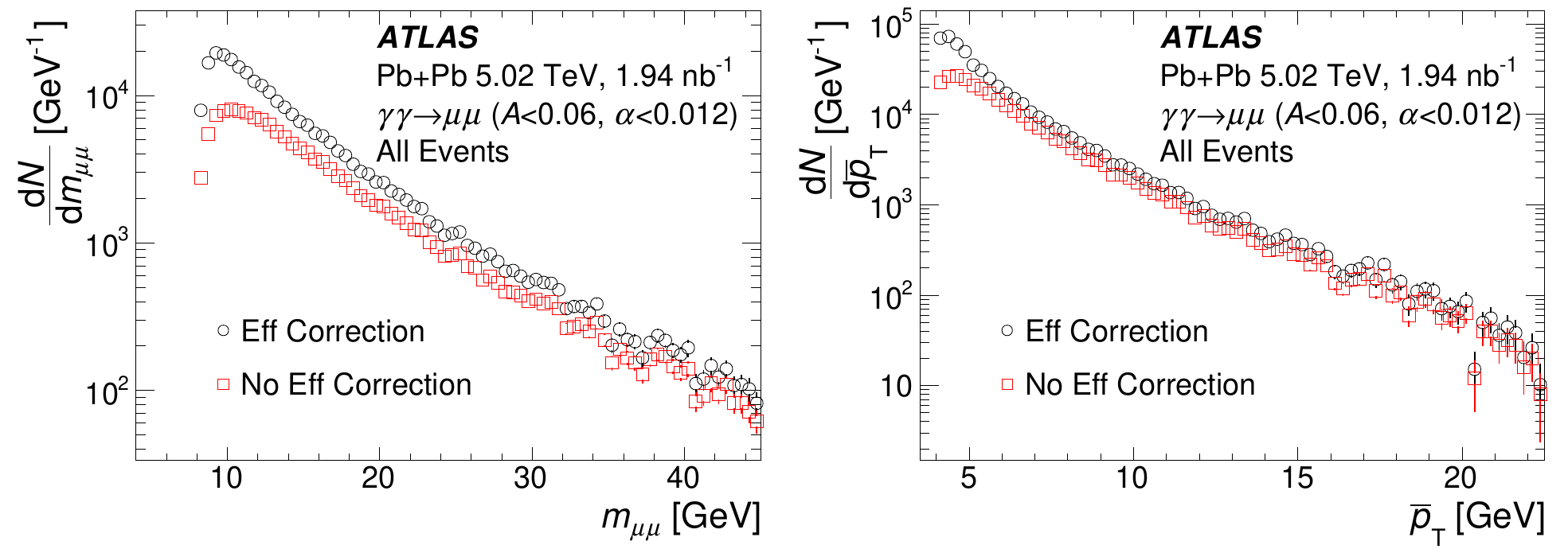}
\caption{
Distributions of the dimuon invariant mass (left) and \ptbar\ (right)
for the dimuons in the signal region before (red) and after (black)
the application of muon reconstruction and trigger efficiency corrections.
\label{fig:mass_ptbar}
}
\end{figure}
 
The efficiency for reconstructing dimuons in the offline analysis is
evaluated using the \starlight\ MC simulation sample.
As with the trigger efficiencies, the pair reconstruction efficiency is taken as the
product of the single-muon reconstruction efficiencies.
The single-muon reconstruction efficiencies are evaluated as a function
of muon \pT\ and $q\eta$.
They are corrected for small data--MC differences observed in previous
measurements~\cite{MUON-2018-03}.
The reconstruction efficiencies have negligible centrality dependence
at mid-rapidity ($|q\eta|<1$), while at forward rapidity
($1<|q\eta|<2.4$) they decrease by $\approx$10\% between peripheral and
central collisions.
 
The pair pointing requirements introduce an inefficiency of a few
percent for \gamgammumu\ pairs. The corrections for this inefficiency
are obtained from simulation and validated using the UPC data sample, where the
\dzp\ and \zzp\ selections remove 1.4\% and 0.5\% of the pairs,
respectively.  The \dzp\ and \zzp\ distributions in the \starlight\ MC
sample generally agree well with those in data, although the simulation
slightly underestimates the \dzp\ resolution for UPC collisions.  To rectify this
disagreement, additional \pT-dependent Gaussian smearing is applied
to the MC single-muon \dz\ values in all events.  With this
adjustment, the MC-evaluated efficiency for \gamgammumu\ pairs to pass
the \dzp\ and \zzp\ vertex selections, \effvtx, agrees well with the
efficiency in UPC events.  For non-UPC collisions, the vertex
selection efficiency decreases by a few percent between the most
peripheral and the most central collisions.

The effect of correcting the dimuon kinematic distributions for the
trigger and muon reconstruction efficiencies is illustrated in
Figure~\ref{fig:mass_ptbar}, where the efficiency-corrected \Mmumu\ and
\ptbar distributions for the Fid-\Aco\ selection are shown with the black points.
The corrections for efficiency increase the yield relative to the measured
distributions by a factor of $\approx$2 for $\Mmumu = 8$~\GeV\ and
$\ptbar = 4$~\GeV. At higher \Mmumu and \ptbar, the corrections
increase the dimuon yield by
$\approx$30\% for the most peripheral collisions (ET0--ET3 or UPC) and
$\approx$50\% in the most central (0--10\%) collisions.

\subsection{Heavy-flavor decay background estimation~\label{sec:template}}
As mentioned above, the dominant background in this measurement results
from muons produced in semileptonic decays of heavy quarks. At
lower \ptbar values, the background is dominated by combinatoric
pairs produced in the decays of uncorrelated heavy quarks. However,
with increasing \ptbar, the background becomes dominated
by correlated pairs. The HF-decay muons can be distinguished from prompt muons by their displaced
secondary vertices that are experimentally observed via the \dz\ and
\zzst\ parameters.
The HF  background surviving the \dzp\ and \zzp\ selections is
estimated using a template-fitting procedure
applied to the \dzp\ distributions.
These distributions are assumed to result from the combination of pure
\gamgammumu\ signal and a background contribution.
The shape of the signal \dzp\ distribution, called the
\textit{signal template} below, is obtained from the
$\starlight{+}\textsc{Hijing}$ MC sample.
The shape of the background \dzp\ distribution, called the
\textit{background template} below, is obtained directly from data using muon pairs that
pass the preselections, but have $A>\AsymCutForAcop$ and
$\Aco>\AcopCutForAsym$.  These last selections effectively eliminate
the \gamgammumu\  contribution, leaving only background pairs.
While the background subtraction is
primarily designed to remove HF-decay contributions, the background
template contains all muon pairs without a
strong back-to-back angular correlation.
 
The template fits are performed over an extended \dzp range,
$\dzp<0.3$~mm, using Poisson log-likelihood fits~\cite{BAKER1984437}
implemented within MINUIT~\cite{James:1994vla}.
They yield a fit signal fraction, \ff, that must be translated to
the analysis \dzp\ range. That is done by evaluating
the ratios of the integrals of the signal and combined templates over $ \dzp < 0.1$~mm:
\begin{equation}
\fsig=\
\frac{ \ff \,
\int_{0~\mathrm{mm}}^{0.1~\mathrm{mm}}\frac{{d}P_{\mathrm{sig}}}{{d}\left(\dzp\right)}\, {d}(\dzp)}{\int_{0~\mathrm{mm}}^{0.1~\mathrm{mm}}\; \frac{{d}
P}{{d}\left(\dzp\right)}\; {d}(\dzp)}.
\label{eq:fsig}
\end{equation}
 
\begin{figure}
\centering
\includegraphics[width=1.0\textwidth]{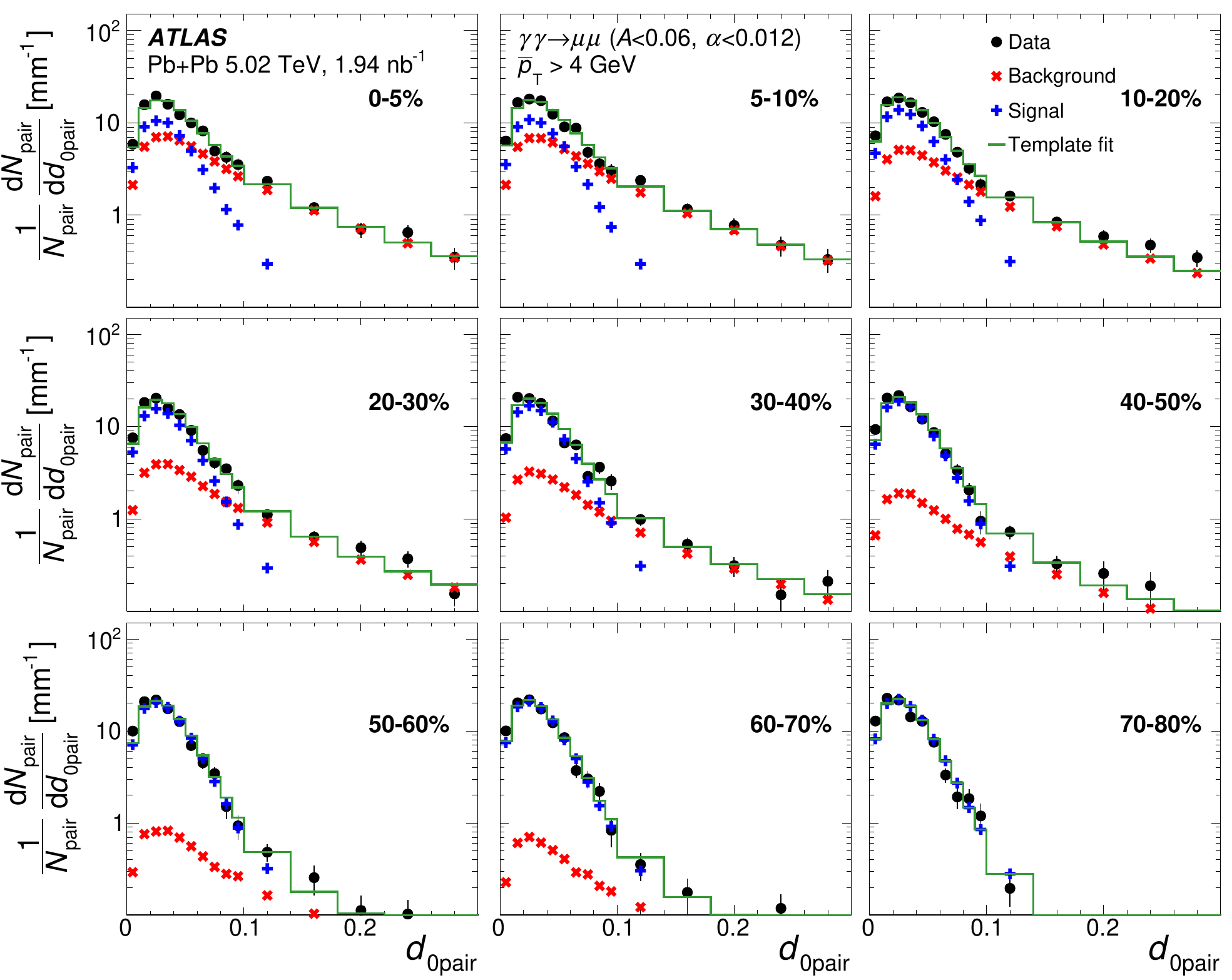}
\caption{
Results of template fits to measured
\ensuremath{{d_0}_{\mathrm{pair}}}\xspace distributions for pairs
passing the muon preselections, the \zzp\ requirement and the
Fid-\Aco\ selection ($A<\AsymCutForAcop$
and $\Aco < \AcopCutForAsym$).
Each panel represents a different centrality interval.
The error bars shown on the data and the templates represent
statistical uncertainties only.
For many of the points, the error bars are smaller than the size of
the marker.
\label{fig:templatefits}
}
\end{figure}
 
The above-described template fits are performed separately for the
Fid-\Aco\ and Fid-\kp\ selections.
The resulting signal fractions are denoted by \fsiga\ and \fsigkp,
respectively.
To allow measurements as a function of \ptbar, separate template fits
are performed in all centrality intervals for pairs in restricted
\ptbar intervals: $4 < \ptbar < 5$~\GeV, $5 < \ptbar < 6$~\GeV,
$6 < \ptbar < 8$~\GeV, and $\ptbar > 8$~\GeV, as well as for an
integrated interval, $\ptbar > 4$~GeV.
Other template fits are performed in intervals of \ymumu, \Dy, and \dphi,
to allow differential measurements as a function of these quantities.
 
Figure~\ref{fig:templatefits} shows examples of the template fits
applied to the \dzp\ distributions in several centrality intervals for
pairs satisfying the preselections, the \zzp\ requirement, and the
Fid-\Aco\ selection.
In all of the analyzed centrality intervals, the template fits
reproduce the data well; the minimum $\chi^2$-equivalent values are consistent with
expectations given the number of degrees of freedom.
The \fsigkp\ values (not shown) are consistent with the \fsiga\ values within the
statistical variations associated with the number of noncoincident
pairs.
 
\begin{figure}
\centering
\includegraphics[width=0.7\textwidth]{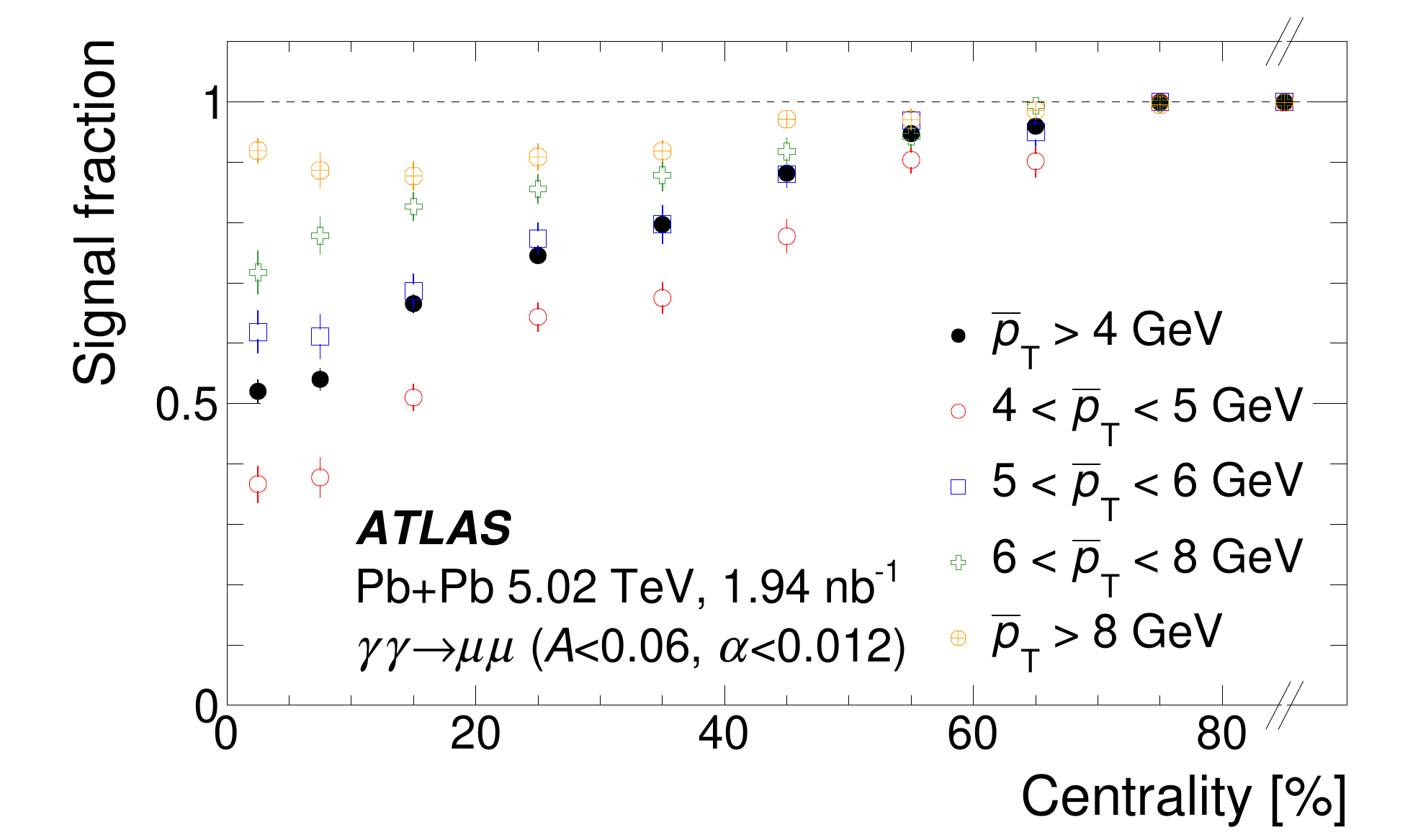}
\caption{
The signal fractions obtained from the template fits and
Eq.~\eqref{eq:fsig} for the Fid-\Aco\ selection, plotted as a
function of centrality for several \ptavg\ intervals.
The rightmost points represent the >80\% centrality interval, which
includes the UPC and Unassigned intervals.
The error bars indicate statistical uncertainties only.
\label{fig:signalfractions}
}
\end{figure}

Figure~\ref{fig:signalfractions} shows the \fsiga\ values obtained
from Eq.~\eqref{eq:fsig} as a function of centrality for different
selections on \ptbar.  Generally, the signal fractions become larger
with increasing \ptbar\ and for more peripheral collisions.  In
particular, for $\ptbar > 4$~\GeV, the signal fraction in the 0--5\%
centrality interval is $\approx$50\% while, for the  UPC selection, it is
consistent with one.  This decrease of the signal
fractions between peripheral and central collisions results from the
geometric enhancement of heavy-quark production relative to \gamgammumu\
processes in more central collisions.
However, with increasing \ptbar, the centrality dependence becomes weaker
such that for the highest \ptbar\ interval, $\ptbar > 8$~\GeV, the
signal fraction varies by only $\approx$12\% as a function
of centrality. This behavior is understood to result from the quenching
of heavy quarks (see Refs.~\cite{HION-2015-06,ATLAS:2021xtw}
and~\cite{Dong:2019byy,} and references therein) that suppresses the
correlated HF-decay background in more central collisions.
 
To evaluate the sensitivity of the above results to the shape of the
background template, three alternative selections for the background
template were applied: $A>0.1 \bm{\wedge} \alpha>0.012$, $0.06<A<0.3
\bm{\wedge} \alpha>0.012$, and $A>0.06 \bm{\wedge} \alpha>0.2$. The
first two significantly change the requirements on the muon momentum
balance while the third places a tighter requirement on the angular
alignment. The results obtained with these alternative background
templates are consistent with the nominal results.
 
The numbers of signal and background pairs,
$N_{\text{sig}}^{\text{fid}}$ and $N_{\text{bkg}}^{\text{fid}}$ for a
given fiducial selection and in a  given centrality and \ptbar\
interval are given by
\begin{align}
N_{\text{sig}}^{\text{fid}}(\text{cent},\ptbar) & =
\fsig^{\text{fid}}(\text{cent},\ptbar)
N^{\text{fid}}(\text{cent},\ptbar), \nonumber \\
N_{\text{bkg}}^{\text{fid}}(\text{cent},\ptbar) & = \left[1 - \fsig^{\text{fid}}(\text{cent},\ptbar)\right]  N^{\text{fid}}(\text{cent},\ptbar),
\label{eq:signal_bkg_yields}
\end{align}
where $\fsig^{\text{fid}}(\text{cent},\ptbar)$ and
$N^{\text{fid}}(\text{cent},\ptbar)$ are, respectively, the signal fraction and
efficiency-corrected number of pairs satisfying a given fiducial selection in the specified
centrality and \ptbar\ intervals.

\begin{figure}[!t]
\centering
\includegraphics[width=1.0\textwidth]{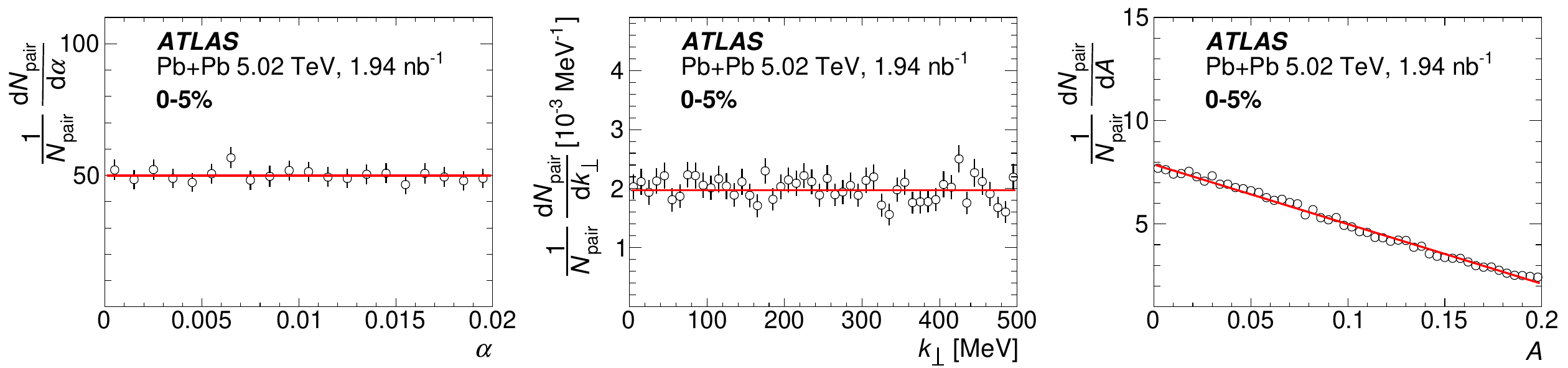}
\caption{
The heavy-flavor \Aco\ (left), \kp\ (middle), and \Amumu\ (right)
distributions for the 0--5\% centrality interval. The \Aco and \kp distributions are obtained by
applying $A > \AsymCutForAcop $ to preselected pairs. The $A$
distribution is obtained by applying $\Aco > \AcopCutForAsym$.
The error bars correspond to statistical uncertainties.
The lines indicate fits to constant functions for \Aco\ and \kp\
and to a linear function for \Amumu.
\label{fig:backgroundshape}
}
\end{figure}

Once the signal fractions $\fsig^{\text{fid}}(\text{cent},\ptbar)$ are determined,
the measured acoplanarity, \kp, and $A$ distributions can be corrected by
subtracting the contribution of background pairs.
The shapes of the background \Aco\ and \kp\ distributions are evaluated
by selecting muon pairs passing the pair preselections and having
$\Amumu > \AsymCutForAcop$.
This asymmetry requirement effectively removes contributions from
\gamgammumu\ pairs leaving, ostensibly, the HF-decay background.
To determine the shape of the background asymmetry
distribution, pairs passing preselections and having
$\Aco > \AcopCutForAsym$ are used.
The resulting differential distributions in \Aco, \kp, and $\Amumu$,
normalized to unit integral, are shown in
Figure~\ref{fig:backgroundshape}.
The \Aco\ and \kp\ distributions are found to be uniform within their
statistical uncertainties, so they are taken to be constants, \Chf, with values
given by
\begin{equation*}
\Chf^{\Aco} = \frac{N_{\text{bkg}}^{\text{fid-\Aco}}}{\AcopCutForAsym}, \;
\Chf^{\kp}  = \frac{N_{\text{bkg}}^{\text{fid-\kp }}}{\KpCutForAsym  },
\end{equation*}
such that the integral of each constant over the fiducial range of the
variable yields the number of background counts obtained from the
template fitting.
The $\Amumu$ distribution has nonzero slope, so it is fitted with a linear
function in each centrality interval.

\subsection{Differential distributions}
\label{sec:backsub}
 
\begin{figure}
\centering
\includegraphics[width=1.0\textwidth]{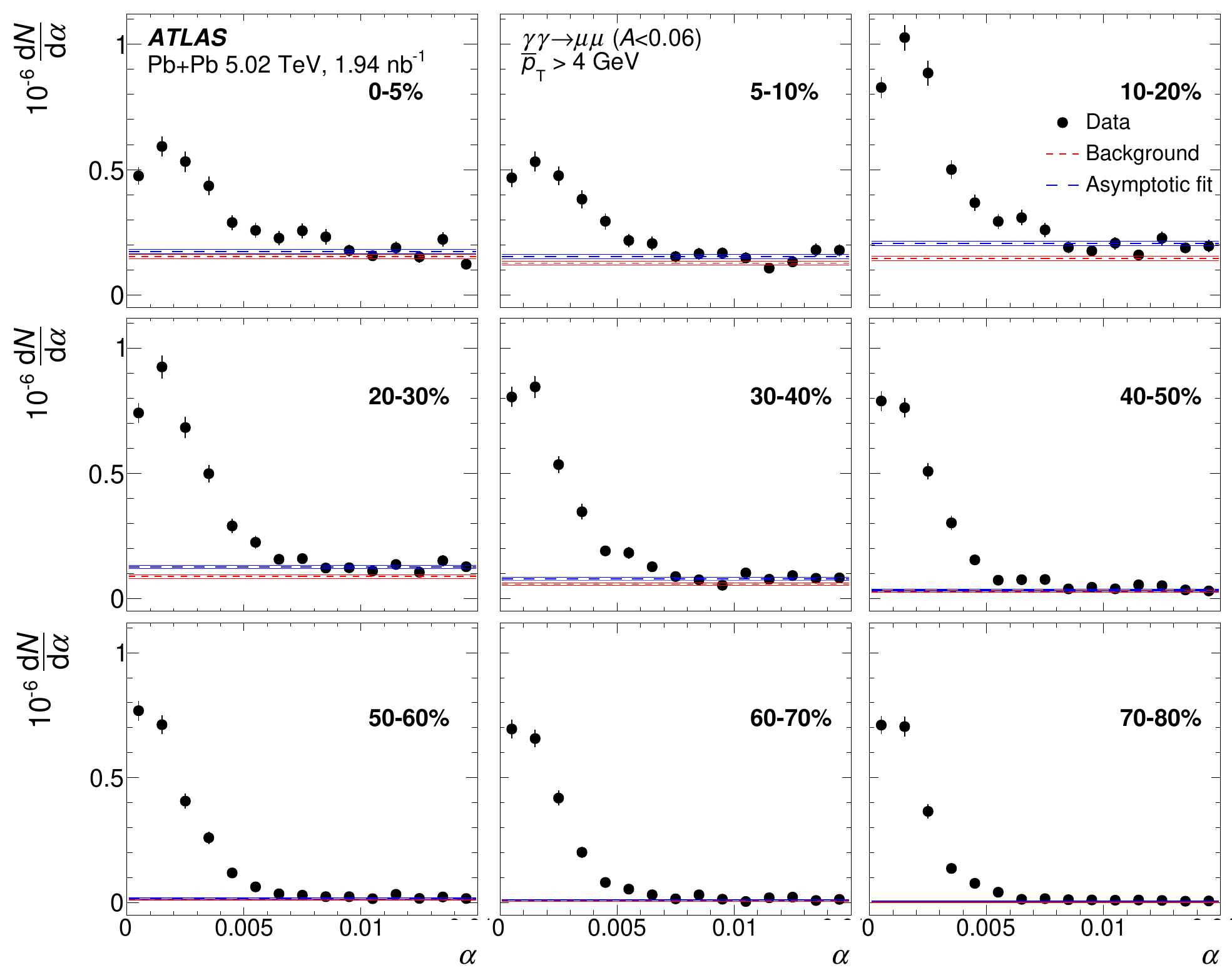}
\caption{
Measured $\Aco$ distributions for pairs passing the full set of
\gamgammumu\ selections and in the Fid-\Aco\ region for nine
centrality intervals.
The error bars on the points indicate statistical uncertainties.
Also shown are the background levels estimated from the \dzp\ template
fitting (red, fine-dashed lines) and from the asymptotic fits
(blue, coarse-dashed lines).
The widths of the lines indicates the statistical uncertainty of the
background estimates.
\label{fig:bkgsub}
}
\end{figure}
 
\begin{figure}
\centering
\includegraphics[width=1.0\textwidth]{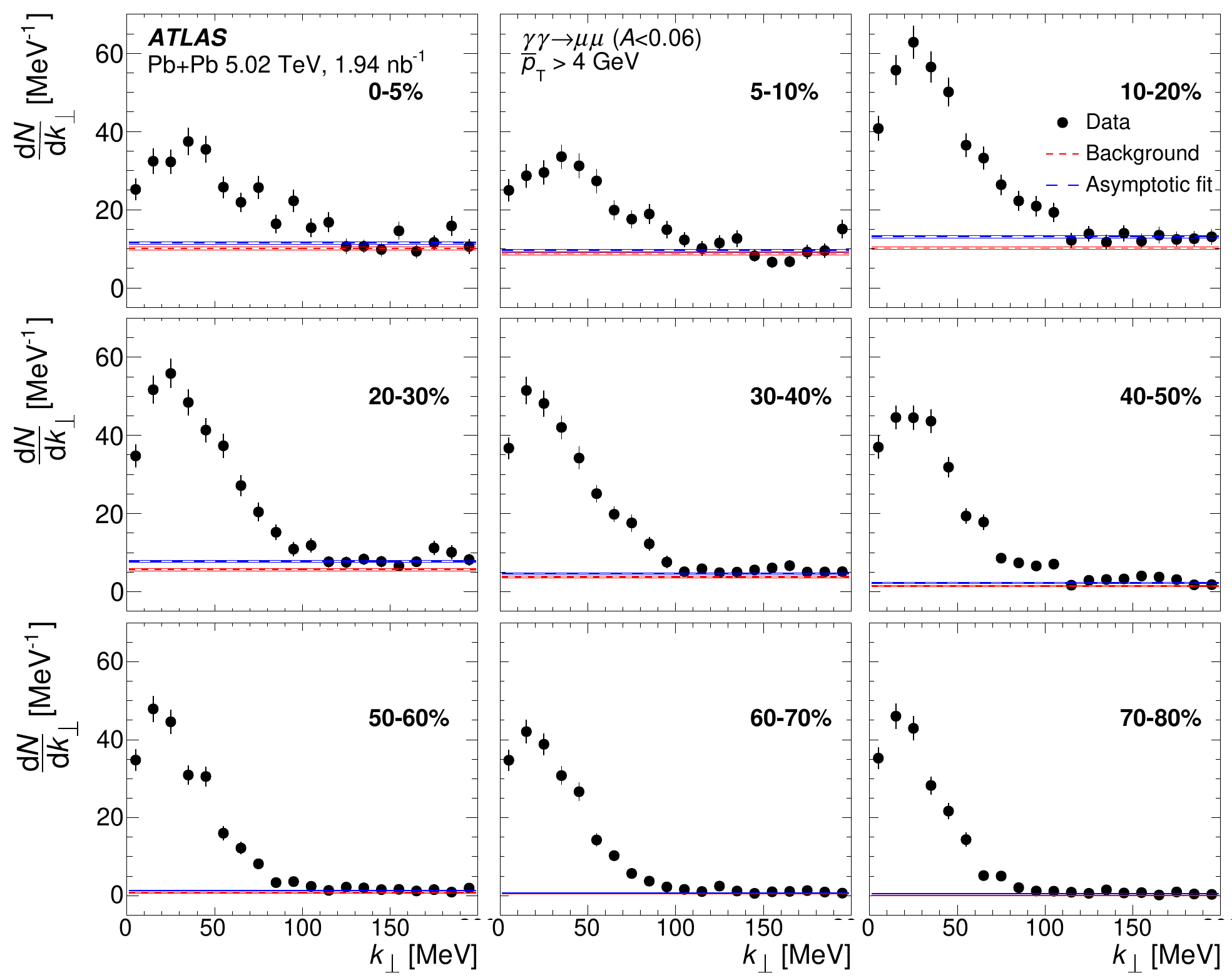}
\caption{
Measured $\kp$ distributions for pairs passing the full set of
\gamgammumu\ selections and in the Fid-\kp\ region for nine
centrality intervals.
The error bars on the points indicate statistical uncertainties.
Also shown are the background levels estimated from the \dzp\ template
fitting (red, fine-dashed lines) and from the asymptotic fits
(blue, coarse-dashed lines).
The widths of the lines indicates the statistical uncertainty of the
background estimates.
\label{fig:bkgsub_kperp}
}
\end{figure}

Figures~\ref{fig:bkgsub} and \ref{fig:bkgsub_kperp} show, for several
centrality intervals, differential dimuon yields versus \Aco\
and \kp, respectively, for the production of dimuons satisfying the
corresponding fiducial selection.
The distributions are plotted over ranges that extend beyond the
fiducial limits in order to emphasize the behavior of the HF-decay
background.
Also shown on the plots are the background levels represented
by the constants, $\Chf^{\Aco}$ and $\Chf^{\kp}$, together with statistical
uncertainties of the constants resulting from the template fit.
The backgrounds provide a good description of the behavior of the \Aco\ and \kp\
distributions at large \Aco\ and \kp, although the data show a slight,
centrality-dependent excess over those HF background estimates.
When viewed over ranges of \Aco\ and \kp\ that extend well beyond the
fiducial bounds, that excess is seen to be uniform versus \Aco\ and
\kp, so it is unlikely to have significant contributions from actual
\gamgammumu\ pairs.

To evaluate the contribution from the observed excess,
the measured \Aco\ and \kp\ distributions are fitted with constant
functions over the ranges $\AcopBkgFitMin < \Aco < \AcopBkgFitMax$
and $\KpBkgFitMin < \kp < \KpBkgFitMax$, respectively.
The results of those constant fits -- referred to as ``asymptotic fits'' in
the remainder of this paper -- are indicated with the blue,
long-dashed lines in Figures~\ref{fig:bkgsub} and
\ref{fig:bkgsub_kperp}.

\subsection{Drell--Yan background estimation~\label{sec:DYback}}
As noted above, Drell--Yan production of prompt dimuons represents a
potential background in this measurement. The relevant DY pairs are
produced in the scattering of quarks and
antiquarks with momentum fractions typically less than $10^{-2}$. Thus, the
DY pair yield is sensitive to the degree of shadowing in nuclear PDFs.
 
\begin{figure}[!t]
\centerline{
\includegraphics[width=0.5\textwidth]{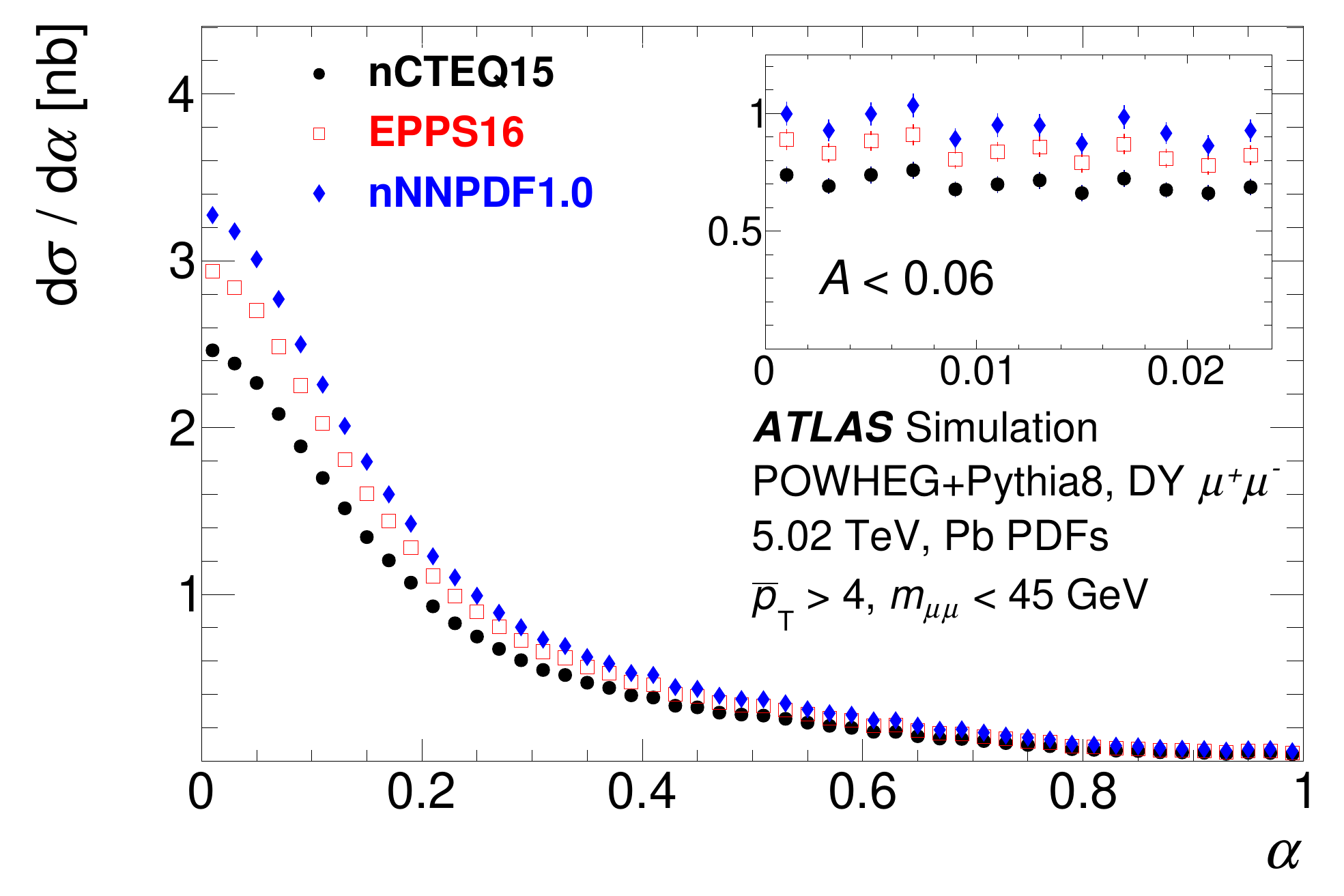}
\includegraphics[width=0.5\textwidth]{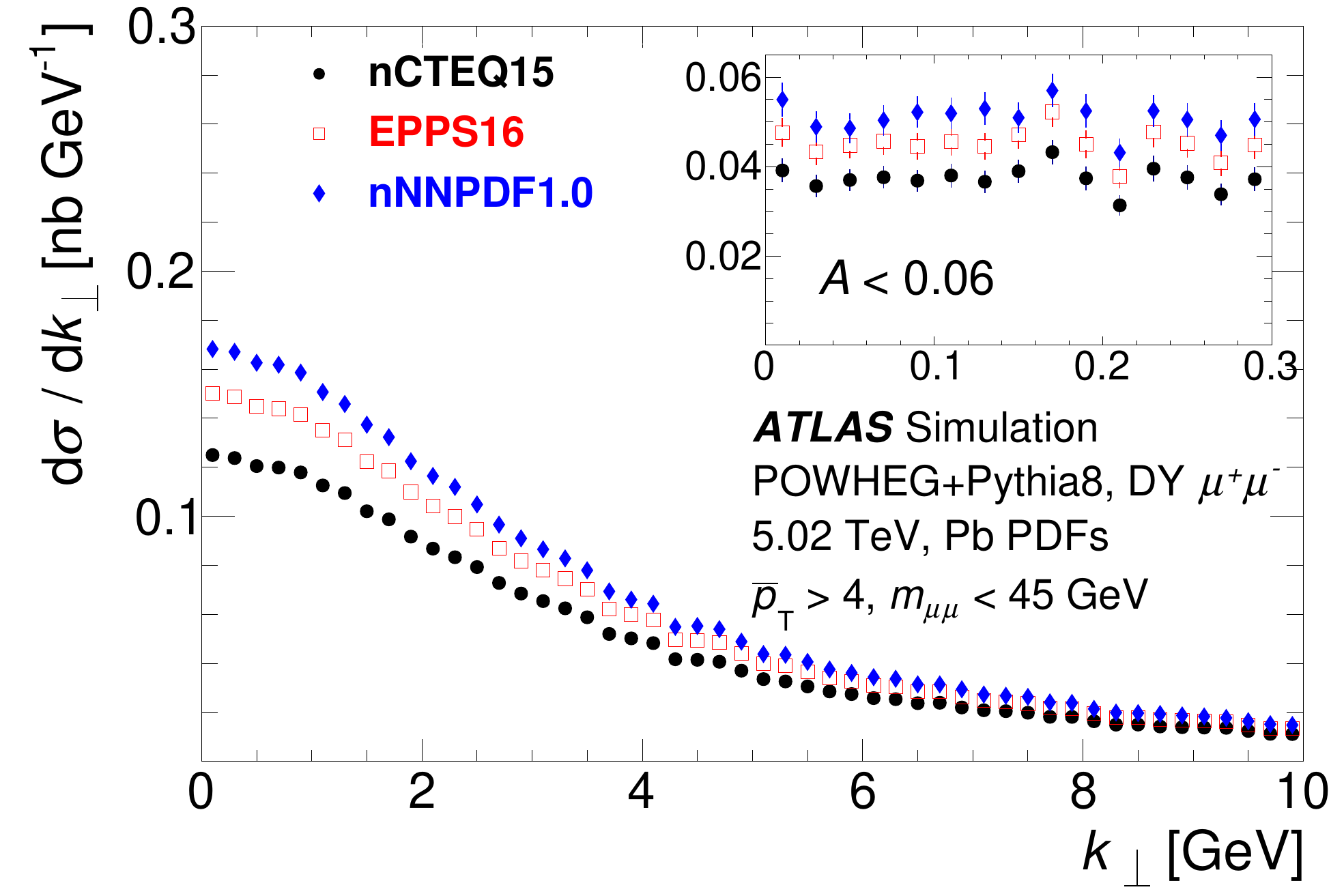}
}
\caption{
Differential cross-sections versus \Aco (left) and \kp\ (right) for
Drell--Yan production of dimuons in $\sqn = 5.02$~\TeV\
nucleon--nucleon collisions from \powheg\ for three
nuclear PDF sets. The error bars indicate statistical uncertainties.
The insets show the differential cross-sections for pairs with $A <
0.06$. They are plotted
over ranges of \Aco or \kp that are twice as wide as the corresponding
fiducial regions. The statistical uncertainties are correlated between
the different nPDF sets because all results derive from the same
set of \powheg\ events.
}
\label{fig:DYDiffPowheg}
\end{figure}

The DY pairs obtained from the \powheg\ MC sample are required to pass
the same kinematic preselections that are applied to the data.
The resulting \powheg\ differential cross-sections as a function of
\Aco and \kp are shown in Figure~\ref{fig:DYDiffPowheg}.
They are plotted over both the full dynamic range populated by the MC
sample  and, in the insets, for pairs  satisfying $A <
\AsymCutForAcop$ over intervals of \Aco and \kp that are twice the
corresponding fiducial ranges. Results are shown for three of the five
nuclear PDFs sets, but the results from all sets differ primarily
in the overall normalization. For the sake of clarity, the effects of PDF uncertainty and
renormalization and factorization scale variations are not shown in
this figure.
The \Aco\ and \kp\ distributions vary substantially as a function of
\Aco\ and \kp\ but are constant, within
statistical uncertainties, over twice the fiducial ranges of the
corresponding variables.
Only a small fraction of the DY pairs passing the preselections used
in this measurement subsequently satisfy the severely restrictive
fiducial selections.
For example, using the nCTEQ15 PDF set, 1.8\% and 1.2\% of the
preselected DY pairs satisfy the Fid-\Aco\ and Fid-\kp\ requirements,
respectively.
 
Table~\ref{table:DYnnsig} lists the effective nucleon--nucleon (NN)
cross-sections, \sigDYeff, for production of
DY dimuons, within the Fid-\Aco\ and Fid-\kp\ fiducial regions,
obtained from \powheg\  for different nuclear PDF sets and for the
\CT[14nnlo] nucleon PDF set.
The table also shows uncertainties obtained by propagating PDF systematic
variations through \powheg\ \cite{Butterworth:2015oua}.

\begin{table}[!b]
\caption{
Effective nucleon--nucleon cross-sections obtained from \powheg\ for the
production of Drell--Yan muon pairs in $\sqn = 5.02$~\TeV\ collisions using
different nuclear PDF sets.
The systematic uncertainties of the fiducial cross-sections obtained by
propagating PDF uncertainties through \powheg\ are also shown. A
separate $\pm 15\%$ uncertainty in the cross-sections, due to
factorization and renormalization scale uncertainties, is not
included in the shown uncertainties.
\label{table:DYnnsig}
}
\centerline{
\begin{tabular}{l|l|l}\hline
PDF set & $\sigDYeffaco~[\text{pb}]$ &
$\sigDYeffkt~[\text{pb}]$  \\ \hline
nCTEQ15   & $12.9  \pm 4.2$ & $7.68 \pm  2.66$\\
EPPS16    & $15.2 \pm 5.7$ & $9.14\pm 3.60$ \\
nNNPDF1.0 & $16.6 \pm 8.7$ & $10.1\pm 5.38 $ \\
nNNPDF2.0 & $17.1 \pm 1.8$ & $10.5\pm 1.15 $ \\
TUJU19    & $17.2 \pm 1.8$ & $10.4\pm 1.6 $ \\ \hline
\CT[14nnlo]  & $24.4 \pm 2.3$ & $15.2\pm 1.4 $ \\ \hline
\end{tabular}
}
\end{table}
 
The \powheg\ DY cross-sections within the fiducial regions used in this
measurement vary by $\approx$30\% between the different nuclear PDF sets,
with the nCTEQ15 PDFs yielding the smallest cross-sections and
nNNPDF2.0 and TUJU19 producing the largest cross-sections. However,
the cross-sections for even those two nuclear PDF sets are smaller
than that obtained from the \CT[14nnlo] set by about 30\% due to nuclear
shadowing. Because the most recent of the implemented nuclear PDF sets, TUJU19 and nNNPDF2.0,
yield consistent results and have smaller uncertainties, the \powheg\ simulations
produced with the nNNPDF2.0 nuclear PDF set are used to estimate the DY background
in this measurement.
 
The number of DY pairs produced in a given centrality interval within
a fiducial region can be estimated using the effective NN
cross-sections obtained above:
\begin{equation}
N^{\text{fid}}_{\text{DY}}(\text{cent}) = \left[\mathcal{L}
\sigma_{\text{had}}^{\PbPb}\right] \Delta \text{cent} \left[\sigDYeff \; \AvgTAA\right],
\label{eq:DY_correction}
\end{equation}
where $\sigma_{\text{had}}^{\PbPb}$ is the total \PbPb\ hadronic
(i.e.\ excluding UPC) cross-section and $\Delta \text{cent}$ represents
the width of a centrality interval expressed as an absolute fraction.
The terms in the first bracket in the equation express the total number
of \PbPb\ collisions sampled for the luminosity used in this
measurement, which is about 15~billion.
When multiplied by $\Delta \text{cent}$, the number of sampled \PbPb\
collisions within the given centrality interval is obtained.
The terms in the second bracket express the per-\PbPb\ collision yield
of DY pairs within the fiducial range of the measurement.
Because the \TAA\ values increase in more central collisions -- they
vary by more than two orders of magnitude between the most peripheral
and most central intervals used in this measurement -- the DY
background will be largest in the most central collisions.
 
Since the \powheg DY \Aco and \kp distributions are uniform within
uncertainties over the fiducial ranges of the measurement,
for the differential measurements of these quantities, the DY
backgrounds are added to the HF backgrounds to yield constants
that are subtracted from the measured distributions:
\begin{equation*}
\Chfdy^{\Aco} = \frac{N_{\text{bkg}}^{\text{fid-\Aco}} + N^{\text{fid-\Aco}}_{\text{DY}}}{\AcopCutForAsym}, \;
\Chfdy^{\kp}  = \frac{N_{\text{bkg}}^{\text{fid-\kp }} + N^{\text{fid-\kp}}_{\text{DY}}}{\KpCutForAsym  }.
\end{equation*}
For the measurement of the asymmetry (\ymumu) distributions, the \powheg\ dimuon
asymmetry (\ymumu) differential cross-sections, scaled according to
Eq.~(\ref{eq:DY_correction}) to produce differential background yields,
are subtracted from the data.

\subsection{Excess observed at large \Aco and \kp}
\label{sec:excess}
\begin{figure}[!t]
\centering
\includegraphics[width=1.0\textwidth]{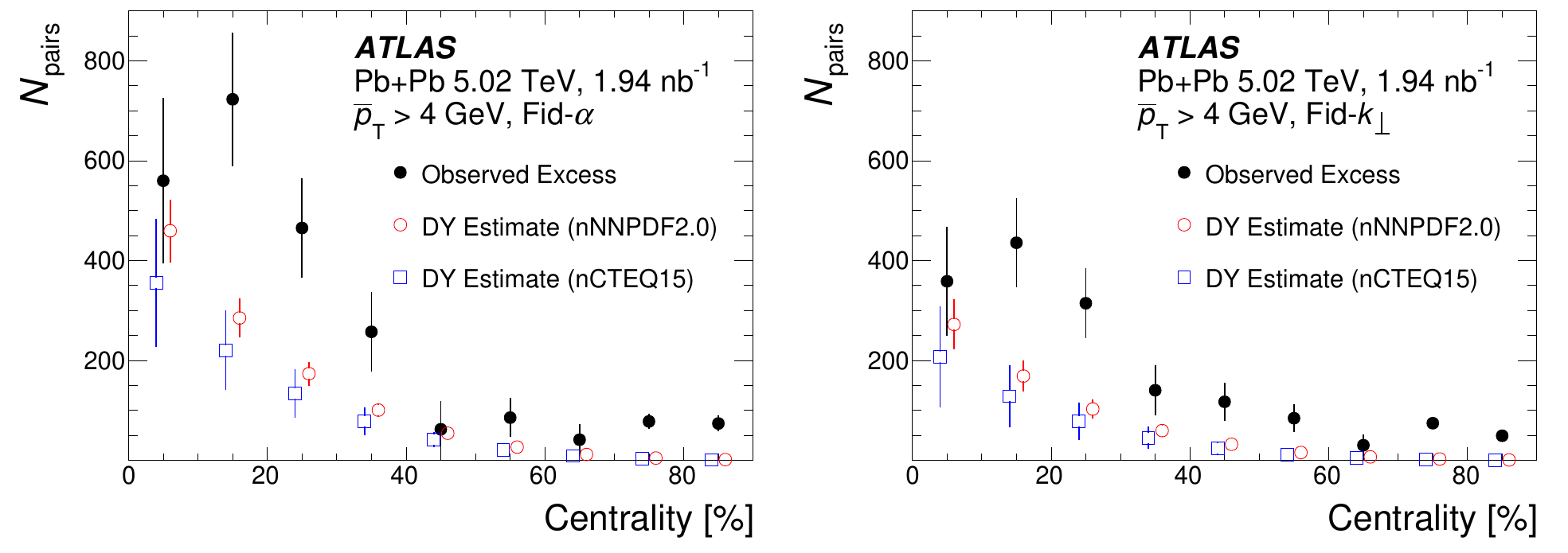}
\includegraphics[width=1.0\textwidth]{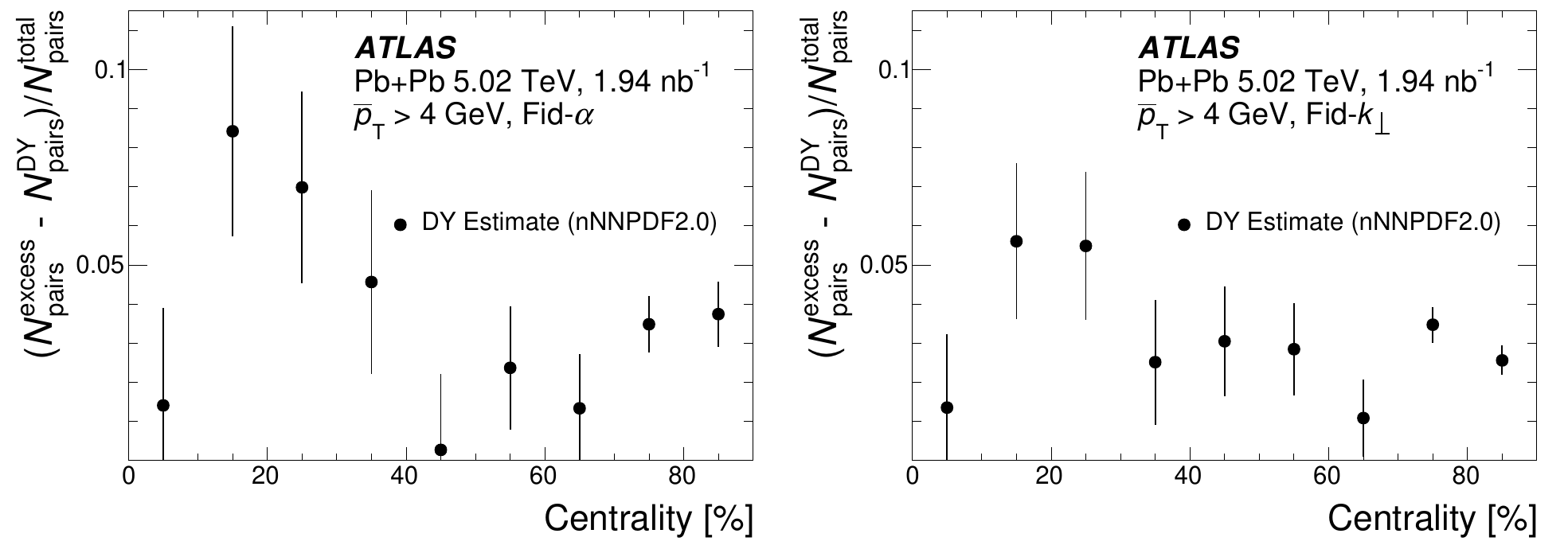}
\caption{
Top row:
comparison of the measured excess background yield for $\ptbar > 4$~\GeV\ with the
estimated DY contribution obtained using the nCTEQ15 and nNNPDF2.0
nuclear PDF sets, which represent the extreme ranges of the
\powheg-predicted rates.
The error bars on the excess background indicate statistical uncertainties,
and the error bars on the DY calculations represent the quadrature sum of the
statistical and systematic uncertainties shown
in Table~\ref{table:DYnnsig}.
The data-points for the DY calculations are shifted slightly
along the $x$-axis, for clarity.
Bottom row:
The differences between the observed background excess and the DY rates
obtained using nNNPDF2.0 PDF set, expressed as a fraction of the fiducial
\gamgammumu\ yields.
The left and right panels correspond to the Fid-\Aco\ and
Fid-\kp\ selections, respectively.
}
\label{fig:DYExcess}
\end{figure}

Figure~\ref{fig:DYExcess} shows, in the top panels, the centrality
dependence of the estimated DY pair yields for the nCTEQ15 and nNNPDF2.0
PDF sets compared with the excess pair yields at large \Aco and \kp
described in Section~\ref{sec:backsub}.  The production of DY pairs
can account for  a substantial fraction of the observed excess, but the \powheg\
simulations systematically underestimate the excess yield
even with the PDF set(s) that predict the largest DY rates.
The differences between the observed background excess and the DY rates
obtained using the nNNPDF2.0 PDF set, expressed as a fraction of the fiducial
\gamgammumu\ yields, are shown in the bottom panels of
Figure~\ref{fig:DYExcess} for the two fiducial selections. No
significant centrality dependence of the excess is evident within
the relatively large uncertainties. Notably, the excess persists in
peripheral events where HF and DY backgrounds are small or negligible.
Averaged over centrality, the excess corresponds to roughly 4\% of the
fiducial yields.
 
Dimuons produced in the decays of $\Upsilon$ states\footnote{There is
zero acceptance for \jpsi decays due to the mass requirement
implicit in the fiducial selections.} represent a plausible
potential background in this measurement. However, the imposed fiducial
selections strongly suppress such a
background by restricting the acceptance to $\Upsilon$ mesons having essentially
zero \pT. The mass distribution shown in Figure~\ref{fig:mass_ptbar} has
an enhancement near the $\Upsilon$ mass, but that enhancement is purely a
result of the fiducial selection. When examined with much finer binning, no
evidence of an excess in the region around the $\Upsilon$ mass is seen for pairs passing
the Fid-\Aco selection, while a significant enhancement is observed in
events passing the preselection and vertex-pointing requirements but
failing the acoplanarity and asymmetry selections.
 
To test for a possible contribution of events associated with
dissociative photon emission by one of the nuclei, the rapidity
distribution of dimuons in the \kp interval $150 < \kp <
300$~\MeV, where the coherent \gamgammumu pairs make little
contribution, was compared with the corresponding distribution for
dissociative processes obtained from SuperChic4 \cite{Harland-Lang:2020veo}. As observed in the
recent ATLAS UPC dimuon measurement \cite{HION-2016-02},
dissociatively produced pairs lie
preferentially at large rapidities because the dissociative
photons tend to be significantly higher in energy than photons produced coherently
by a nucleus. However, the shape of the measured rapidity distribution for dimuons
within the above \kp interval was consistent, within
uncertainties, with that for pairs having $\kp < 150$~\MeV. Thus, the
excess background does not appear to result from dissociative
processes that involve breakup of a nucleon.
 
The absence of a clear dissociative component in non-UPC collisions does not
contradict the observation in Ref.~\cite{HION-2016-02} that
dissociative events comprise 15\% of the so-called \textit{XnXn} sample
which requires at least one neutron in each ZDC. In UPC
collisions, nuclei emit neutrons primarily as a result of Coulomb
excitation processes \cite{Bertulani:2005ru} that have a low probability to break up
one, or especially both, nuclei. For example, in the mass region used
for the measurement in Ref.~\cite{HION-2016-02} the per-nucleus single-breakup
probability was estimated to be 23\% while the double-breakup
probability is 5\%. Since dissociative processes substantially increase
the likelihood that the photon-emitting nucleus
emits neutrons, dissociative events are significantly enhanced in UPC
\textit{XnXn} events. However, in non-UPC \PbPb\ collisions, neutrons are
produced with unit probability for all except the most peripheral
collisions. Thus, the \textit{XnXn} requirement does not enhance the rate for
dissociative processes, with the result that the dissociative
contribution is, at most, a few percent of the pairs over the kinematic range considered in
Ref.~\cite{HION-2016-02}. Of those, about 40\% fall within the
Fid-\Aco range of this measurement. Such a small contribution would not be visible
in the presence of the much larger HF and DY backgrounds, except
possibly in the very most peripheral collisions.
 
Since the origin of the excess observed in the data is not known,
it is not subtracted as part of this measurement. However, to
account for the possibility that it represents an unidentified
background, it is treated as a systematic uncertainty of the
background subtraction procedure.
 
\subsection{Background-subtracted \Aco, \kp, and $A$ distributions}
 
\begin{figure}[h]
\centering
\includegraphics[width=1.0\textwidth]{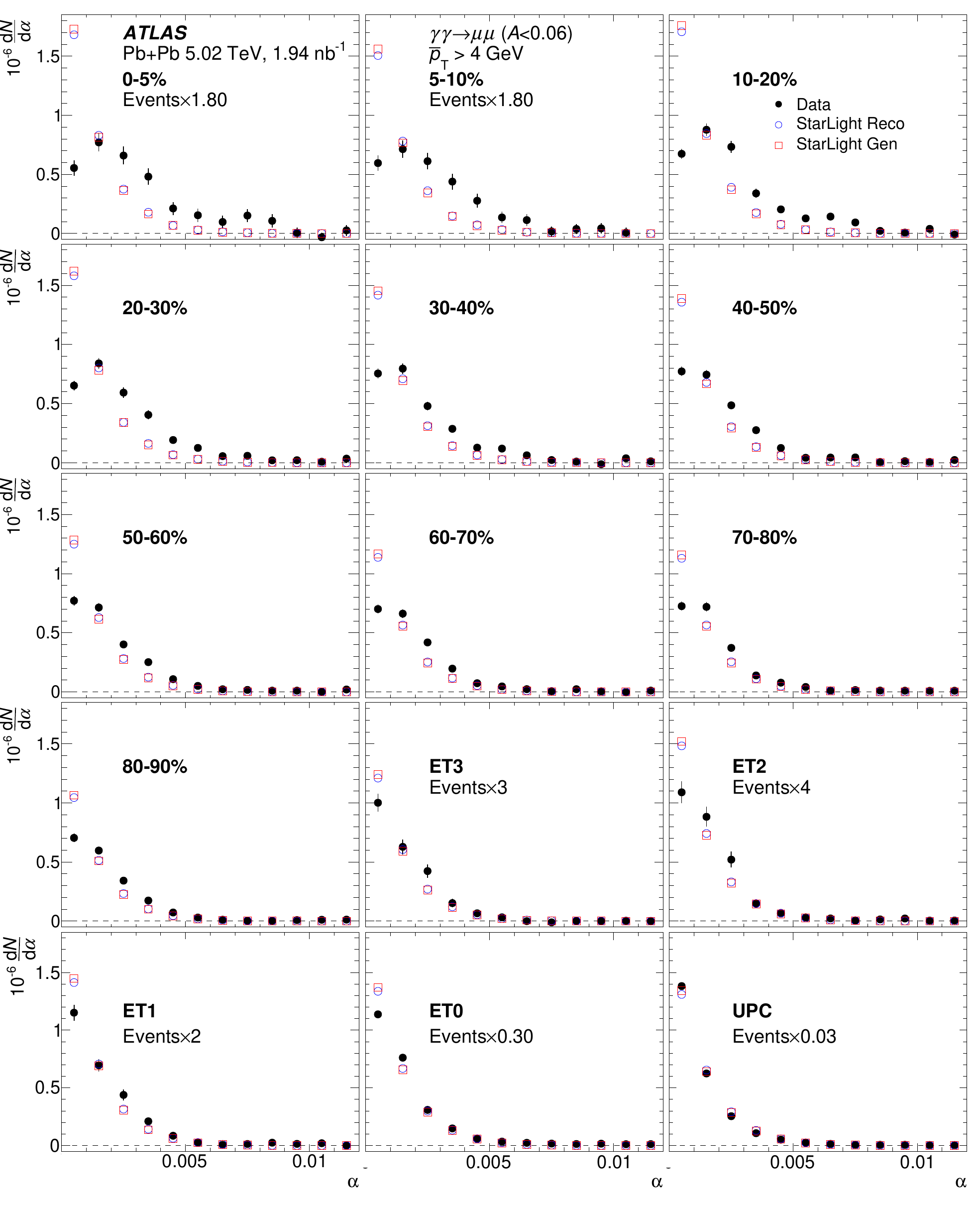}
\caption{
\HFDY\ background-subtracted \Aco\ distributions for pairs satisfying
the Fid-\Aco selection and having
$\ptavg>4~\GeV$ in different centrality intervals from the most
central 0--5\% (top left) to the UPC interval (bottom right).
For a few panels the distributions are scaled to allow a common
$y$-axis range for the plots.
The scale factors are stated on the panels.
Also shown for comparison are the generated and reconstructed
distributions obtained from the \starlight\ simulation samples.
The \starlight\ generated and reconstructed distributions are scaled to match the
the corresponding data distributions over the $\Aco<\AcopCutForAsym$ interval.
\label{fig:Acop_afterSubtraction}
}
\end{figure}

The \HFDY\ background-subtracted acoplanarity distributions for
\gamgammumu\ pairs having $\ptbar>4~\GeV$ are shown in
Figure~\ref{fig:Acop_afterSubtraction} for several
centrality intervals.
The error bars indicate statistical uncertainties only.
The figures also show the distributions from the \starlight\ MC
sample for both the generated and reconstructed muon pairs.
The generated and reconstructed distributions differ only slightly due to the
excellent angular resolution of the inner detector.
The data and reconstructed MC distributions agree well in the UPC and most peripheral
non-UPC centrality intervals, but in more central collisions the data
systematically deviate from the MC predictions, with the data having wider \Aco\
distributions and a suppression at the smallest \Aco\ values.

\begin{figure}[h]
\centering
\includegraphics[width=1.0\textwidth]{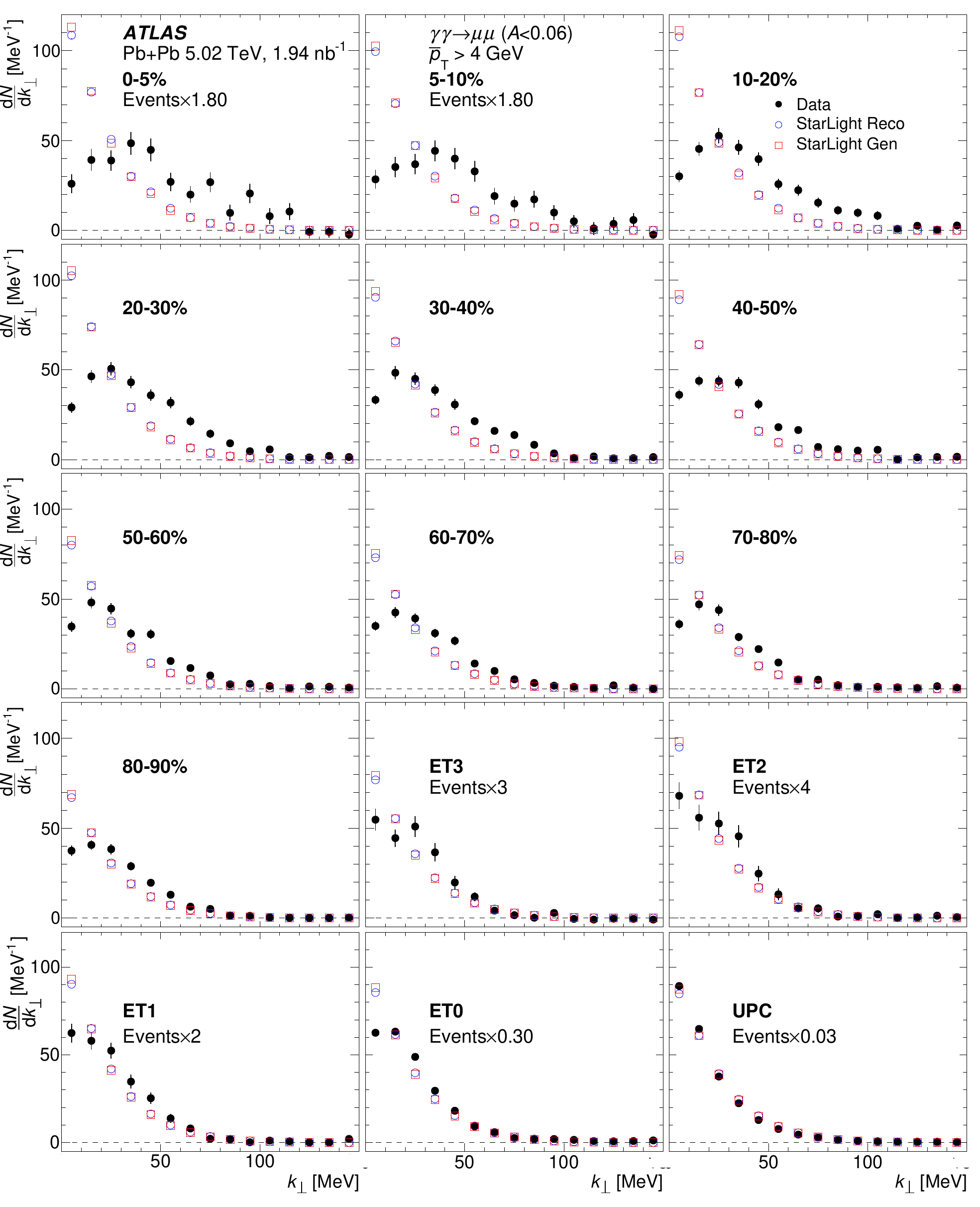}
\caption{
\HFDY\ background-subtracted $k_{\perp}$ distributions for pairs
satisfying the Fid-\kperp selection and having
$\ptavg>4~\GeV$ in different centrality intervals from the most
central 0--5\% (top left) to the UPC interval (bottom right).
For a few panels the distributions are scaled to allow a common
$y$-axis range for the plots.
The scale factors are stated on the panels.
Also shown for comparison are the generated and reconstructed
distributions obtained from the \starlight\ simulation samples.
The \starlight\ generated and reconstructed distributions are scaled to match the
the corresponding data distributions over the $\kp<\KpCutForAsym$ interval.
\label{fig:AcopWeighted_afterSubtraction}
}
\end{figure}

Results similar to those shown in Figure~\ref{fig:Acop_afterSubtraction},
but for the \HFDY-subtracted \kp\ distributions, are presented in
Figure~\ref{fig:AcopWeighted_afterSubtraction}.
Like the \Aco\ distributions, the \kp\ distributions broaden and
show a depletion near $\kp = 0$ with decreasing centrality percentile.
However, the depletion becomes more significant than that observed in
the \Aco\ distributions and becomes apparent at larger centrality
intervals.
Similar to the \Aco\ results, the MC generated and reconstructed
distributions barely differ while the reconstructed MC and data distributions have
very different behavior except in the most peripheral
centrality intervals.
 
\begin{figure}[h]
\centering
\includegraphics[width=1.0\textwidth]{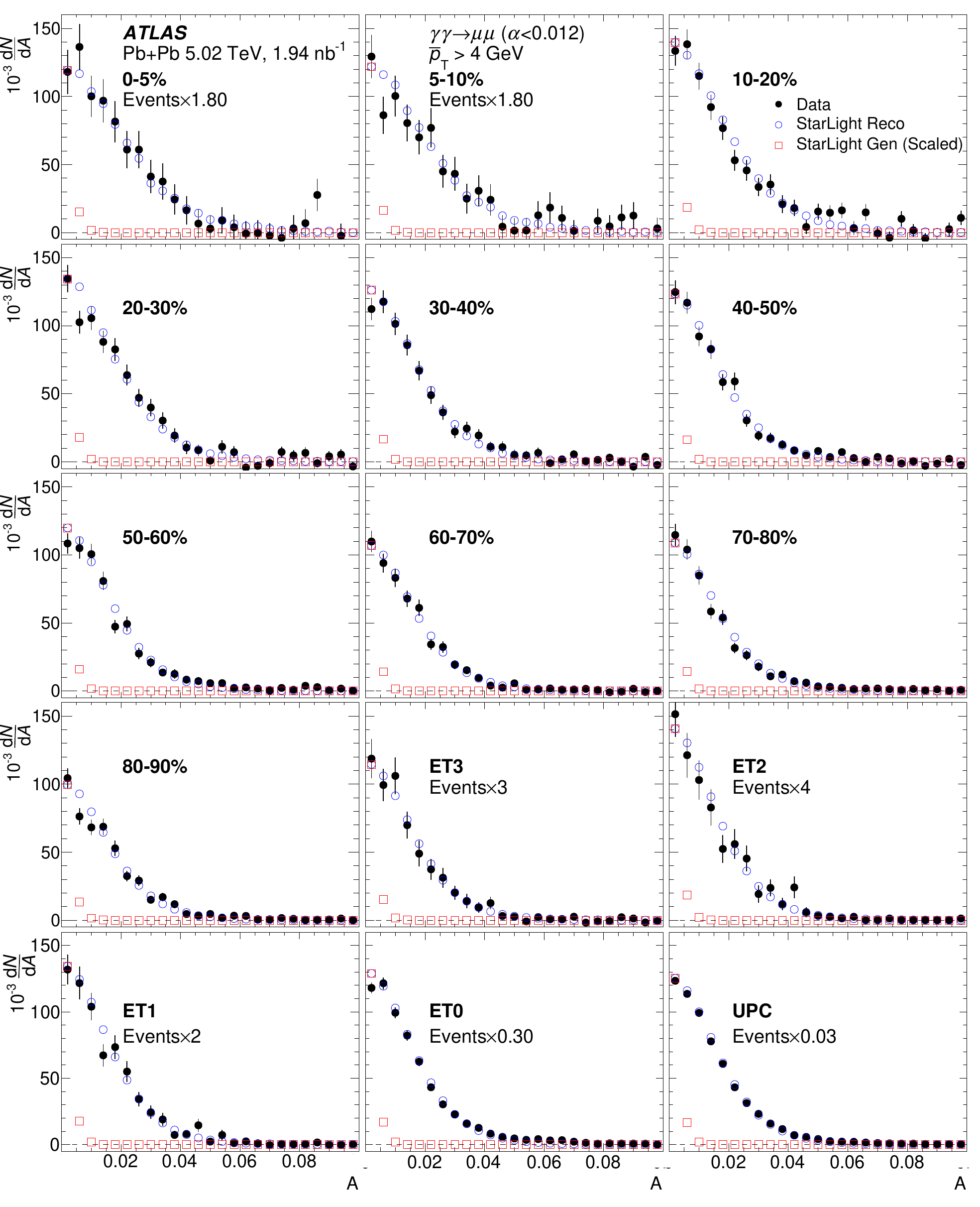}
\caption{
\HFDY\ background-subtracted asymmetry distributions for pairs having
$\ptavg>4~\GeV$ in different centrality intervals from the most
central 0--5\% (top left) to the UPC interval (bottom right).
In a few of the panels, the distributions are scaled to allow a common
$y$-axis range for the plots.
The scale factors are stated on the panels.
Also shown for comparison are the generated and reconstructed
distributions obtained from the \starlight\ simulation samples.
The \starlight\ reconstructed distributions are scaled to match the
the corresponding data distributions over the $0<\Amumu<0.1$ interval.
The \starlight\ generated distributions, which are much narrower than
the measured distributions or the data, are scaled vertically to allow
direct comparison with the other distributions.
}
\label{fig:Asym_afterSubtraction}
\end{figure}

Figure~\ref{fig:Asym_afterSubtraction} shows \HFDY\
background-subtracted \Amumu\ distributions compared with the generated
and reconstructed \starlight\ MC distributions. The data agree well
with the reconstructed \starlight\ \Amumu\ distributions in all centrality
intervals. This provides further evidence that the muon
pairs in this measurement result from \gamgammumu processes.
The resolution of the \Amumu\ measurement, demonstrated by the
difference between the MC generated and reconstructed distributions,
is poor relative to that of the \Aco\ and \kp\ measurements and is
not sufficient to probe effects at transverse momentum scales of 100~\MeV.

\FloatBarrier
 
\subsection{Unfolding}

Iterative Bayesian unfolding~\cite{DAgostini:1995} implemented using the
ROOUnfold~\cite{Adye:2011gm,} framework is used to unfold the
\Aco\ and \kp\ distributions.  Although the resolution in each of these
variables is comparable to or smaller than the histogram bin widths
used in Figures~\ref{fig:Acop_afterSubtraction} and
\ref{fig:AcopWeighted_afterSubtraction}, the observed suppression at
small \Aco\ and \kp\ values may be influenced by the finite resolution
in these quantities.  Also, the distributions in the most peripheral
centrality intervals are sufficiently steep that the finite \Aco\ or
\kp\ resolution noticeably affects their shape.  Response matrices in
both \Aco\ and \kp\ are produced from the \starlight\ MC
sample. Because the \starlight sample contains few pairs at large
\Aco\ or \kp, those regions of the response matrices are augmented
using parameterizations of the pair \Aco\ and \kp\ response.  Different
response matrices are produced for all combinations of centrality and
\ptbar\ intervals used in the measurement.
\begin{figure}[!tb]
\centering
\includegraphics[width=1.0\textwidth]{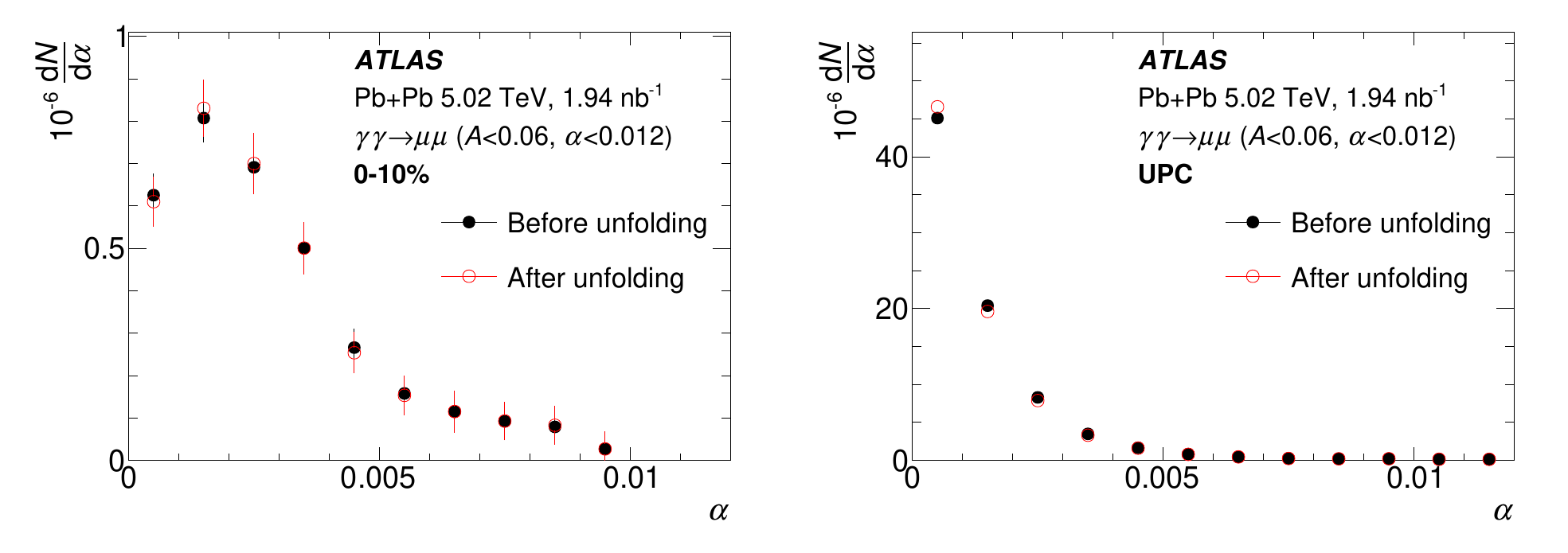}
\includegraphics[width=1.0\textwidth]{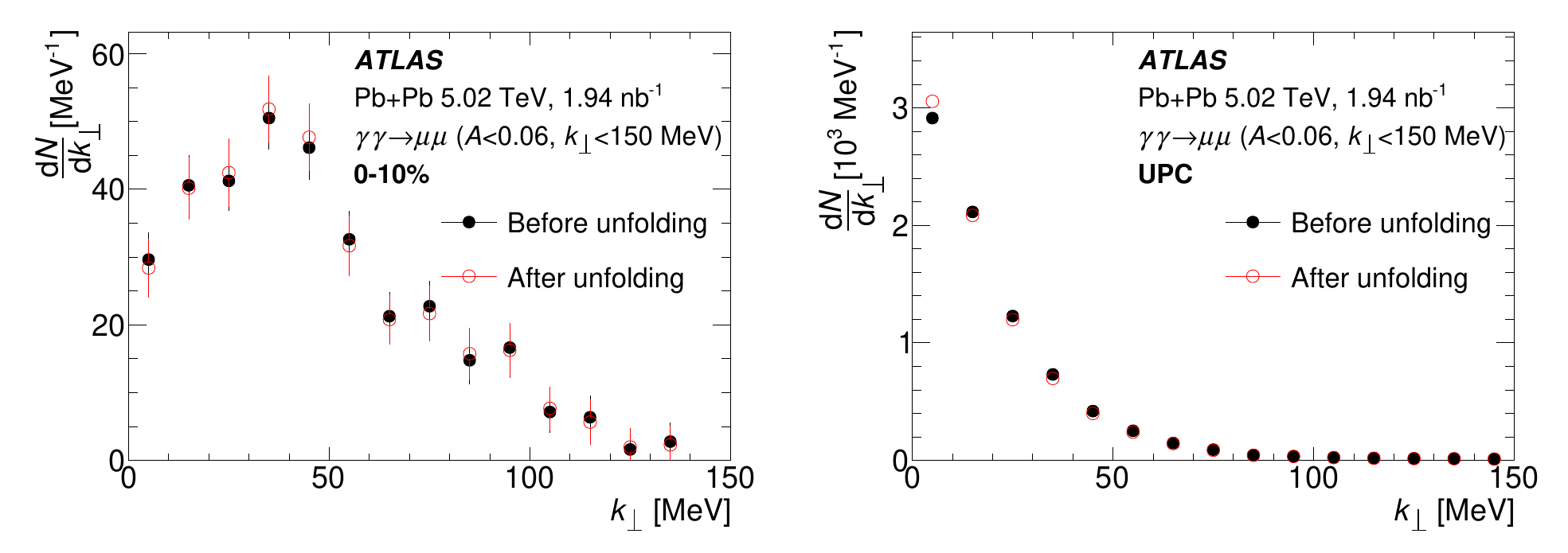}
\caption{
The \Aco\ (top) and \kp\ (bottom) distributions of dimuons before
and after unfolding.
The left and right panels correspond to the 0--10\% and UPC
intervals, respectively.
\label{fig:unfold_example}
}
\end{figure}
 
Although the response matrix for the unfolding is nearly diagonal, the
steep derivative in the \Aco and \kp distributions at small values
means that the unfolding takes a few iterations to converge. Thus,
a conservative choice to use three iterations was made. The statistical
uncertainties  of the unfolded results, obtained using
pseudoexperiments on the data and on the response matrix, are
essentially unchanged from the distributions before unfolding. The
reweighting of the MC \kp and \Aco distributions to match the data
has negligible impact on the unfolded distributions.
 
The \Aco\ and \kp\ distributions are unfolded prior to the
\HFDY\ background subtraction.
Since the background is flat in both \Aco\ and \kp, it is unaffected
by the unfolding, so the \HFDY\ subtraction is the same for
the unfolded distributions as it is in the measured distributions
shown above.
 
Figure~\ref{fig:unfold_example} shows a comparison of the
\HFDY-subtracted \Aco\ (top row) and \kp\ (bottom row)
distributions for the 0--10\% (left) and UPC (right) intervals.
For the 0--10\% centrality interval, the effect of the unfolding is only
easily visible at the smallest \Aco\ and \kp\ values.
For the UPC interval, the unfolding sharpens the distributions and
increases the yield at the smallest \Aco\ and \kp\ values.
 
No attempt is made to unfold the asymmetry distributions, as the $A$
resolution precludes sensitivity to effects at the momentum scales
observed in the \Aco\ and \kp\ distributions.
However, an efficiency correction, obtained from the \starlight\ MC sample, is
applied to account for a loss of pairs resulting from the
$A < \AsymCutForAcop$ selection due to migration in $A$.
The use of the \starlight\ asymmetry distribution is acceptable for this purpose because of the
insensitivity of the asymmetry to the observed differences between the
data and MC samples.

% End of text imported from the .//tex/analysis.tex input file

\section{Observables and systematic uncertainties}
\label{sec:systematics}

% The next lines are included from the .//tex/systematics.tex input file
 
\subsection{Observables\label{sec:observables}}
Cross-sections for \gamgammumu\ production are obtained in intervals of centrality
and \ptbar\  using:
\begin{equation}
\sigma_{\text{cent}}\fidup(\ptbar) = \frac{1}{\mathcal{L}}
\left[N_{\text{sig}}\fidup(\text{cent},\ptbar) - N\fidup_{\mathrm{DY}}(\text{cent},\ptbar)\right],
\label{eq:CS}
\end{equation}
where $\mathcal{L}$ represents the integrated luminosity, and
$N_{\text{sig}}$ represents the number of background-subtracted pairs (see Eq.~\eqref{eq:signal_bkg_yields} and Eq.~\eqref{eq:DY_correction}).
The cross-sections in Eq.~\eqref{eq:CS} can be evaluated using the
Fid-\Aco\ or Fid-\kp\ fiducial selections.
However, the two fiducial selections give results that agree within statistical
and systematic uncertainties.
For this reason, the cross-sections are presented in this \papertype
only for the Fid-\kp\ selection.
 
To reduce systematic uncertainties resulting from overall normalization
factors, e.g.\ due to efficiency corrections and luminosity,
``\NormYields'' ($Y$) are calculated according to
\begin{equation}
Y(\text{cent},\ptbar) =
\frac{\sigma_{\text{cent}}\fidup(\ptbar)}{\sigma_{\text{tot}}\fidup(\ptbar)},
\label{eq:yielddef}
\end{equation}
where $\sigma_{\text{cent}}\fidup(\ptbar)$ is defined in
Eq.~\eqref{eq:CS} and $\sigma_{\text{tot}}\fidup(\ptbar)$ is
the total cross-section, obtained by summing
$\sigma_{\text{cent}}\fidup(\ptbar)$ over all centrality
intervals, including the UPC interval.

Differential cross-sections for \gamgammumu\ production
are obtained from the unfolded \Aco\ and \kp\ distributions.
In the \kp\ case, the differential cross-section is calculated as
\begin{equation*}
\frac{{d}\sigma}{{d}\kp} = \frac{1}{\mathcal{L}}
\left(\frac{\Delta N}{\Delta\kp} - \Chfdy^{\kp}\right),
\end{equation*}
where $\Delta N$ represents the number of unfolded
counts in a given \kp\ interval of width $\Delta \kp$,
and other factors are defined in
Section~\ref{sec:analysis}.
A similar formula applies to the calculation of
${d}\sigma/{d}\Aco$.
These cross-sections are obtained for each combination of centrality
and \ptbar\ intervals included in the analysis using the corresponding fiducial selection.

\subsection{Systematic uncertainties}
 
This section discusses estimates of the systematic uncertainties in the
cross-sections, \NormYields, and \kp\ distributions.
The uncertainties in the \Aco\ distributions mirror those in the \kp\ distributions and are not
discussed separately.
The following sources of systematic uncertainty are considered:
 
\begin{itemize}
\item \textbf{Luminosity:} The systematic uncertainty of the combined
2015 and 2018 \PbPb\ integrated luminosity is 1.5\%, obtained using the
LUCID-2 detector~\cite{LUCID2} for the primary luminosity
measurement~\cite{DAPR-2011-01}.
 
\item \textbf{Muon working point: }
The analysis is performed using the ``medium'' muon selection
requirements~\cite{MUON-2018-03}.
The measurements are repeated using the ``tight'' working
point~\cite{MUON-2018-03}, which results in an increase in the muon
purity, but reduces the muon-pair yields by about 20\%.
After applying an efficiency correction appropriate for the alternative working point,
the cross-sections and \NormYields\ differ by 7\% and 2\% respectively. This variation
is included as a systematic uncertainty that
covers possible contributions from misidentified muons.
For the differential cross-sections, the change in the muon working
point mainly introduces point-by-point statistical fluctuations with
no systematic trend beyond the $7\%$ change in the overall normalization.
Thus, no additional systematic uncertainty is applied to the \Aco and
\kperp differential cross-sections or yields.

\item \textbf{Trigger efficiency: } The trigger efficiencies are
evaluated directly from a data sample large enough for the
statistical uncertainties in the efficiencies to be
negligible. However, a systematic uncertainty is applied to cover the effects of possible
centrality-dependent differences between the composition of muons used for the efficiency
measurement and that of muons used in the measurement. It is evaluated from the following difference:
\[ \delta\varepsilon^{\text{cent}} =
\left|\left(\varepsilon^{\text{cent}}_{\text{medium}} -
\langle\varepsilon_{\text{medium}}\rangle\right) -
\left(\varepsilon^{\text{cent}}_{\text{tight}} - \langle\varepsilon_{\text{tight}}\rangle\right)\right|,\]
where the averages are taken over centrality intervals.
The systematic uncertainty essentially compares the efficiencies obtained
using ``medium'' and ``tight'' muon selections in a given centrality interval
while removing average changes due to the different muon populations.
This uncertainty is within $3\%$ across most of the centrality range,
but increases to $6\%$ for the 0--10\% most central collisions.
 
\item \textbf{Reconstruction efficiency: } The effects of the muon
reconstruction efficiency are accounted for by applying weights --
the inverse product of the single-muon efficiencies -- to each
muon pair used in the analysis. The efficiencies are functions of
\pt, $q\eta$, and collision centrality. Uncertainties in the
efficiencies produce uncertainties in the measured
distributions and in the cross-sections and yields.  These uncertainties are evaluated
by varying the reconstruction efficiencies within their
own uncertainties~\cite{MUON-2018-03} and by evaluating the resulting
changes in the measurements.  This procedure yields a ${\approx}2\%$
systematic uncertainty in the cross-sections and a ${\lesssim}0.5\%$ uncertainty
in the \NormYields.  For the differential measurements, the
variation of the reconstruction efficiency introduces no systematic
effects beyond a change in the normalization which is already
accounted for in the cross-section.

\item \textbf{\dzp\ and \zzp\ selections: }
A potential systematic uncertainty associated with the vertex-pointing requirement
is assessed by varying the \dzp\ requirement from its default value of $\dzp<0.1$~mm,
to 0.08 and 0.14~mm, and the \zzp\ requirement to $\zzp<0.22$~mm (from 0.2~mm)
and including the resulting variation as an uncertainty. The
uncertainty covers both the  efficiency  for signal
muons to pass the selections and the effect of the vertex-pointing
requirements on the HF background subtraction. The former is most
relevant in the UPC and peripheral centrality intervals where the
uncertainty is small or negligible. The latter is relevant
in more central collisions. For the cross-section and \NormYields measurement,
the largest systematic uncertainty, ${\approx}2\%$, is obtained in the
0--5\% centrality interval.
Varying the \dzp and \zzp\ requirements introduces no systematic
variation of the dimuon yields with \Aco or \kp.

\item \textbf{Background shape parameterization: }
The corrected \Aco\ and \kp distributions are obtained by subtracting
a background that is taken to be constant, consistent with fits to the
distributions in Figure~\ref{fig:backgroundshape}.
The sensitivity of the results to the assumption of a constant background
is evaluated by parameterizing the background by linear functions
and performing the subtraction using the resulting function.  The
deviation from the default analysis result is
included as a systematic uncertainty.
This uncertainty applies only to the differential cross-sections.
It is at most ${\approx}0.2\%$ and is  negligible compared to the
other uncertainties.
 
\item \textbf{Signal template variation for \dzp\ fits:} The HF
background levels estimated from the template fits depend on the
shape of the signal template that, in the default analysis, is
obtained from MC simulation. An alternative, data-driven approach uses for
the signal template the \dzp distribution measured in the UPC
sample, which is composed almost entirely of \gamgammumu\ pairs. An
additional, centrality-dependent smearing determined from the MC
simulation is applied to the UPC \dzp distribution to account for
the poorer $d_0$ resolution in the more central collisions.
A $\approx$1\% variation between the results obtained from the default
analysis and from the alternative signal templates is included as a
systematic uncertainty of the cross-sections and the \NormYields.
For the differential cross-sections, the variation of the template
produces an additive change in the background, and thus, in the
background-subtracted results. The corresponding systematic uncertainty
effectively accounts for the sensitivity of the
signal fractions to the uncertainty in the shape of the
signal template.
 
\item \textbf{Excess background:}
As discussed in Section~\ref{sec:excess}, the difference between the
excess observed at large \Aco or \kp after HF background subtraction and the \powheg DY estimate is
taken as a systematic uncertainty to cover possible unidentified
background sources. It is evaluated separately in each kinematic bin
in the measurement, except for the measurements as a function of
\ymumu and \Dy, where the low pair yields at largest \ymumu and \Dy
values preclude a statistically viable determination. For those measurements,
the excess fraction is evaluated averaged over \ymumu and \Dy and then
applied in each bin.
This uncertainty is treated one-sided (downwards only), and
is typically within 4\% for the cross-section and 2\%
for the \NormYields.
 
\end{itemize}
The final uncertainties in the cross-sections are obtained by summing the
individual uncertainties listed above in quadrature. The resulting
uncertainty in the cross-sections vary slightly with centrality but are typically $\approx$8\%.

% End of text imported from the .//tex/systematics.tex input file

\section{Results}
\label{sec:results}

% The next lines are included from the .//tex/results.tex input file
\subsection{Theoretical calculations}
The results of the measurements presented here are compared with three
sets of theoretical calculations: \starlight, QED
calculations based on a generalization of the EPA
method, and calculations using a fully
quantum-mechanical treatment of the photon transverse momentum
distribution through Wigner distributions.
For all three, the calculations rely on distributions of impact
parameters populated by \PbPb\ collisions in each centrality interval
that are obtained from a Glauber MC simulation~\cite{Loizides:2014vua}
that also provides the basis for the ATLAS centrality calibration.
 
While \starlight\ was originally designed for UPC processes, it was
recently updated to allow the calculation of \gamgammumu\ processes
over fixed impact parameter ranges, although there is no treatment of
the impact parameter dependence of the photon transverse momenta.
Since it has already been established that the non-UPC \Aco\ and \kp\
distributions disagree with \starlight, only the predictions for
the total dimuon production cross-sections are compared with data.
 
The authors of Ref.~\cite{Zha:2018tlq,} provided self-normalized
dimuon acoplanarity distributions in the calibrated \PbPb\ centrality
bins used in this measurement. These results are referred to as
``QED'' in the plots and in the discussion below. The authors of
Ref.~\cite{Klein:2020jom} have provided self-normalized
dimuon \Aco and \kp\ distributions. These results are referred to
as ``PWF'' in the rest of this paper.
 
\subsection{Cross-sections and relative yield measurements}
 
\begin{figure}[h]
\centering
\includegraphics[width=0.5\textwidth]{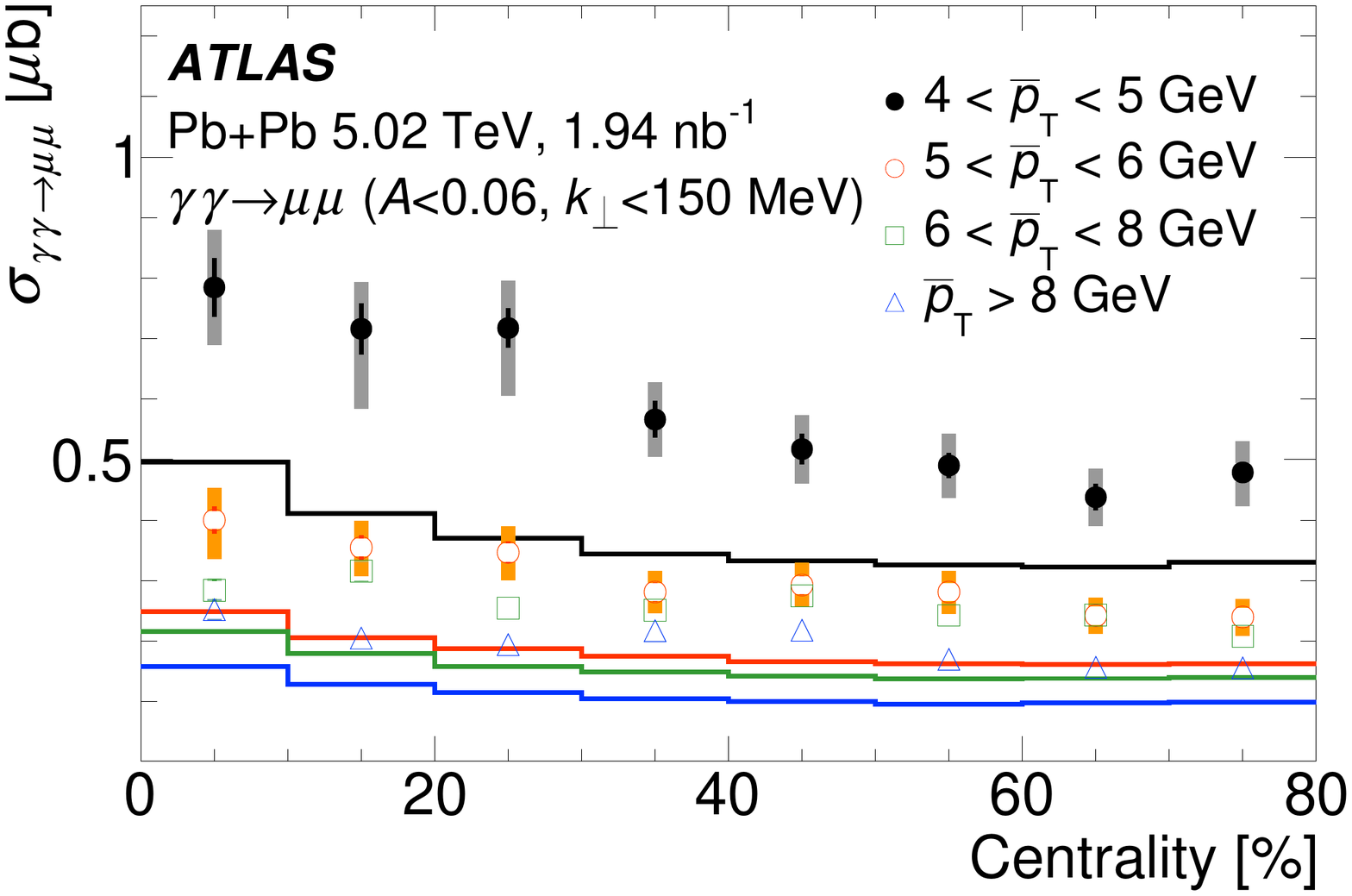}
\includegraphics[width=0.5\textwidth]{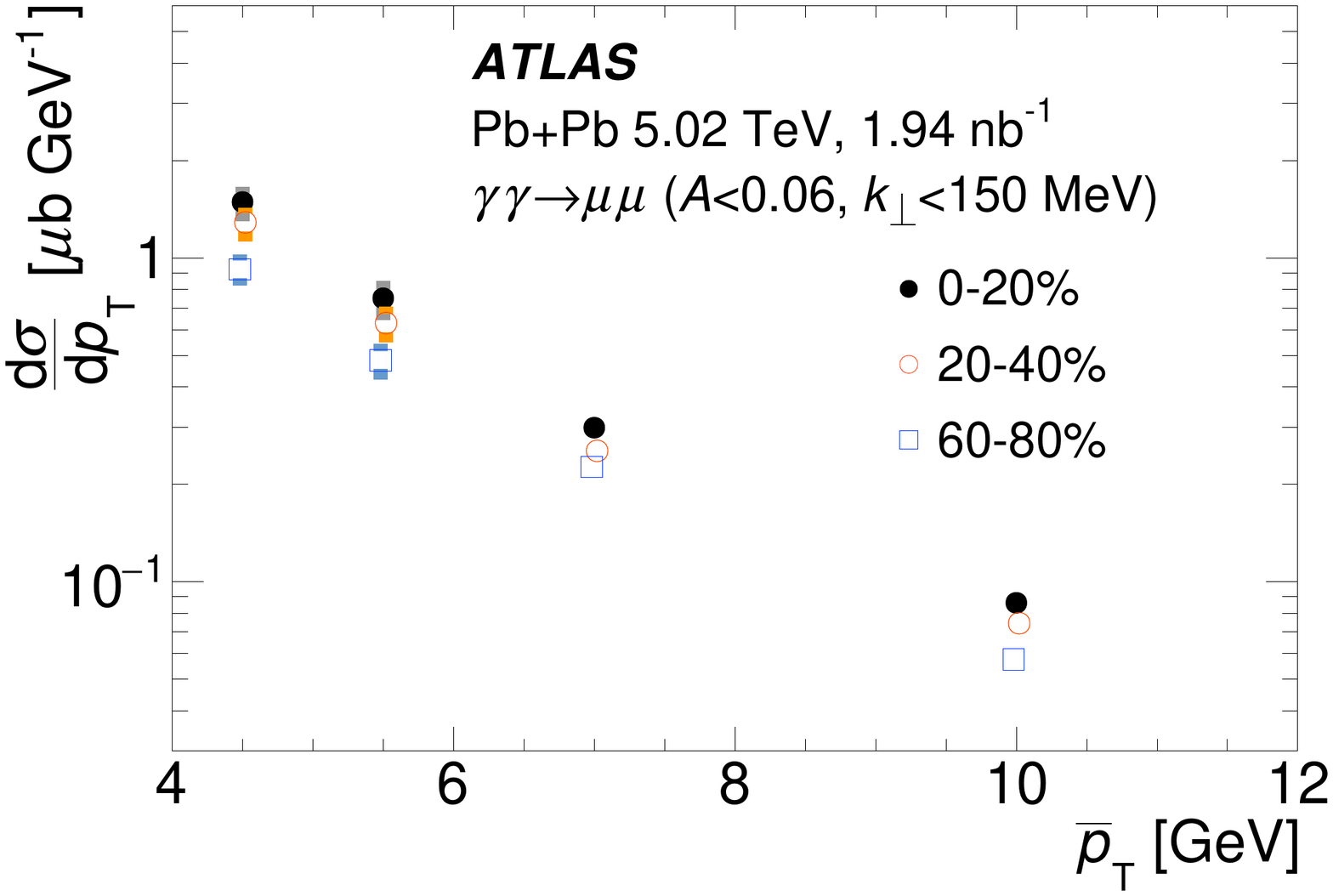}
\caption{
The cross-sections for \gamgammumu production in non-UPC
\PbPb\ collisions as a function of centrality (top) and \ptbar (bottom).
The error bars indicate statistical (lines) and
systematic uncertainties (shaded boxes).
The solid lines in the left panel represent
the results of \starlight calculations (see text) for the given
\ptbar interval.
The data points in the right panel for the 0--20\%  and 60--80\%
centrality intervals are staggered, for clarity.
\label{fig:CrossSections}
}
\end{figure}

The top panel of Figure~\ref{fig:CrossSections} shows the measured \gamgammumu cross-sections as a
function of centrality for four different \ptbar\ intervals.
Also shown for comparison are calculations from the \starlight\
event generator.
The cross-sections decrease slightly from central to peripheral events.
This decreasing trend is qualitatively reproduced by \starlight,
but the predicted cross-sections from \starlight\ are considerably
different from the measurements.
It has been observed~\cite{HION-2016-02} that
\starlight\ predicts too low a \gamgammumu yield in UPC
collisions.
One possible explanation~\cite{Azevedo:2019fyz} for the \starlight deficit in UPC
dimuon production is its model's truncation of the single-nucleus photon
flux at transverse distances smaller than the nuclear radius.
That truncation may have greater impact on \gamgammumu processes
in non-UPC nuclear collisions, where the neglected regions have
larger geometric overlap. In the bottom panel of Figure~\ref{fig:CrossSections},
the measured cross-sections in three different centrality intervals
show a rapid decrease with \ptbar.

The total cross-sections for \gamgammumu production in the Fid-\Aco and
Fid-\kp fiducial regions but with no centrality or ZDC requirement after
subtraction of the HF and DY backgrounds are
\begin{align*}
\sigma_{\gamgammumu}^{\text{Fid-}\Aco} & = 61.9\pm 0.2\text{(stat)} - 5.6\text{(syst)} + 5.2\text{(syst)}~\mu\text{b},\\
\sigma_{\gamgammumu}^{\text{Fid-}\kp}  & = 61.3\pm 0.2\text{(stat)} - 5.2\text{(syst)} + 5.0\text{(syst)}~\mu\text{b}.
\end{align*}
The cross-sections in
the two fiducial regions differ by only 1\% because they are dominated by the
ultraperipheral contribution, for which the \Aco and \kp restrictions
remove only a small fraction of the pairs. These cross-sections are larger than that reported in the recent UPC
measurement~\cite{HION-2016-02}, primarily due to the fact that the present measurement does not
impose a minimum mass requirement of 10~\GeV\ and also because it
includes \gamgammumu pairs in non-UPC events. The total cross-section
obtained using the same fiducial selections as the UPC measurement is
consistent with the result reported in Ref.~\cite{HION-2016-02}.
 
\begin{figure}[h]
\centering
\includegraphics[width=0.7\textwidth]{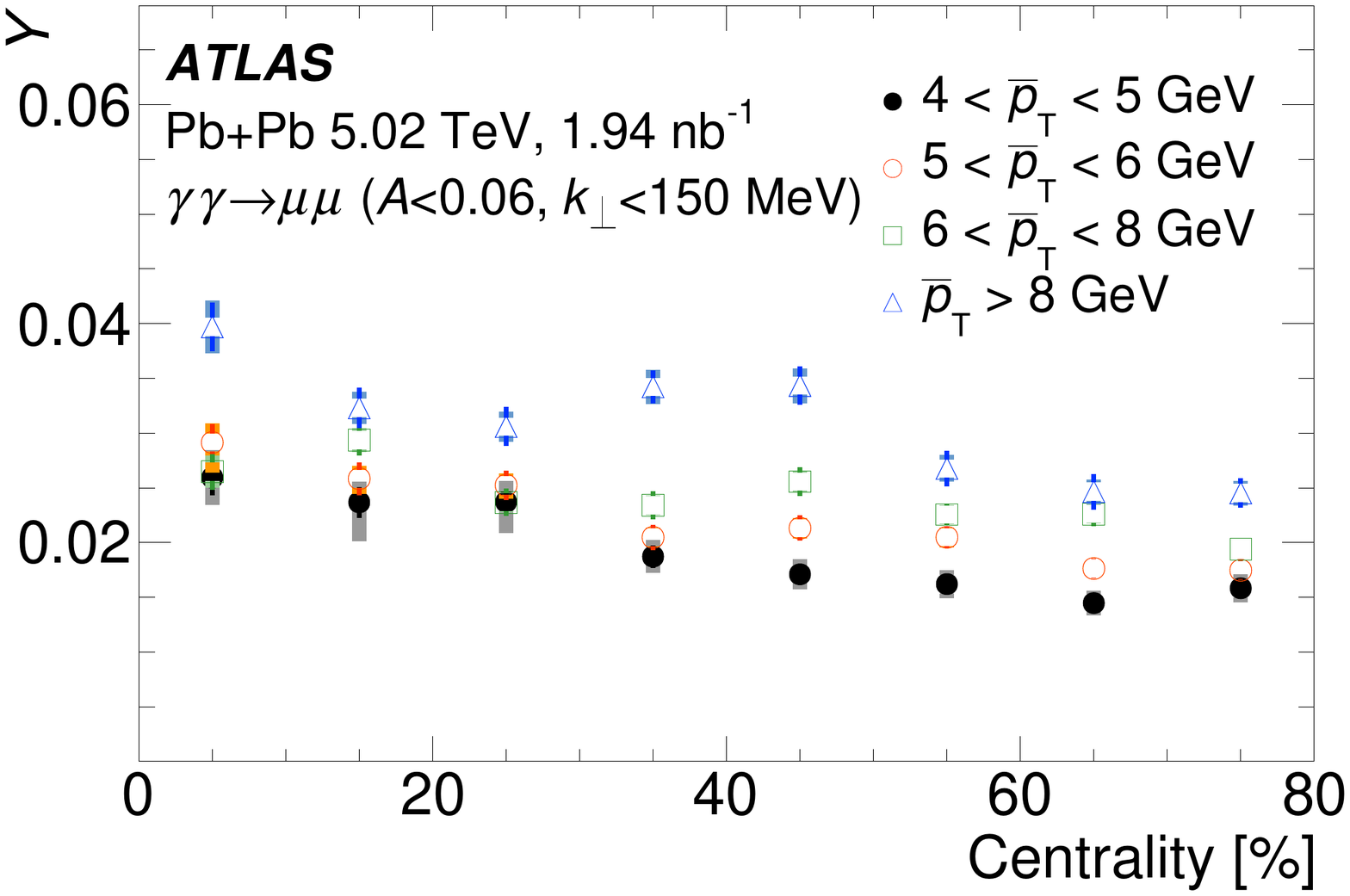}
\caption{
Measured \NormYields\ of \gamgammumu\ pairs passing the Fid-\kp
selection as a function of centrality for four \ptbar\ intervals.
The error bars indicate statistical (lines) and
systematic (shaded boxes) uncertainties.
\label{fig:yields_afterSubtraction1}
}
\end{figure}

\begin{figure}[h]
\centering
\includegraphics[width=0.5\textwidth]{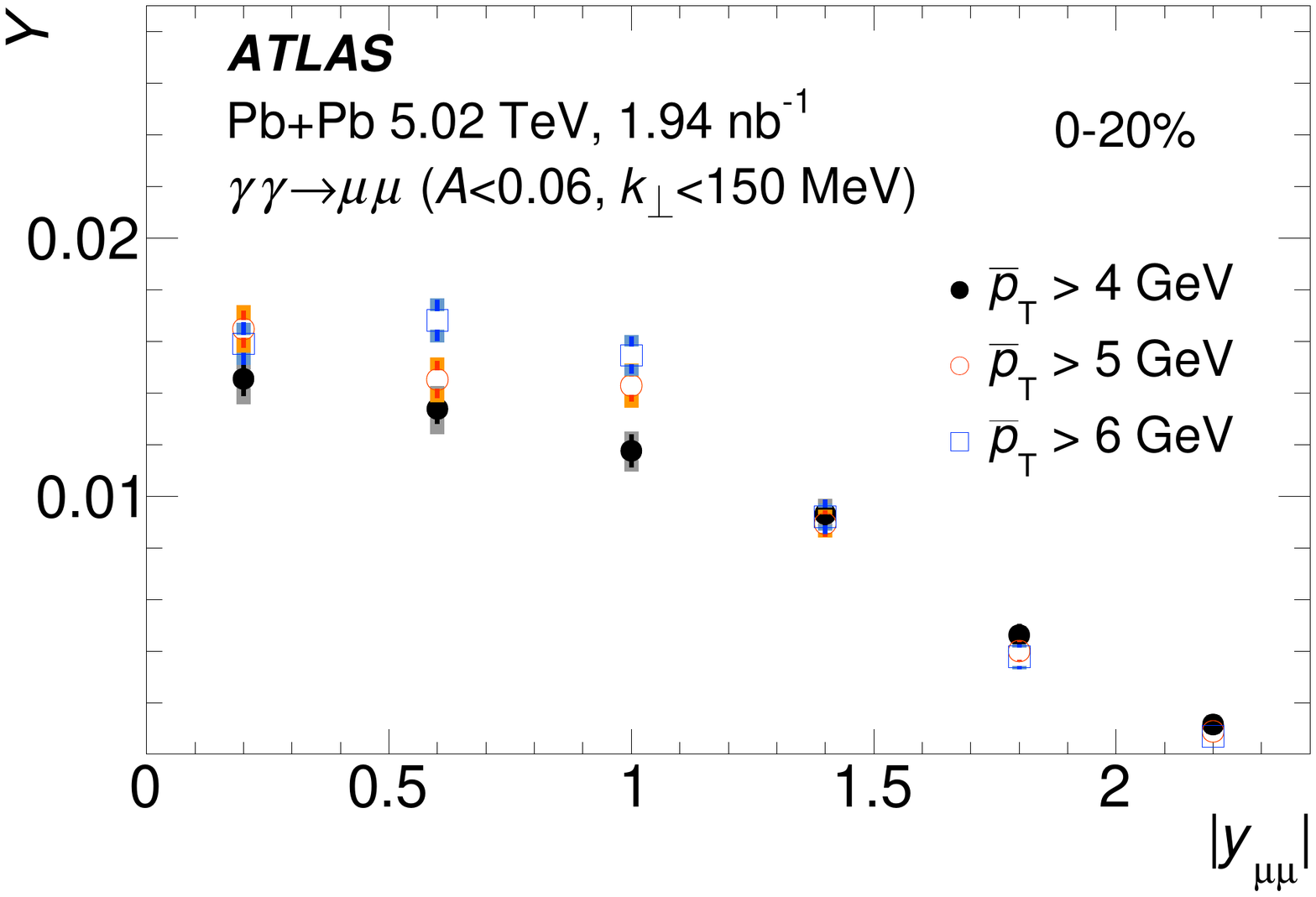}
\includegraphics[width=0.5\textwidth]{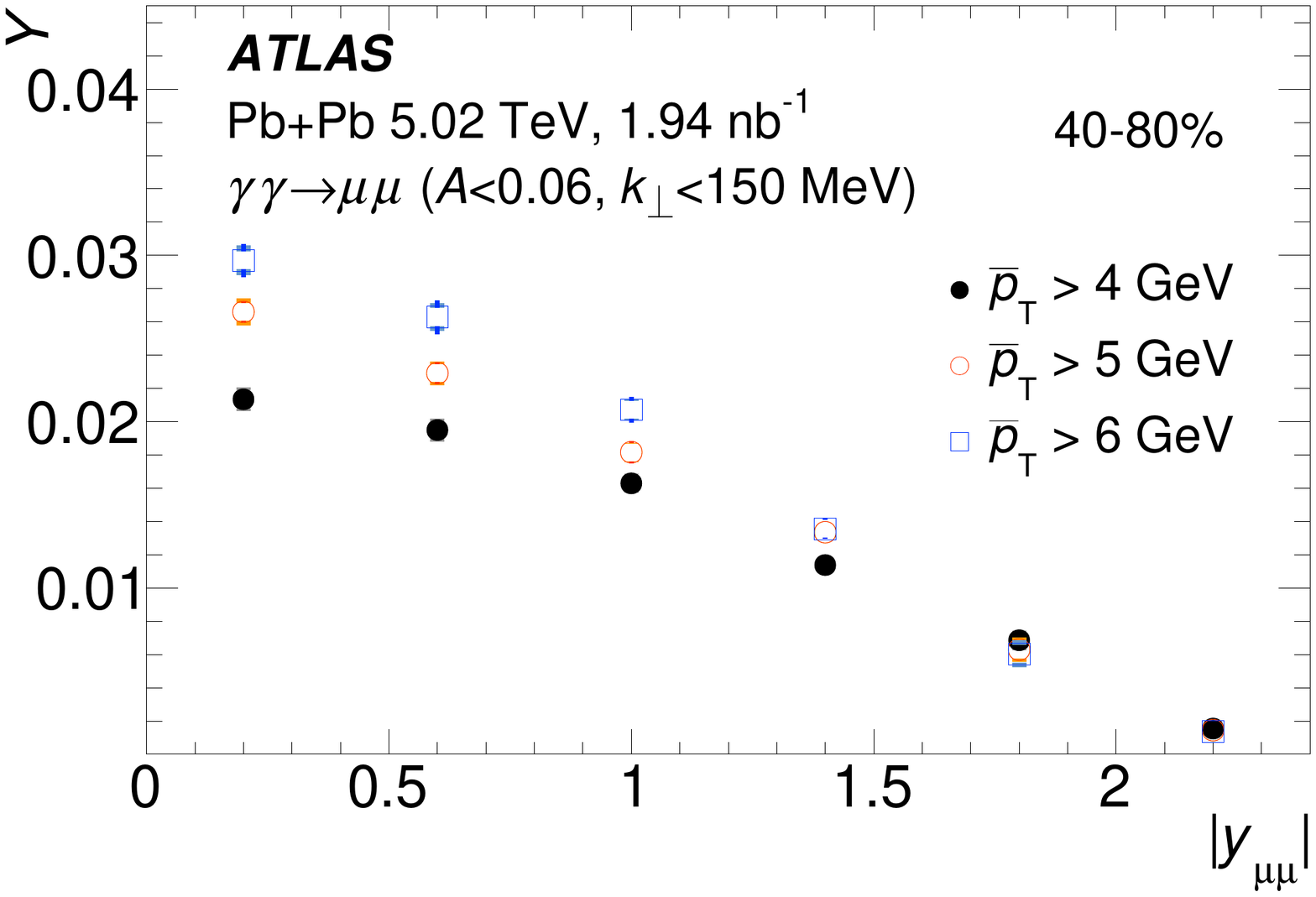}
\caption{
Measured \NormYields\ of \gamgammumu\ pairs passing the Fid-\kp
selection as a function of the pair rapidity, \mbox{$\ymumu$}\xspace.
The widths of the \mbox{$\ymumu$}\xspace intervals along the
$x$-axis are 0.4 units.
The top and bottom panels correspond to the 0--20\% and 40--80\%
centrality intervals, respectively.
The error bars indicate statistical (lines) and
systematic (shaded boxes) uncertainties.
For the 40-80\% interval, both sets of errors are
typically smaller than the symbols.
\label{fig:yields_afterSubtraction_rapidity_dep}
}
\end{figure}

Figure~\ref{fig:yields_afterSubtraction1} shows the \NormYields\  of
\gamgammumu\ pairs, $Y$, calculated according to
Eq.~\eqref{eq:yielddef},  in various centrality intervals and for the four \ptbar\
ranges presented in Figure~\ref{fig:CrossSections}.
For a given \ptbar\ interval, it is observed that the $Y$ values
increase from peripheral to central events, i.e.\ the fraction of
\gamgammumu\ pairs produced in central events per centrality
percentile is larger than in peripheral events.
It is also observed that at higher \ptbar, a larger fraction of
\gamgammumu\ pairs is produced in more central events.
This is compensated for by a reduction of the \NormYields\
in the \UPC\ interval (not shown in
Figure~\ref{fig:yields_afterSubtraction1}), since by construction the
\NormYields\ add up to one (see Eq.~\eqref{eq:yielddef}).
This result indicates that the dimuon transverse momentum spectra
become  harder with decreasing \PbPb\ collision impact parameter.

Figure~\ref{fig:yields_afterSubtraction_rapidity_dep} shows the
\ymumu dependence of the \NormYields\ for two
centrality and three \ptbar\ ranges.
The \NormYields\ decrease with rapidity, approaching zero
at $\ymumu=2.4$.
This behavior primarily results from the $\eta$ acceptance of the two
muons used to form the pair: $|\eta|<2.4$, which restricts the pair
rapidity to  $\ymumu< 2.4$.
However, the \ymumu\ dependent trends in the \NormYields\ can be compared
between the different \ptbar\ intervals, since the acceptance effects
are similar for the different \ptbar\ intervals.
For $\ymumu{<}1.2$ a stronger \ptbar dependence is observed,
with the \NormYields\ increasing with \ptbar.
While at forward rapidities, the \NormYields\ are consistent,
within uncertainties, for the three \ptbar intervals shown
in Figure~\ref{fig:yields_afterSubtraction_rapidity_dep}.

\subsection{\Aco and \kp distributions}
\begin{figure}[!hp]
\centering
\includegraphics[width=1.0\textwidth]{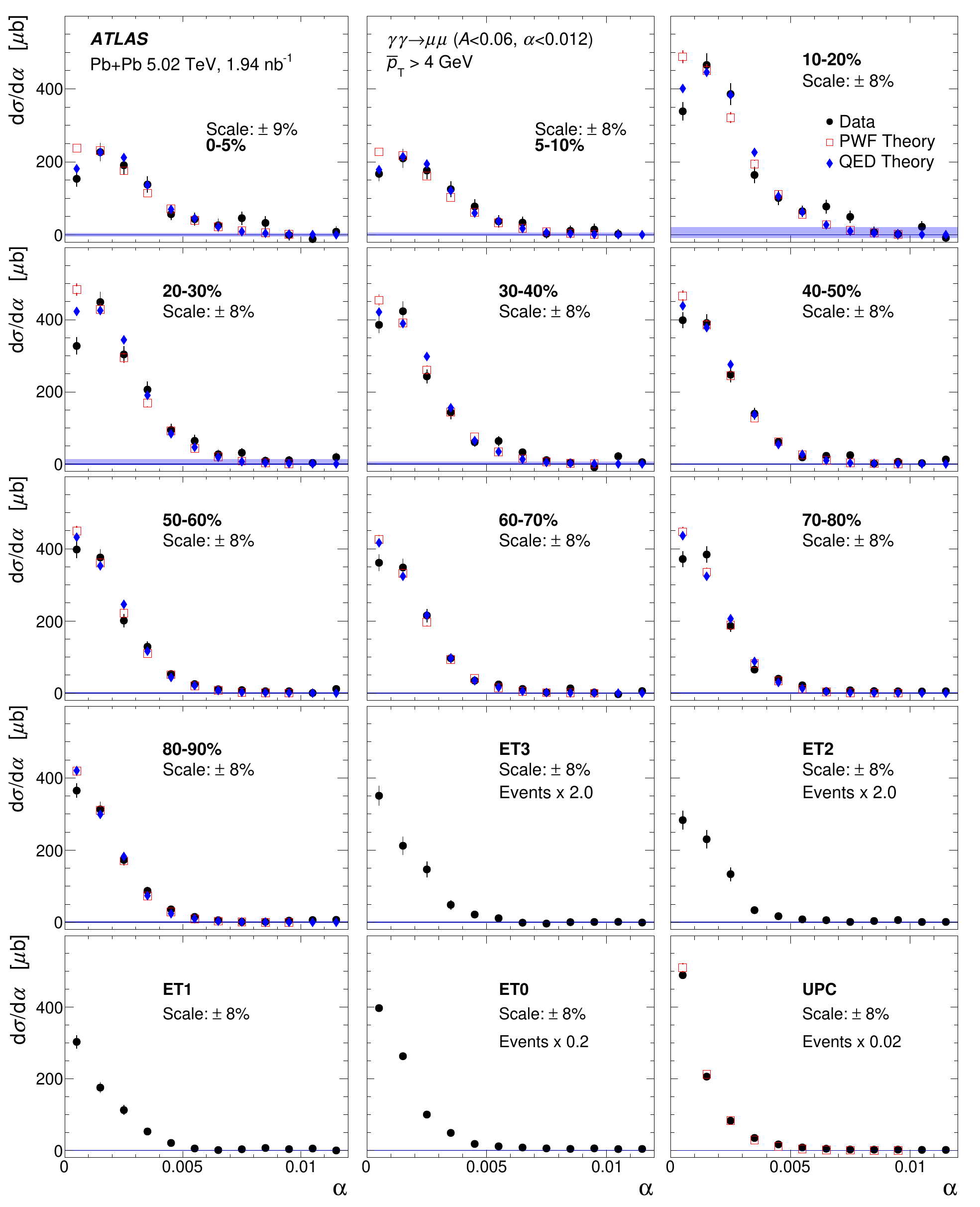}
\caption{
Differential cross-sections as a function of \Aco\ for \gamgammumu\
pairs passing the Fid-\Aco\ selection.
Each panel represents a different centrality or \sumETFCal interval,
with the last panel representing the UPC interval.
The error bars indicate combined statistical and systematic uncertainties, excluding
the background subtraction uncertainties, which are indicated by a
shaded band at ${d}\sigma/{d}\Aco = 0$,
and overall normalization uncertainties, which are quoted on each panel as ``Scale''.
Also shown are the results of the PWF theoretical calculations
(see text).
}
\label{fig:final_acoDistsCent}
\end{figure}
 
\begin{figure}[!hp]
\centering
\includegraphics[width=1.0\textwidth]{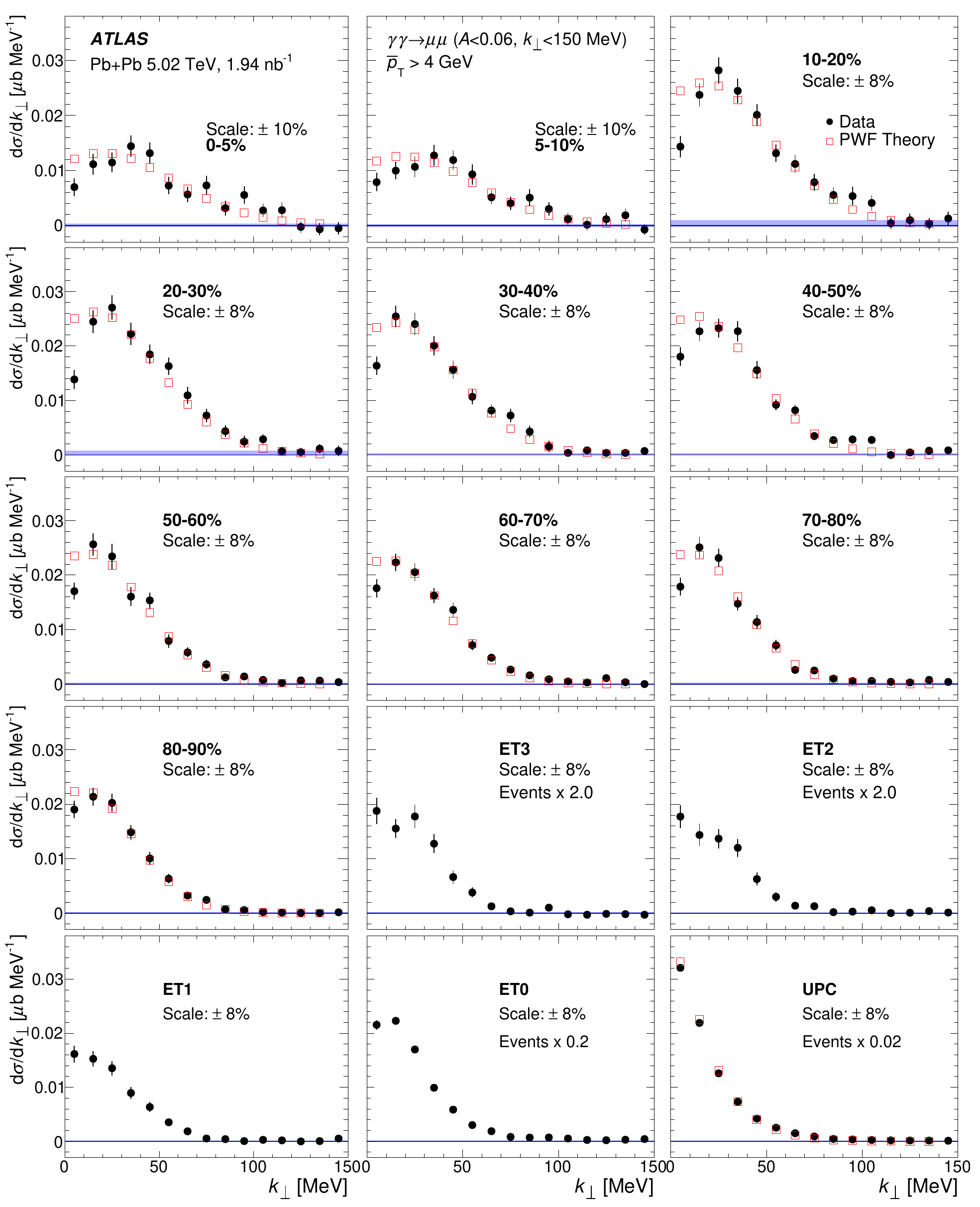}
\caption{
Differential cross-sections as a function of \kp\ for \gamgammumu\
pairs passing the Fid-\kp\ selection.
Each panel represents a different centrality or \sumETFCal interval,
with the last panel representing the UPC interval.
The error bars indicate combined statistical and systematic uncertainties, excluding the
background subtraction uncertainties, which are indicated by shaded
bands at ${d}\sigma/{d}\kperp = 0$, and
overall normalization uncertainties, which are quoted on each panel as ``Scale''.
Also shown are the results of the QED and PWF theoretical calculations
(see text).
\label{fig:final_ktDistsCent}
}
\end{figure}
 
Figures~\ref{fig:final_acoDistsCent} and \ref{fig:final_ktDistsCent}
show two of the primary results of this measurement: the
differential cross-sections as a function of \Aco\ and \kp,
respectively, in different \PbPb\ centrality intervals, including the
ET0--ET3 and UPC intervals.
The main observation in the previous ATLAS measurement of non-UPC
\gamgammumu\ production~\cite{HION-2018-11}, namely a
centrality-dependent broadening of the \Aco\ distributions,
is confirmed.
A similar broadening is observed in the \kp\ distributions.
It is interesting to note that
the \Aco and \kperp distributions in the most peripheral ET0--ET3
intervals are visibly broadened compared to the UPC interval.
 
In addition to the broadening, a centrality-dependent suppression of the
dimuon yield at small \Aco\ and \kp\ values is seen in the data.
This suppression was present, but not statistically significant, in the
results of the previous measurement~\cite{HION-2018-11}.
The suppression is greater in the \kp\ distributions for reasons that
are made clear below.
The PWF and QED calculations reproduce the measured distributions in the
UPC interval reasonably well.
The calculations also qualitatively reproduce the centrality-dependent
broadening observed in the data.
In addition, both calculations predict a centrality-dependent
depletion at small \Aco\ (or \kp).
However, the depletion observed in the data is better reproduced by the
QED calculations, which describe well the observed suppression in the \Aco
distributions in the most central collisions; agreement in the middle
centrality intervals is not as good. In Ref.~\cite{Zha:2018tlq,}, the
depletion at small values of acoplanarity is attributed to quantum
interference effects. It is not, as yet, understood why the PWF
calculations show less suppression.

\begin{figure}[!hp]
\centering
\includegraphics[width=1.0\textwidth]{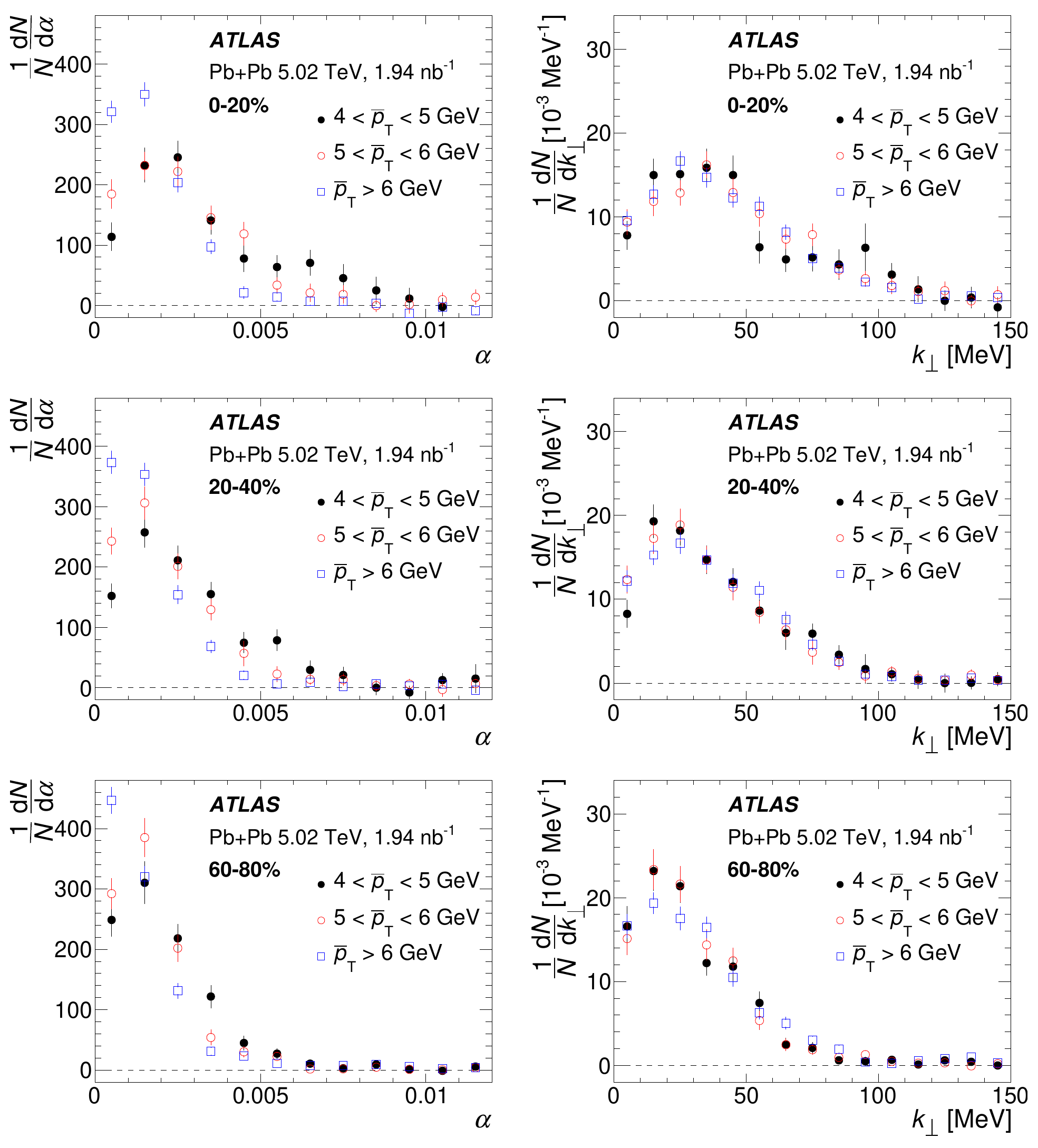}
\caption{
Background-subtracted \Aco\ (left column) and \kp\ (right column)
distributions measured in different \ptbar\ intervals for three
centrality intervals: 0--20\%, 20--40\%, and 60--80\%.
The distributions are normalized to have unit integral.
The error bars indicate statistical and systematic uncertainties
added in quadrature.
\label{fig:AcopKperpPtDependence}
}
\end{figure}
Assuming that the broadening and depletion in the \Aco or \kp
distributions results from a physical process having an
associated transverse momentum scale, $q_{\text{T}}$, then, as argued in
Ref.~\cite{HION-2018-11}, the modifications of the \Aco distributions,
naively, should scale with the ratio of that scale to the muon momentum,
$q_{\text{T}}/\ptbar$. In contrast, the \kp distributions, which effectively
probe the component of the dimuon $\vec{p}_{\text{T}}$ perpendicular
to the dimuon axis, should only depend on the scale, $q_{\text{T}}$.
To test this hypothesis, Figure~\ref{fig:AcopKperpPtDependence} shows
distributions of \Aco (left panels) and \kp (right panels)
for three different \ptbar\ intervals and three different
centrality intervals  chosen to improve the statistical significance of the
data. The \Aco distributions show a clear dependence on \ptbar
such that  at higher \ptbar\ the distributions become narrower and show less suppression
at $\Aco = 0$. Thus, integrating over \ptbar\ will smooth out some of the
centrality-dependent features in the data. In contrast, the \kp
distributions show no significant dependence on \ptbar. Thus, an
integration over \ptbar will have little or no impact on the shape of
the \kp distributions. For this reason, the \kp variable should be
preferred for future measurements and theoretical calculations.

\FloatBarrier
\subsection{Characterizing the centrality dependence of \kperp distributions}
In Ref.~\cite{HION-2018-11}, the broadening of the acoplanarity
distributions was characterized by a transverse momentum
scale, obtained from the RMS of the \Aco\ distribution.
The observation that the \Aco\ and the \kp\ distributions have the
most-probable value shifted away from zero means that the RMS does
not fully capture the modifications observed in the data.
Nonetheless, the moments provide a model-independent means to quantify
modifications of the \Aco\ and \kp\ distributions.
 
\begin{figure}[!b]
\centering
\includegraphics[width=1.0\textwidth]{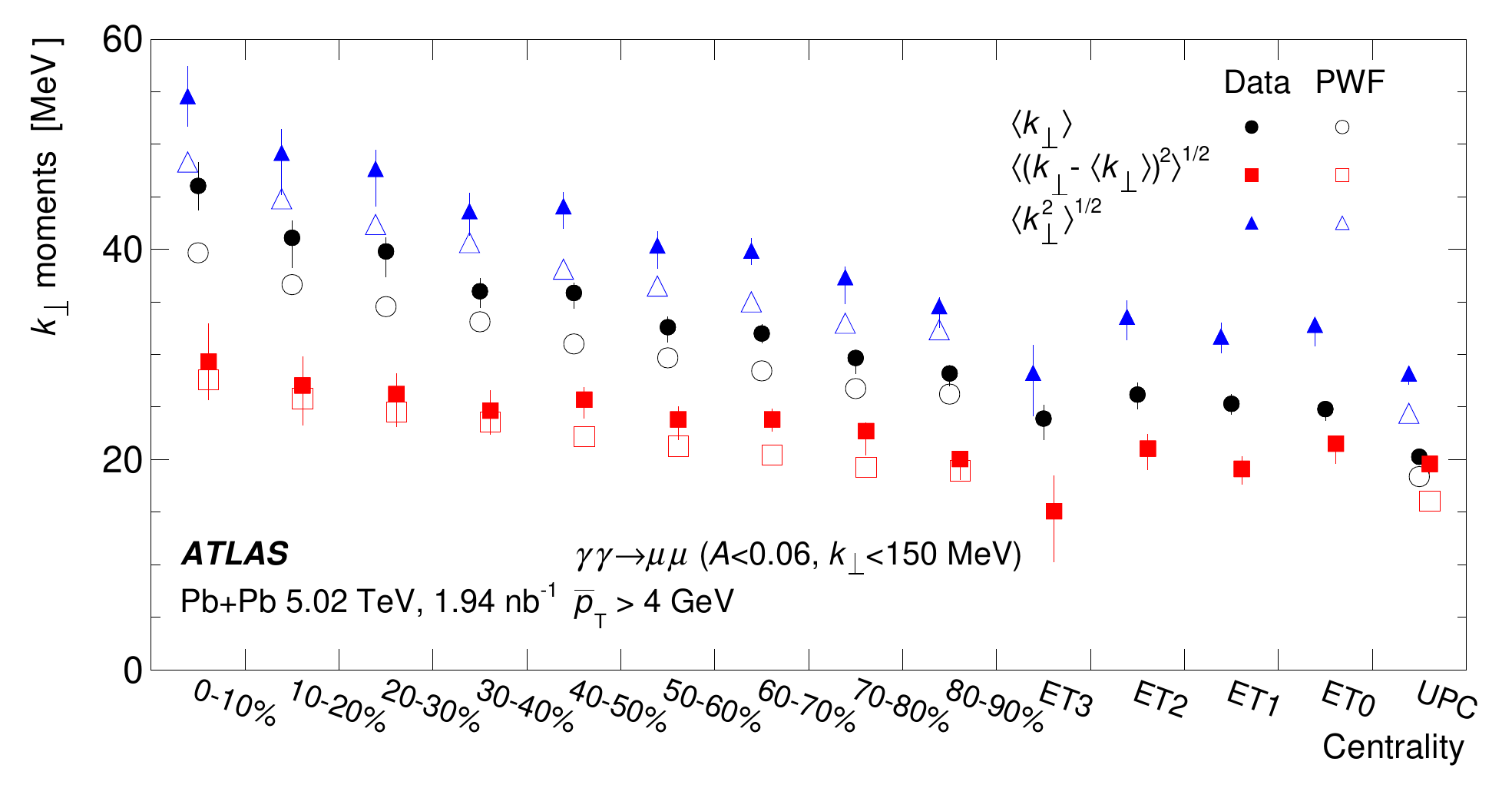}
\caption{
Moments of the \gamgammumu \kperp\ distributions as a function of
centrality compared with the PWF predictions.
The error bars indicate combined statistical and systematic
uncertainties. In some bins, the statistical uncertainties, determined
both by the number of counts and by their \kperp distribution, dominate.
\label{fig:moments}
}
\end{figure}

Figure~\ref{fig:moments} presents moments of the \kp\ distribution as a
function of centrality for the $\ptbar > 4$~\GeV\ selection.
The moments are calculated from the \kp\ distributions prior to the
background subtraction and then corrected, analytically, to remove the
contribution from the constant background. Results are shown for the
mean, the RMS, and the standard deviation of the \kp
distributions.
 
A significant increase in the mean is observed between the UPC and the
four ET0--ET3 intervals, and then there is a further steady increase in the more central
\PbPb\ collisions. Similar behavior is observed in the RMS values, but
the standard deviation shows a much smaller increase between UPC
collisions and the most central collisions. The measured moments are
compared in the figure with those obtained from the PWF calculation. The
PWF predictions reproduce many of the trends seen in the data, but the
mean and RMS values systematically lie below the data.
 
\begin{figure}[!tb]
\centering
\includegraphics[width=1.0\textwidth]{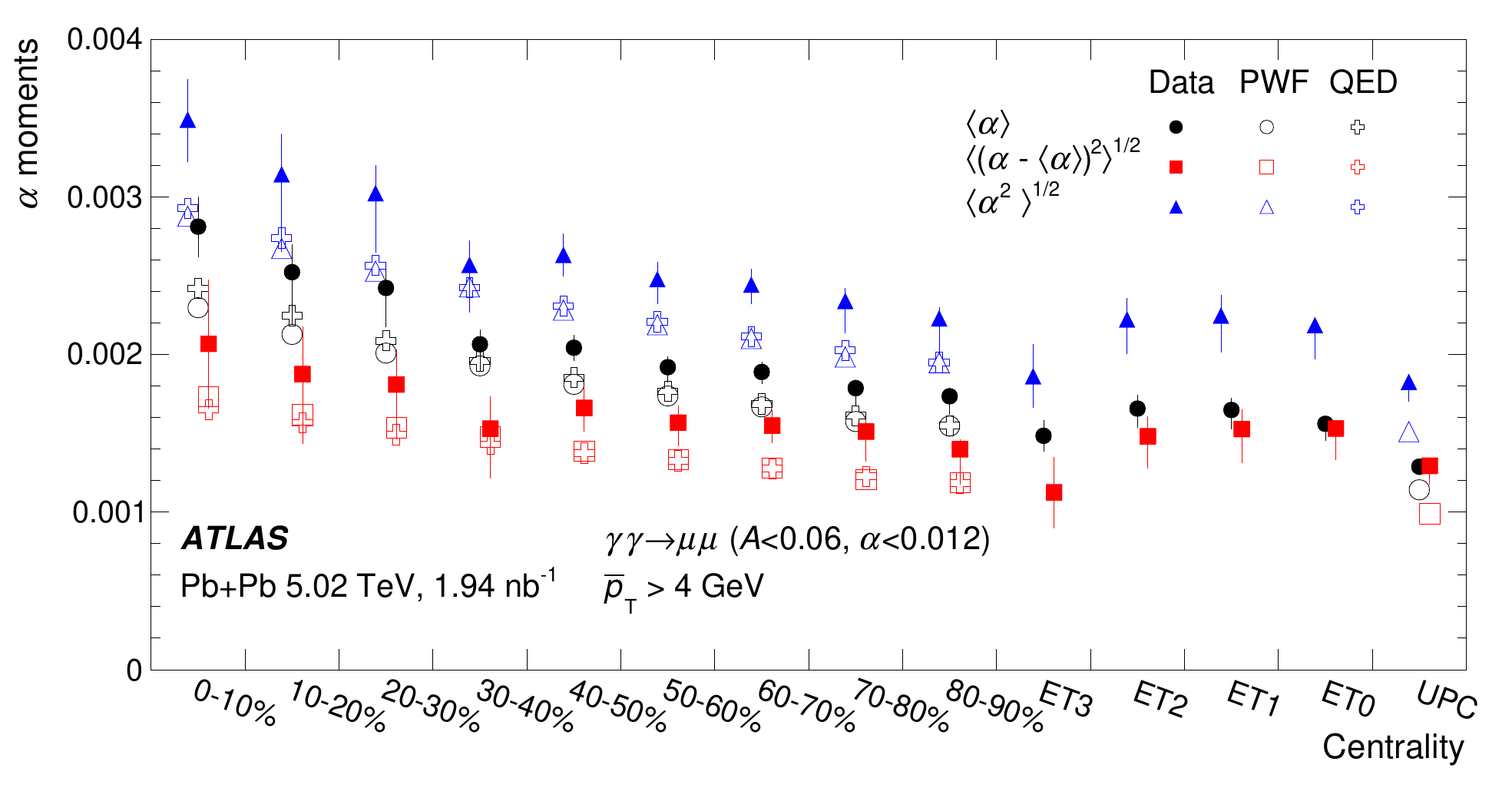}
\caption{
Moments of the \gamgammumu \Aco\ distributions as a function of
centrality compared with the QED and PWF predictions.
The error bars on the data points indicate combined statistical and
systematic uncertainties. In some bins the statistical uncertainties, determined
both by the number of counts and by their \Aco distribution, dominate.
\label{fig:momentsAco}
}
\end{figure}

Figure~\ref{fig:momentsAco} shows moments of the \Aco\ distributions
compared with both the PWF and QED calculations.
The moments calculated for the
QED and PWF predictions are in excellent agreement except for the most
central collisions where the QED results are slightly higher. Both
calculations show a systematic difference from the data in
more peripheral collisions, especially the 60--70\% and 70--80\%
intervals. However, in those intervals, both calculations show weaker
suppression near $\Aco = 0$ than is observed in the data.

\FloatBarrier
\subsection{Rapidity gap and event-plane dependence of \kperp}
\label{sec:ktrapgapdphi}
 
\begin{figure}[t]
\centerline{
\includegraphics[width=0.5\textwidth]{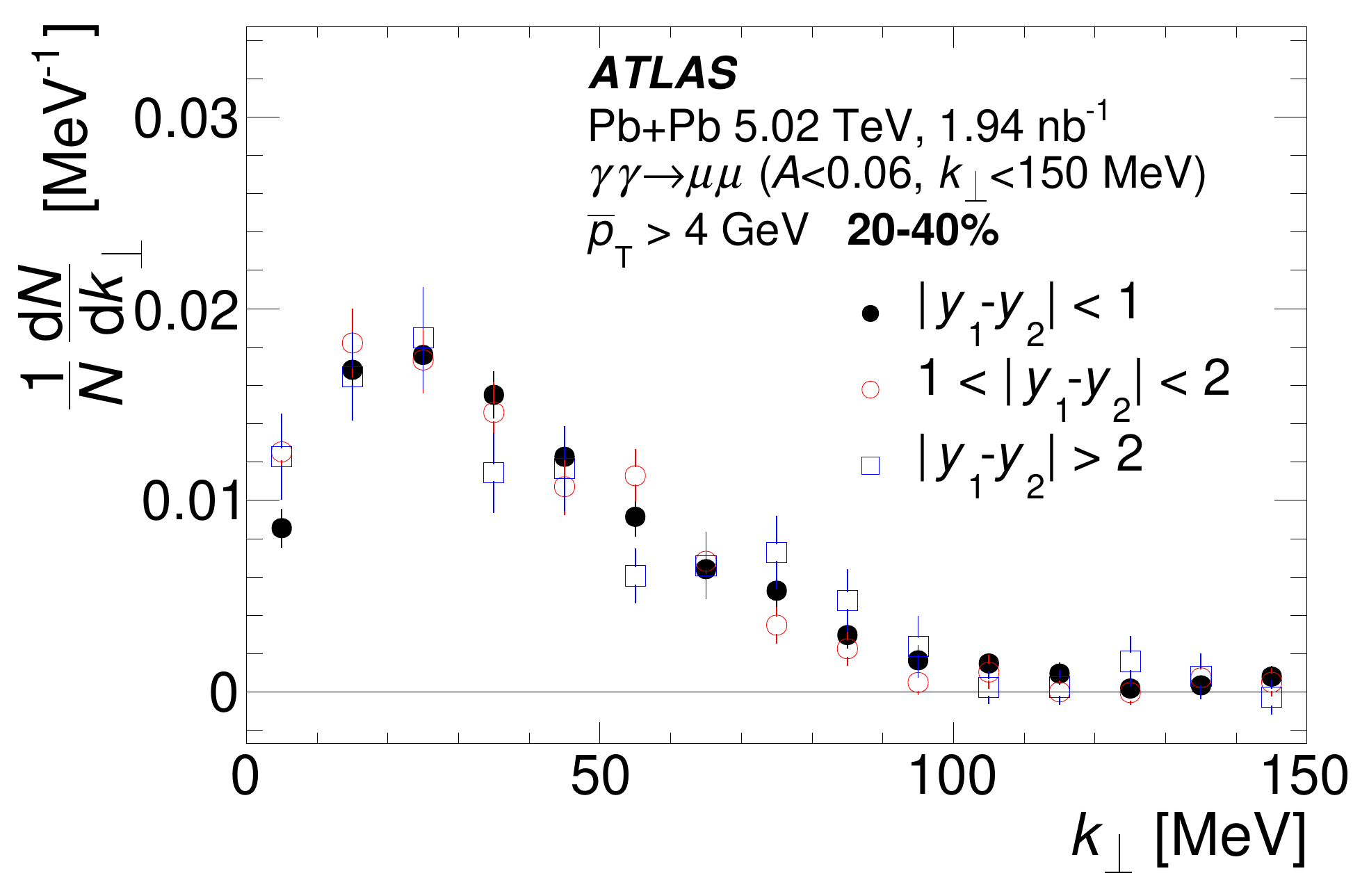}
\includegraphics[width=0.5\textwidth]{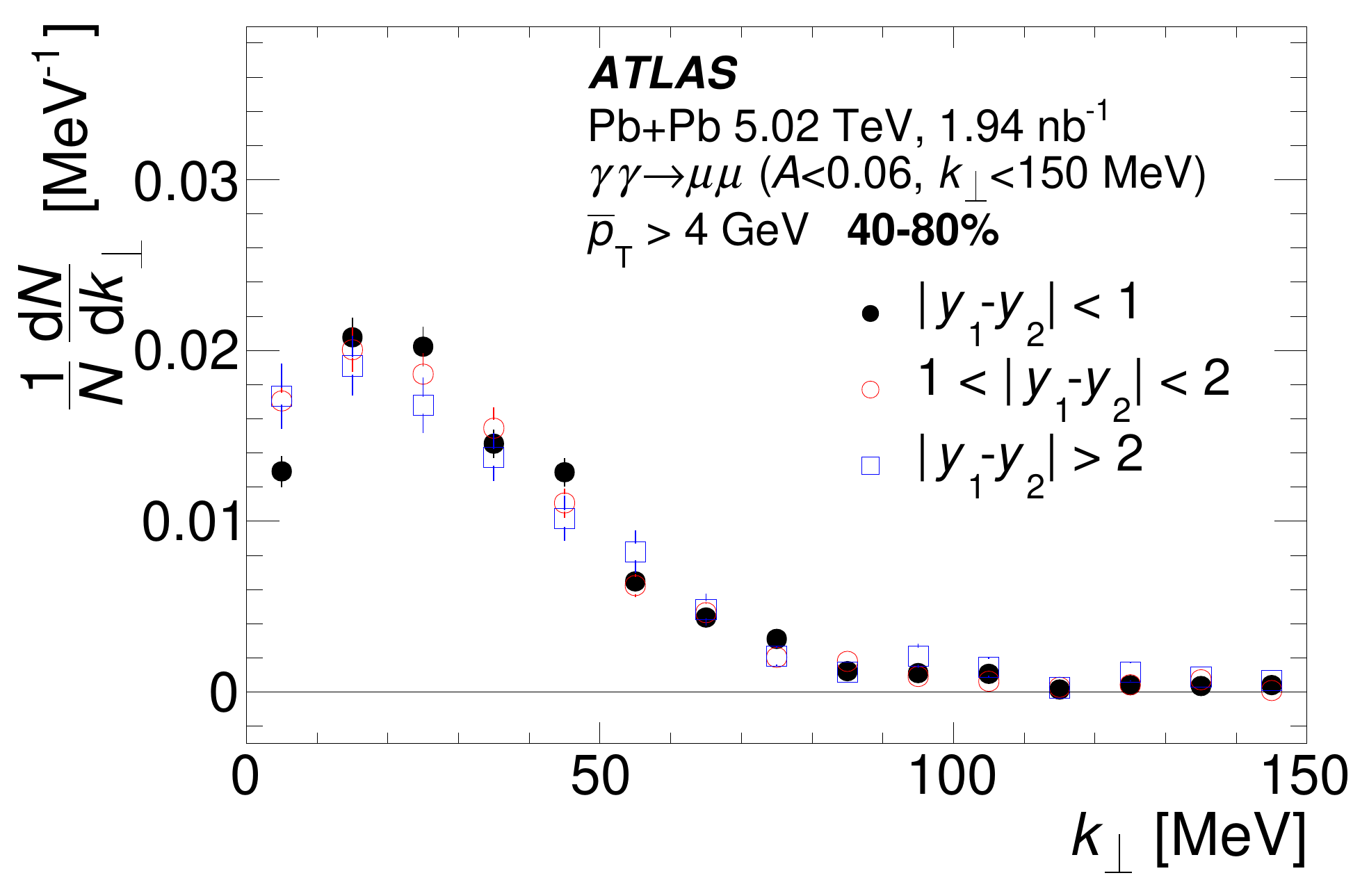}
}
\caption{
Dimuon \kp distributions  in  three intervals of \Dy\ in the
20--40\% (left) and 40--80\% (right) centrality intervals,
respectively. The distributions are self-normalized to allow the
distributions to be directly compared. The error bars show combined statistical
and systematic uncertainties.
\label{fig:rapgap}
}
\end{figure}
 
A specific prediction from Ref.~\cite{Klein:2018fmp} is that if the
centrality-dependent modifications of the \Aco\ distributions result
from the muons being deflected in magnetic fields generated during the
\PbPb\ collision, then the broadening should vary as the hyperbolic
tangent of  the rapidity difference between the two muons, $\Dy \equiv |y_1 - y_2|$, where $y_1$ and $y_2$
are the rapidities of  the two muons in the pair.
Figure~\ref{fig:rapgap} shows the \kp\ distributions for two centrality
intervals: 20--40\% and 40--80\%, and for three different \Dy\ ranges:
$\Dy \leq 1$, $1 < \Dy \leq 2$, and $\Dy > 2$. The results are presented
in self-normalized form to allow them to be compared directly, even though the yields vary with \Dy.
At most, a weak variation with \Dy is observed, but with a
dependence opposite to what would be expected from magnetic field effects. Namely, the suppression near $\kp =
0$ is greater for smaller \Dy, while a $\tanh{\Dy}$ dependence would
have the opposite behavior and should vary more rapidly with \Dy.
 
A second way to observe potential effects of magnetic fields on the
\gamgammumu pairs is to study the dependence of the \kp\ distributions on
the orientation of the dimuons in the transverse plane relative to
the direction of the second-order event plane. In \PbPb\ collisions
of intermediate centrality, the second-order event plane angle,
$\Psi_2$, is understood to be well-correlated with the direction of
the impact parameter vector, which is, in turn, correlated with the direction of
any magnetic fields generated during a \PbPb\ collision.
 
\begin{figure}[!b]
\centering
\includegraphics[width=0.5\textwidth]{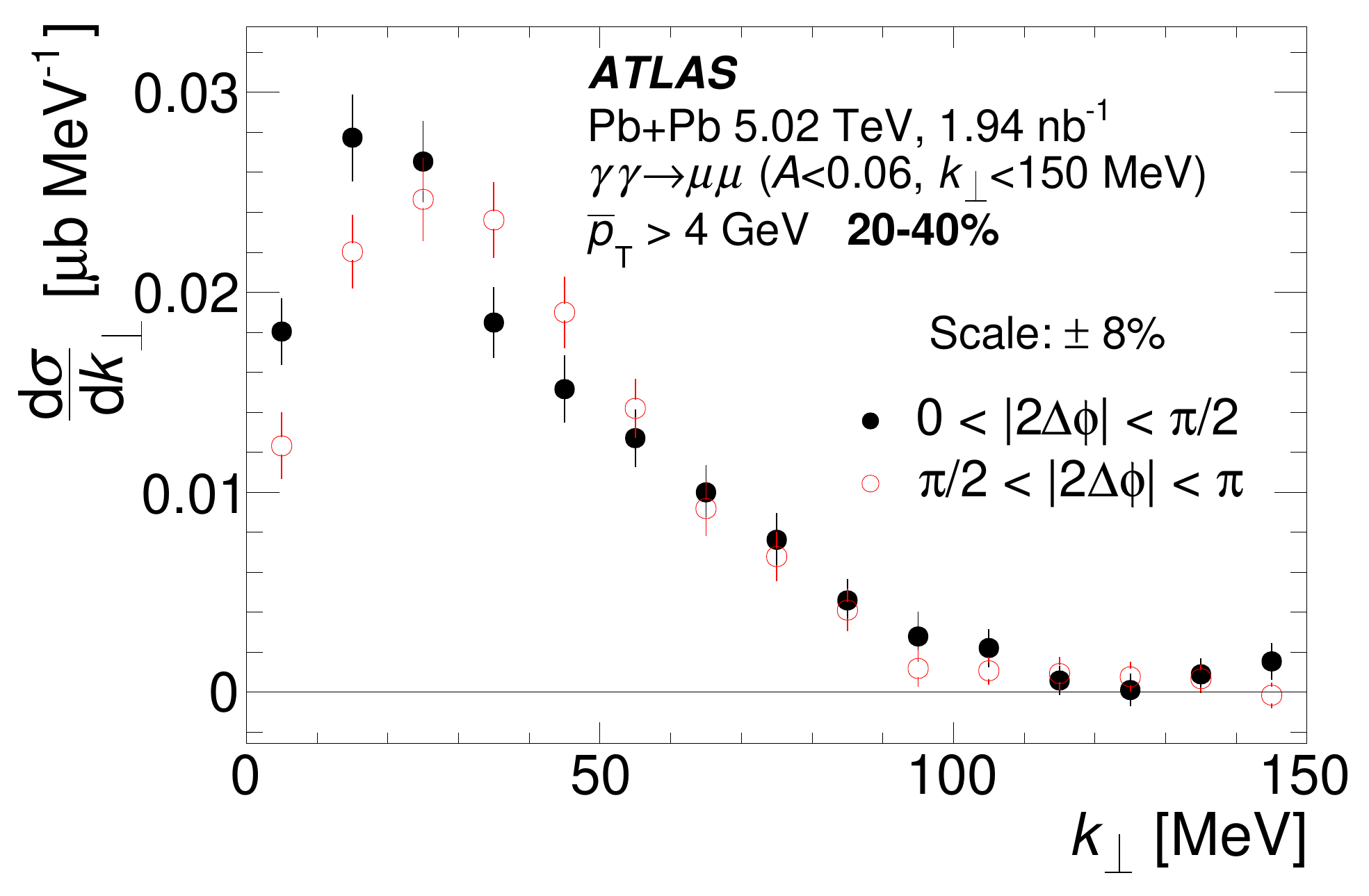}
\includegraphics[width=0.5\textwidth]{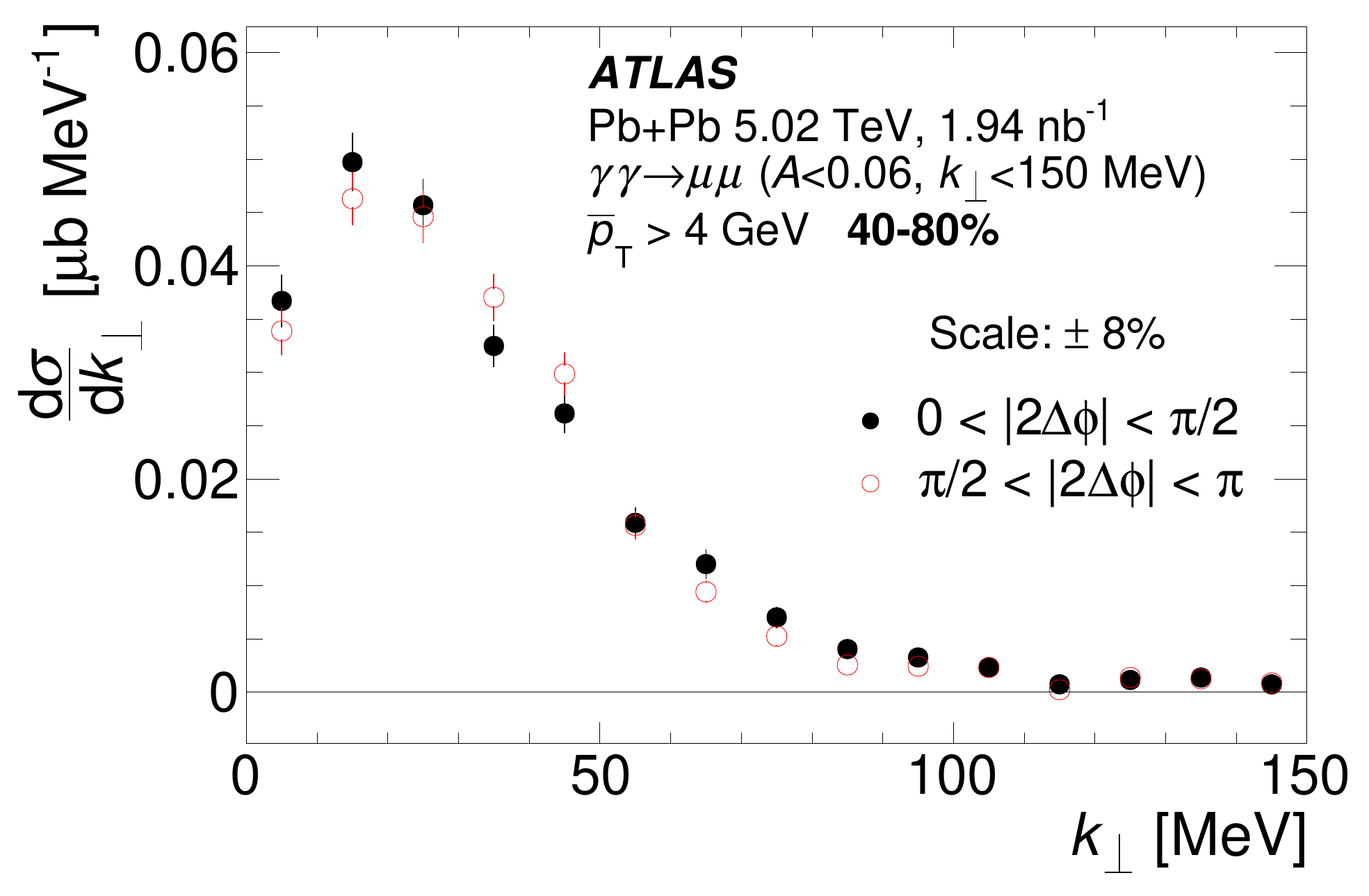}
\caption{
Differential cross-sections for \gamgammumu\ production as a function of \kp for two
intervals of \dphi: $0 < \dphi < \pi/2$ and $\pi/2 < \dphi < \pi$ in the
20--40\% (top) and 40--80\% (bottom)
centrality intervals, respectively.
The error bars show combined statistical
and systematic uncertainties.
\label{fig:yields_afterSubtraction_ep_angle_dep}
}
\end{figure}
 
Dimuon yields were measured as a function of the
quantity $\dphi \equiv \left|2\left(\phi_{\mu\mu}-\Psi_2\right)\right|$,
where $\phi_{\mu\mu}$ represents the azimuthal orientation of the
dimuon, and $\Psi_2$ is the second-order event plane angle obtained
from the FCal using methods applied in previous ATLAS flow
analyses~\cite{HION-2016-01,HION-2019-03}.
The angle, $\phi_{\mu\mu}$, is calculated by rotating one of the muon azimuthal
angles by $\pi$ and then averaging the result with the other muon,
\[ \phi_{\mu\mu} = \frac{1}{2}\left(\phi_1 + \pi +\phi_2\right).\]
Figure~\ref{fig:yields_afterSubtraction_ep_angle_dep}
shows differential cross-sections as a function of \kp\ in two intervals of
\dphi, $\dphi < \pi/2$ and $\pi/2 \leq  \dphi < \pi$,
for two centrality intervals.
A significant variation with \dphi is seen in the depletion near $\kp =0$
for the 20--40\% centrality interval.
Otherwise, the distributions are indistinguishable at larger \kp values,
which means that the overall broadening of the dimuon \kp distributions
is the same in the two \dphi intervals.
 
\begin{figure}[!t]
\centerline{
\includegraphics[width=0.5\textwidth]{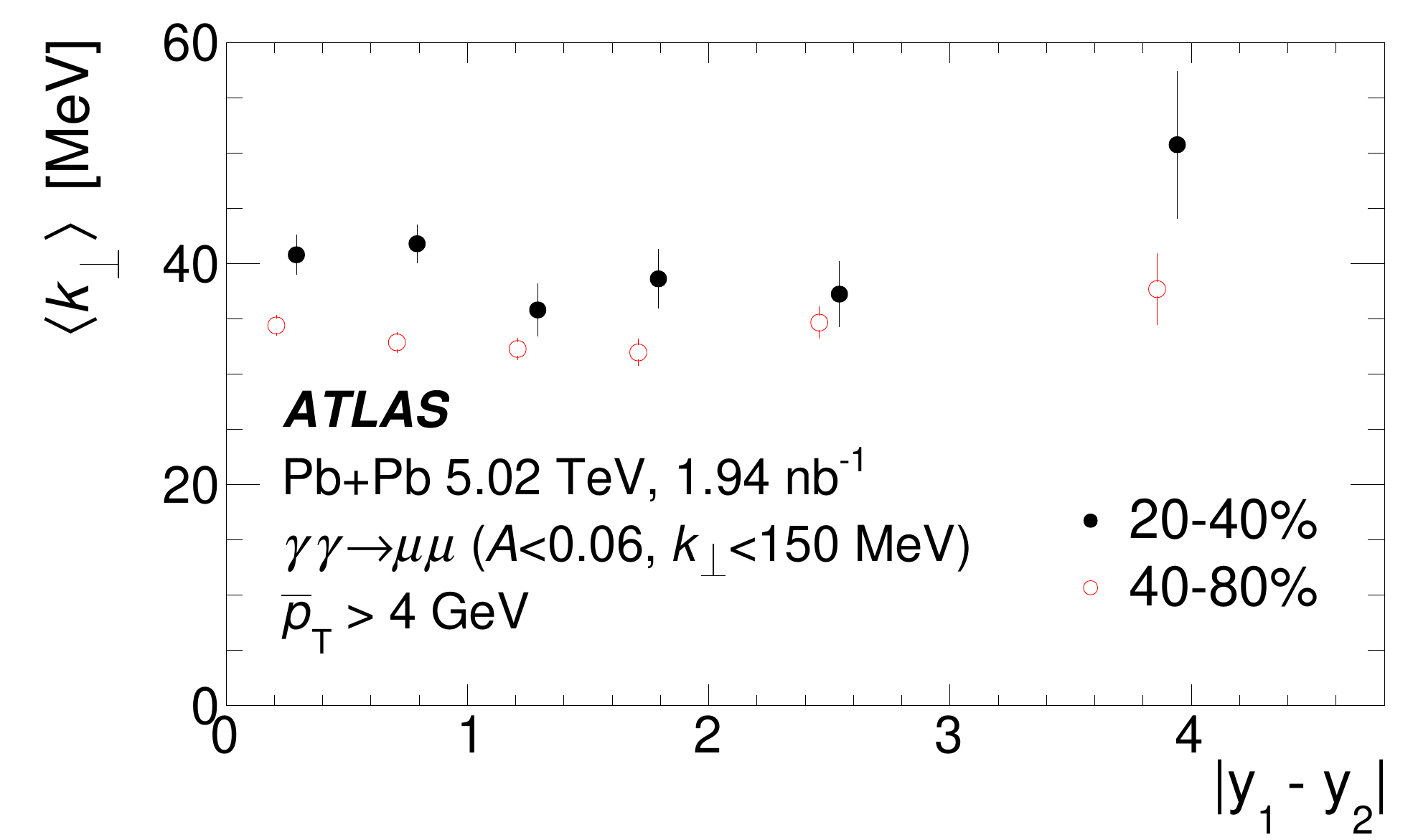}
\includegraphics[width=0.5\textwidth]{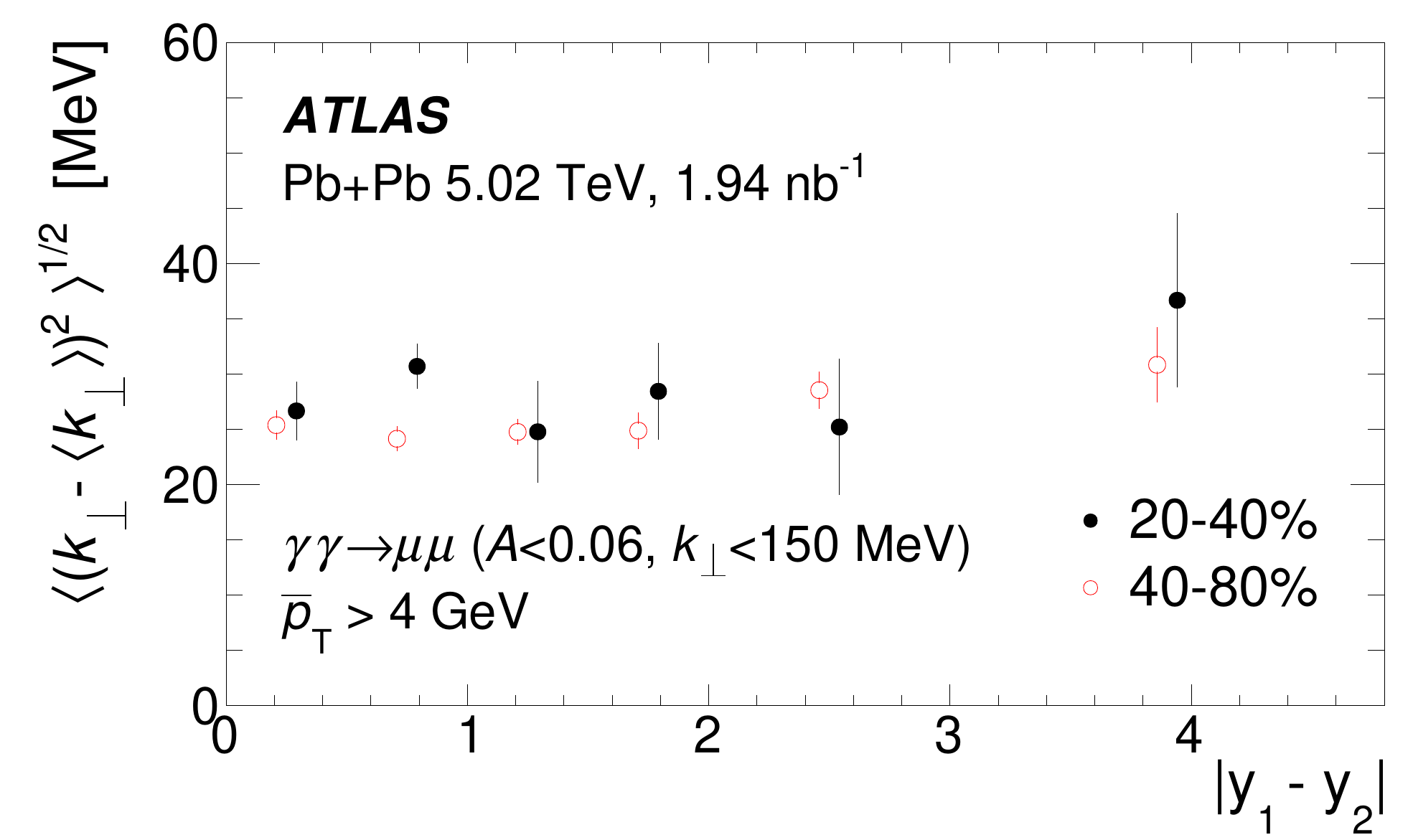}
}
\centerline{
\includegraphics[width=0.5\textwidth]{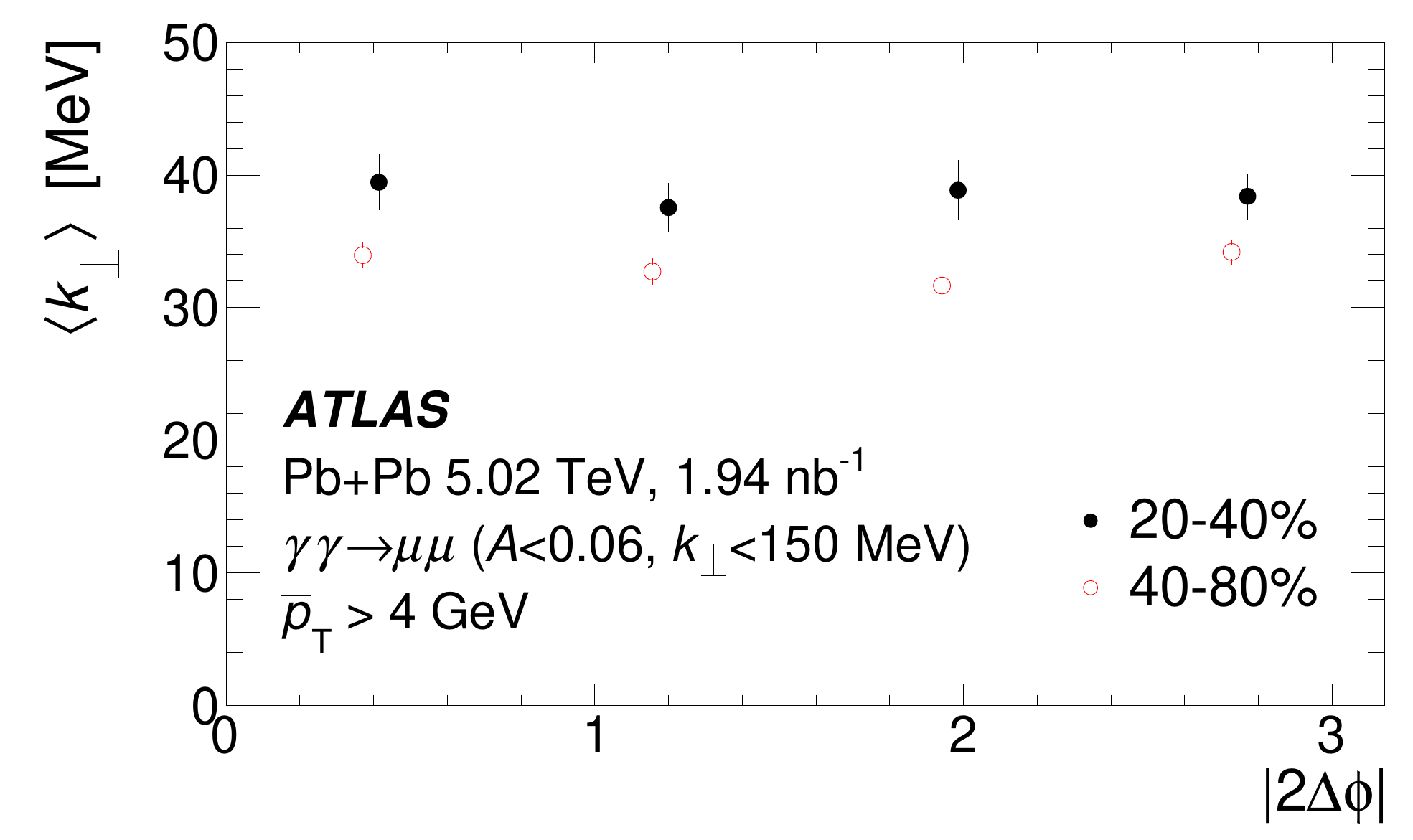}
\includegraphics[width=0.5\textwidth]{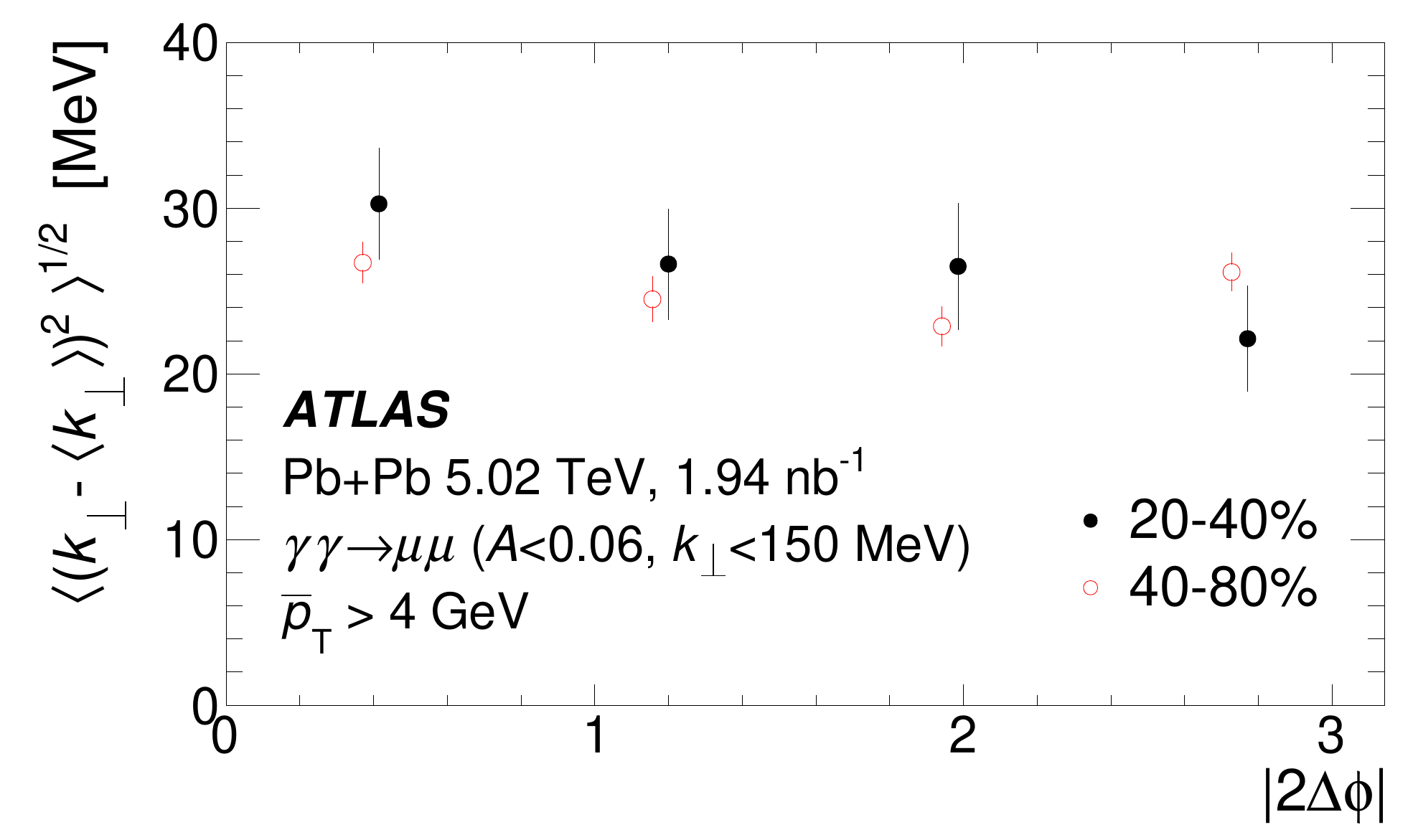}
}
\caption{
The average (left) and the standard deviation (right) of the \gamgammumu
\kp distributions in the 20--40\% and 40--80\% centrality intervals as
a function of  \Dy (top) and \dphi (bottom).
Results are shown for \Dy\ intervals of width 0.5 for $\Dy < 2.0$,
and two additional intervals covering $2.0 \leq \Dy < 3.0$ and $3.0 \leq \Dy < 4.8$.
The \dphi\ results are shown for four equal intervals covering $[0,\pi]$.
The error bars show combined statistical and systematic uncertainties.
The points are staggered for clarity of presentation.
}
\label{fig:kpmoments_ep}
\end{figure}
 
To further explicate the results presented in
Figures~\ref{fig:rapgap} and
\ref{fig:yields_afterSubtraction_ep_angle_dep},
Figure~\ref{fig:kpmoments_ep} shows results for the
mean and standard deviation of the dimuon \kp distributions as a
function of \Dy (top) and \dphi (bottom) using finer bins in both
variables.
The average \kp values have, at most, a
weak dependence on \Dy\ or \dphi, while the standard
deviations of the \kp distributions are constant within
uncertainties.
Thus, the mechanism responsible for the broadening of the \kp
distributions does not appear to depend on either \Dy or the direction of
the muons relative to the impact parameter vector in the
transverse plane. The absence of variation with \Dy\ by values $> 4$~\MeV\
of either the mean or standard deviation of the \kp distributions
rules out magnetic broadening as a significant contribution to the observed
modifications of the \kp distributions.  However, the mechanism
responsible for the suppression at $\kp = 0$ may vary with \Dy or
\dphi. Indeed, the \kp distributions in Figures~\ref{fig:rapgap} and
\ref{fig:yields_afterSubtraction_ep_angle_dep} indicate such
dependence, although the suppression near $\kp = 0$ is not easily seen
in the calculated moments.

% End of text imported from the .//tex/results.tex input file

\FloatBarrier
\section{Conclusion}
\label{sec:conclusion}

% The next lines are included from the .//tex/conclusions.tex input file
ATLAS has measured dimuon production via \gamgam\ scattering
processes in non-ultraperipheral \PbPb\ collisions at
$\sqn=5.02$~\TeV.
The measurements use data from the 2015 and 2018
\PbPb\ runs at the LHC corresponding to an integrated luminosity
of $1.93~\mathrm{nb}^{-1}$.
Backgrounds, dominated by heavy-flavor decays, are evaluated using
template fits to the distribution of muon-pair \dz\ values. A much smaller
background from DY processes, estimated using \powheg
calculations implemented with nNNPDF2.0 nuclear PDFs, is subtracted from the data.
Cross-sections and \NormYields\ of \gamgammumu\ pairs are measured
as a function of pair rapidity, \ptbar, and centrality.
The cross-sections vary weakly with centrality, decreasing from central
to peripheral collisions. The \starlight model, which was recently
augmented to allow evaluation of cross-sections for
(e.g.) \gamgammumu production within restricted
impact parameter intervals,  substantially
underestimates the measured cross-sections.

Measurements of the \Aco\ and \kp\ distributions show a significant
centrality dependence consistent with the results in
Ref.~\cite{HION-2018-11}.
However, with the improved statistical precision
of this measurement, an additional depletion is observed in the
acoplanarity and \kperp distributions near zero values of the
corresponding quantities.
 
The \Aco\ distributions are observed to vary strongly with \ptbar, while
distributions of \kp\ are observed to be approximately independent of
\ptbar, as would be expected if the broadening is determined by a
physics effect with an intrinsic transverse momentum scale.
The \kp distributions are, thus, better suited for studying the
modifications of the dimuon alignment if the study integrates over any finite
interval of \ptbar.
 
Comparisons of the measured \Aco\ and \kp\ distributions with two
theoretical calculations are presented.
Both the QED and the photon Wigner function-based calculations
quantitatively reproduce the centrality-dependent
broadening observed  in the data.
While both calculations also predict a suppression near $\Aco = 0$
and/or $\kp = 0$, the QED calculations describe
the data more accurately; the PWF results show significantly
less depletion than observed in the data.
 
Measurements of the \kp\ distributions as a function of \Dy and
\dphi suggest that the broadening of the \Aco and \kp
distributions does not have significant contribution from interactions
of the muons with magnetic fields generated in the quark--gluon plasma.
Specifically, the data show no dependence of the broadening on \Dy, while a
$\tanh{\Dy}$ dependence is predicted.
Similarly, no dependence of the standard deviation of the \kp distributions
on the orientation of the muon pair relative to the second-order
event plane is observed.
This result would be inconsistent with magnetic broadening in a field
whose direction in the transverse plane is correlated with the
impact parameter vector.
While the overall broadening of the distributions does not depend
on \Dy or \dphi, the magnitude of the depletion at small \kp appears
to depend on both quantities.

% End of text imported from the .//tex/conclusions.tex input file

\section*{Acknowledgments}

% The next lines are included from the .//acknowledgements/Acknowledgements.tex input file

We thank CERN for the very successful operation of the LHC, as well as the
support staff from our institutions without whom ATLAS could not be
operated efficiently.
 
We acknowledge the support of
ANPCyT, Argentina;
YerPhI, Armenia;
ARC, Australia;
BMWFW and FWF, Austria;
ANAS, Azerbaijan;
CNPq and FAPESP, Brazil;
NSERC, NRC and CFI, Canada;
CERN;
ANID, Chile;
CAS, MOST and NSFC, China;
Minciencias, Colombia;
MEYS CR, Czech Republic;
DNRF and DNSRC, Denmark;
IN2P3-CNRS and CEA-DRF/IRFU, France;
SRNSFG, Georgia;
BMBF, HGF and MPG, Germany;
GSRI, Greece;
RGC and Hong Kong SAR, China;
ISF and Benoziyo Center, Israel;
INFN, Italy;
MEXT and JSPS, Japan;
CNRST, Morocco;
NWO, Netherlands;
RCN, Norway;
MEiN, Poland;
FCT, Portugal;
MNE/IFA, Romania;
MESTD, Serbia;
MSSR, Slovakia;
ARRS and MIZ\v{S}, Slovenia;
DSI/NRF, South Africa;
MICINN, Spain;
SRC and Wallenberg Foundation, Sweden;
SERI, SNSF and Cantons of Bern and Geneva, Switzerland;
MOST, Taiwan;
TENMAK, T\"urkiye;
STFC, United Kingdom;
DOE and NSF, United States of America.
In addition, individual groups and members have received support from
BCKDF, CANARIE, Compute Canada and CRC, Canada;
PRIMUS 21/SCI/017 and UNCE SCI/013, Czech Republic;
COST, ERC, ERDF, Horizon 2020 and Marie Sk{\l}odowska-Curie Actions, European Union;
Investissements d'Avenir Labex, Investissements d'Avenir Idex and ANR, France;
DFG and AvH Foundation, Germany;
Herakleitos, Thales and Aristeia programmes co-financed by EU-ESF and the Greek NSRF, Greece;
BSF-NSF and MINERVA, Israel;
Norwegian Financial Mechanism 2014-2021, Norway;
NCN and NAWA, Poland;
La Caixa Banking Foundation, CERCA Programme Generalitat de Catalunya and PROMETEO and GenT Programmes Generalitat Valenciana, Spain;
G\"{o}ran Gustafssons Stiftelse, Sweden;
The Royal Society and Leverhulme Trust, United Kingdom.
 
The crucial computing support from all WLCG partners is acknowledged gratefully, in particular from CERN, the ATLAS Tier-1 facilities at TRIUMF (Canada), NDGF (Denmark, Norway, Sweden), CC-IN2P3 (France), KIT/GridKA (Germany), INFN-CNAF (Italy), NL-T1 (Netherlands), PIC (Spain), ASGC (Taiwan), RAL (UK) and BNL (USA), the Tier-2 facilities worldwide and large non-WLCG resource providers. Major contributors of computing resources are listed in Ref.~\cite{ATL-SOFT-PUB-2021-003}.

% End of text imported from the .//acknowledgements/Acknowledgements.tex input file

\printbibliography
 
\clearpage
% ATLAS Collaboration author list
% Reference date of HION-2020-10 is 2022-02-17
% Author list last updated on date 02-MAY-23
% Data extracted on 02-May-2023 for paper reference HION-2020-10
% at 8:42pm
 
\begin{flushleft}
\hypersetup{urlcolor=black}
{\Large The ATLAS Collaboration}

\bigskip

\AtlasOrcid[0000-0002-6665-4934]{G.~Aad}$^\textrm{\scriptsize 101}$,
\AtlasOrcid[0000-0002-5888-2734]{B.~Abbott}$^\textrm{\scriptsize 119}$,
\AtlasOrcid[0000-0002-7248-3203]{D.C.~Abbott}$^\textrm{\scriptsize 102}$,
\AtlasOrcid[0000-0002-1002-1652]{K.~Abeling}$^\textrm{\scriptsize 55}$,
\AtlasOrcid[0000-0002-8496-9294]{S.H.~Abidi}$^\textrm{\scriptsize 29}$,
\AtlasOrcid[0000-0002-9987-2292]{A.~Aboulhorma}$^\textrm{\scriptsize 35e}$,
\AtlasOrcid[0000-0001-5329-6640]{H.~Abramowicz}$^\textrm{\scriptsize 150}$,
\AtlasOrcid[0000-0002-1599-2896]{H.~Abreu}$^\textrm{\scriptsize 149}$,
\AtlasOrcid[0000-0003-0403-3697]{Y.~Abulaiti}$^\textrm{\scriptsize 116}$,
\AtlasOrcid[0000-0003-0762-7204]{A.C.~Abusleme~Hoffman}$^\textrm{\scriptsize 136a}$
% ...
\bigskip
\\

% ...
$^{29}$Physics Department, Brookhaven National Laboratory, Upton NY; United States of America.\\
% ...
$^{35}$$^{(a)}$Facult\'e des Sciences Ain Chock, R\'eseau Universitaire de Physique des Hautes Energies - Universit\'e Hassan II, Casablanca;$^{(b)}$Facult\'{e} des Sciences, Universit\'{e} Ibn-Tofail, K\'{e}nitra;$^{(c)}$Facult\'e des Sciences Semlalia, Universit\'e Cadi Ayyad, LPHEA-Marrakech;$^{(d)}$LPMR, Facult\'e des Sciences, Universit\'e Mohamed Premier, Oujda;$^{(e)}$Facult\'e des sciences, Universit\'e Mohammed V, Rabat;$^{(f)}$Institute of Applied Physics, Mohammed VI Polytechnic University, Ben Guerir; Morocco.\\
% ...
$^{55}$II. Physikalisches Institut, Georg-August-Universit\"{a}t G\"ottingen, G\"ottingen; Germany.\\
% ...
$^{101}$CPPM, Aix-Marseille Universit\'e, CNRS/IN2P3, Marseille; France.\\
$^{102}$Department of Physics, University of Massachusetts, Amherst MA; United States of America.\\
% ...
$^{116}$Department of Physics, New York University, New York NY; United States of America.\\
% ...
$^{119}$Homer L. Dodge Department of Physics and Astronomy, University of Oklahoma, Norman OK; United States of America.\\
% ...
$^{136}$$^{(a)}$Departamento de F\'isica, Pontificia Universidad Cat\'olica de Chile, Santiago;$^{(b)}$Millennium Institute for Subatomic physics at high energy frontier (SAPHIR), Santiago;$^{(c)}$Instituto de Investigaci\'on Multidisciplinario en Ciencia y Tecnolog\'ia, y Departamento de F\'isica, Universidad de La Serena;$^{(d)}$Universidad Andres Bello, Department of Physics, Santiago;$^{(e)}$Instituto de Alta Investigaci\'on, Universidad de Tarapac\'a, Arica;$^{(f)}$Departamento de F\'isica, Universidad T\'ecnica Federico Santa Mar\'ia, Valpara\'iso; Chile.\\
% ...
$^{149}$Department of Physics, Technion, Israel Institute of Technology, Haifa; Israel.\\
$^{150}$Raymond and Beverly Sackler School of Physics and Astronomy, Tel Aviv University, Tel Aviv; Israel.\\
% ...
$^{a}$ Also Affiliated with an institute covered by a cooperation agreement with CERN.\\
$^{b}$ Also at Borough of Manhattan Community College, City University of New York, New York NY; United States of America.\\
$^{c}$ Also at Bruno Kessler Foundation, Trento; Italy.\\
$^{d}$ Also at Center for High Energy Physics, Peking University; China.\\
$^{e}$ Also at Centro Studi e Ricerche Enrico Fermi; Italy.\\
$^{f}$ Also at CERN, Geneva; Switzerland.\\
% ...

\end{flushleft}

% Created with Glance <Atlas.Glance@cern.ch>

\end{document}